\documentclass[aps,prb,twocolumn,superscriptaddress,longbibliography,10pt]{revtex4-2}
\usepackage{graphicx}
\usepackage{hyperref}
\usepackage{amssymb}
\usepackage{siunitx}
\usepackage{dcolumn}
\usepackage{float}
\usepackage{booktabs}
\usepackage{amsmath} 
\usepackage[utf8]{inputenc}
\usepackage{lmodern}
\usepackage{color}
\definecolor{red}{rgb}{1.0,0.0,0.0}
\usepackage{makecell}
\usepackage{marginnote}

\usepackage{natbib}

\newcommand{\e}{\mathrm{e}}

\DeclareMathAlphabet{\bi}{OML}{cmm}{b}{it}

\def\ba{\begin{aligned}}
	\def\ea{\end{aligned}}
\def\be{\begin{equation}}
	\def\ee{\end{equation}}
\def\bearr{\begin{eqnarray}}
	\def\eearr{\end{eqnarray}}

\def\l{\left}
\def\r{\right}

\usepackage{tikz}

\usepackage{xcolor}
\pagecolor{white}

\usepackage{xcolor}
\usepackage{physics}
\color{black}

\hypersetup{
	colorlinks=true,
	linkcolor=blue,
	citecolor=blue,
	urlcolor=blue
}

\begin{document}
	\title{Photodriven germanium hole qubit}
	\bigskip
	\author{Bashab Dey and John Schliemann\\
		Institute of Theoretical Physics, University of Regensburg, Regensburg, Germany}
	\begin{abstract}
 Hole qubits in germanium quantum dots are promising candidates for coherent control and manipulation of the spin degree of freedom through electric dipole spin resonance. We theoretically
study the time dynamics of a single heavy-hole qubit in a laser-driven planar germanium quantum
dot confined laterally by a harmonic potential in presence of linear and cubic Rashba spin-orbit
couplings and an out-of-plane magnetic field. We obtain an approximate analytical formula of the
Rabi frequency using a Schrieffer-Wolff transformation and establish a connection of our model with
the ESDR results obtained for this system. For stronger beams, we employ different methods such
as unitary transformation and Floquet theory to study the time evolution numerically. We observe
that high radiation intensity is not suitable for the qubit rotation due to the presence of high frequency noise superimposed on the Rabi oscillations. We display the Floquet spectrum and highlight
the quasienergy levels responsible for the Rabi oscillations in the Floquet picture. We study the
interplay of both the types of Rashba couplings and show that the Rabi oscillations, which are
brought about by the linear Rashba coupling, vanish for typical values of the cubic Rashba coupling
in this system.
	\end{abstract}
	
	\maketitle

    \section{Introduction}

Qubit is the basic unit of quantum information. 
The search for materials where qubits can be fast-operated, efficiently controlled and well shielded from the environment has been at the forefront of research. Solid state spin qubits, such as in Si, Ge and III-V semiconductor heterostructures, have attracted immense attention in recent times as nanometer-scale quantum devices can be lithographically fabricated onto them, creating an isolated environment for the spins to achieve long coherence times \cite{buckard-rmp,chatterjee-review,david-spintronics,zhang-rev}. Furthermore, technological advancements in microelectronics and production of high quality Si hold great prospect of building a scalable semiconductor platform to develop quantum computers which may host millions of qubits \cite{kane,fowler}.

The Zeeman-split electronic spin states are suitable candidates for single and multi-qubit operations, as proposed by Loss and DiVincenzo \cite{LD}. Single qubit gates such as $\textit{Pauli}$ gates have been realized in Si and GaAs by inducing Rabi oscillations between the spin-up and -down states with the help of electron spin resonance (ESR) \cite{esr-koppens,koppens-echo,esr-pla,esr-veld1,esr-veld2}. 
However, ESR offers experimental roadblocks from the point of view of scaling as magnetic fields are difficult to localize in miniature landscapes. On the other hand, electrical driving is easier to implement locally through the application of ac voltage across the \textit{in-situ} gate electrodes. Modulation of $g$-tensor \cite{kato-gfactor,salis-gfactor-coherence,deacon-gfactor-InAs,pingenot-gfactor-InGaAs,ferr-gfactor-aniso}, slanting-Zeeman field ESR \cite{Ladri-slanting,brunner-2qubit-edsr,yoneda-99pc,zajac-cnot} and electric dipole spin resonance (EDSR) \cite{nowack-electric,Nadj-electric} have proved to be reliable techniques for achieving coherent qubit control through pure electrical drives. Among these, EDSR is of particular interest as it harnesses the spin-orbit coupling (SOC) of the material to perform the qubit rotations. The mechanism of EDSR has been extensively studied over the years in quantum wells \cite{rashba-edsr}, planar quantum dots with electron- \cite{edsr-loss-orig} or hole-qubits \cite{edsr-loss-hole}, TMD monolayers \cite{edsr-mos2} and double quantum dots \cite{sherman}.
 The electronic spin-qubits in 2D GaAs quantum dot \cite{nowack-electric} and InSb nanowire \cite{Nadj-electric} have successfully exhibited single-spin EDSR. 
 
The electron spin qubits are prone to decoherence and relaxations due to interaction with phonons \cite{phonon1,phonon2,phonon3,phonon4}  in presence of SOC and contact-hyperfine interaction with the sea of nuclear spins \cite{hyperfine1,hyperfine2,hyperfine3}. The latter can be substantially minimized in group IV semiconductors such as Si and Ge as they can be engineered into nuclear-spin free materials by isotopic purification \cite{Si-isotope, Si-isotope2,Ge-isotope}. In recent times, hole spin-qubits have emerged as viable alternatives to the electronic counterparts \cite{hole-review}. The suppressed hyperfine interaction due to the $p$-nature of the hole wave function leads to longer coherence times \cite{hole-hyperfine1,hole-hyperfine2,hole-hyperfine3,hole-hyperfine4}. Moreover, the valley degeneracy which stands as an obstacle in using Si electrons as  spin-qubits \cite{valley}, is absent for holes. The most attractive feature is, however, the stronger SOC of holes as compared to that of conduction electrons, which facilitates faster EDSR \cite{edsr-loss-hole}. However, it can also lead to stronger decoherence through spin-phonon interactions \cite{hole-loss}. 

Although Si might appear to be the natural choice for hole spin-qubits, it is Ge that provides some of the most desirable features for qubit control \cite{Ge-review}. The smaller effective mass of holes in Ge \cite{hole-mass} relaxes nanofabrication requirements, as the quantum dots are larger than in Si. Since Ge is heavier than Si, it has stronger SOC \cite{ge-heavy} which is desirable for faster qubit operations. Single-hole qubit rotations have been successfully demonstrated in planar Ge quantum wells and nanowires using EDSR \cite{ge-hole-exp1,ge-hole-exp2,ge-hole-exp3,ge-hole-exp4}. The  2D holes of Ge exhibit $p$-$cubic$ Rashba SOC \cite{winkler,winkler2} consisting of cubic- $(\propto p_+ p_- p_+ \sigma_+ + \text{h.c.})$ and spherically-symmetric terms $(\propto p_+^3 \sigma_- + \text{h.c.})$, out of which the latter dominates. The Dresselhaus SOC is absent due to bulk-inversion symmetry of the crystal. Several theoretical studies have attributed the EDSR in planar Ge quantum dots to the cubic-symmetric component of the SOC \cite{strained-Ge,optimal-Ge} in presence of an out-of-plane magnetic field. The electrical operation of planar Ge hole spin qubits in an in-plane magnetic field has also been studied theoretically \cite{in-plane}.  However, it has been recently argued that the cubic-symmetric component is negligibly small and a $p$-$linear$ \textit{direct} Rashba SOC, which exists in [001]-oriented Ge/Si quantum wells, is indeed responsible for the EDSR \cite{emergent}. Its origins are attributed to the local $C_{2v}$ interface \cite{PhysRevB.54.5852,PhysRevB.92.165301,PhysRevB.69.115333,PhysRevB.89.075430} and is deduced by performing atomistic pseudopotential method calculations \citep{emergence,emergent}.  Further studies have also claimed the existence of a $p$-linear Dresselhaus SOC in these heterostructures originating from heavy-hole/light-hole mixings \cite{Dresselhaus}. A specific kind of $p$-linear Rashba SOC can also be induced by moving the dot across inhomogeneous strain fields, which along with $g$-factor modulations results in faster Rabi rotations \cite{special-rashba}. Inhomogenous and inseparable electric fields can also lead to a different type of SOC mechanism that supports hole manipulation with in-plane magnetic fields \cite{inhomogeneous}.

In the last two decades, there have been significant developments in intense-ultrafast laser spectroscopy \cite{ultrafast1,ultrafast2,ultrafast3}, which gave birth to Floquet engineering. The previous works on EDSR with hole spin-qubits dealt with the application of electric pulses through gate electrodes only. The study of EDSR with laser pulses or optical EDSR is still missing. Secondly, the EDSR problem was only treated perturbatively and the effects of stronger electric fields were not addressed. Thirdly, the previous works did not take into account the effect of simultaneous presence of both linear and cubic Rashba SOC in determining the nature of EDSR. In this work, we make a comprehensive study of all the above aspects by studying EDSR of
2D HH states of Ge driven by a circularly polarized laser beam. We consider the dominant forms of SOC in this system viz. the $p$-linear and the spherically symmetric component of $p$-cubic Rashba SOC. We show that the effect of laser field is equivalent to that of the usual gate-driven EDSR setup. For small linear Rashba parameter and weak driving (perturbative limit), we perform a Schrieffer-Wolff transformation to obtain approximate analytical expressions of the Rabi frequency and Rabi transition probabilities. For realistic system parameters, the Rabi frequency turns out to be
of the order of megahertz. We discuss the dependence
of the maximum transition probability and width of the
resonances on the magnetic field strength, driving frequency and Rashba parameter. For stronger driving and
larger Rashba strengths, we resort to numerical methods.
In presence of either of the Rashba couplings, we exploit
the ‘rotational’ symmetries of the system to derive a uni-
tary transformation that converts the driven Hamiltonian into a static one, making the numerical computation of the time-evolved state easier. We observe a high frequency component superimposed on the resonant Rabi
oscillations for larger amplitudes of radiation, which the
perturbation theory does not capture. When both the
Rashba couplings are present, we use Floquet theory to
numerically calculate the time evolution. We study the
interplay of both the couplings and observe that large
and realistic values of the cubic Rashba coupling drives
the system out of resonance and effectively destroys the
Rabi oscillations. We also show that the Rabi frequency
is equal to quasienergy gap between two adjacent Floquet
levels, which increases with the radiation amplitude.

The paper is organized as follows. In Sec. \ref{mod}, we discuss the theoretical model of the planar Ge quantum dot. In Sec. \ref{drive}, we derive the interaction Hamiltonian of the laser
beam with the hole qubit. In Sec. \ref{evol-gauge}, we study relation between time evolution of the driven system in different gauges. Section \ref{evol-anal} deals with analytical formalism to obtain the Rabi frequency of the system when the SOC and drive are treated perturbatively. In Sec. \ref{evol-numerical}, we discuss the numerical methods such as unitary transformation and Floquet theory use to solve the Schrodinger equation. In Sec. \ref{results}, we present and analyse the results of our study for realistic system parameters and laser strengths in presence of either or both types of SOC in this system. Finally, we conclude our results in Sec. \ref{conc}.

	\section{Model}\label{mod}
	
\begin{figure}
		\centering
		\includegraphics[trim={1cm 1cm 1cm 2cm},clip,width=11cm]{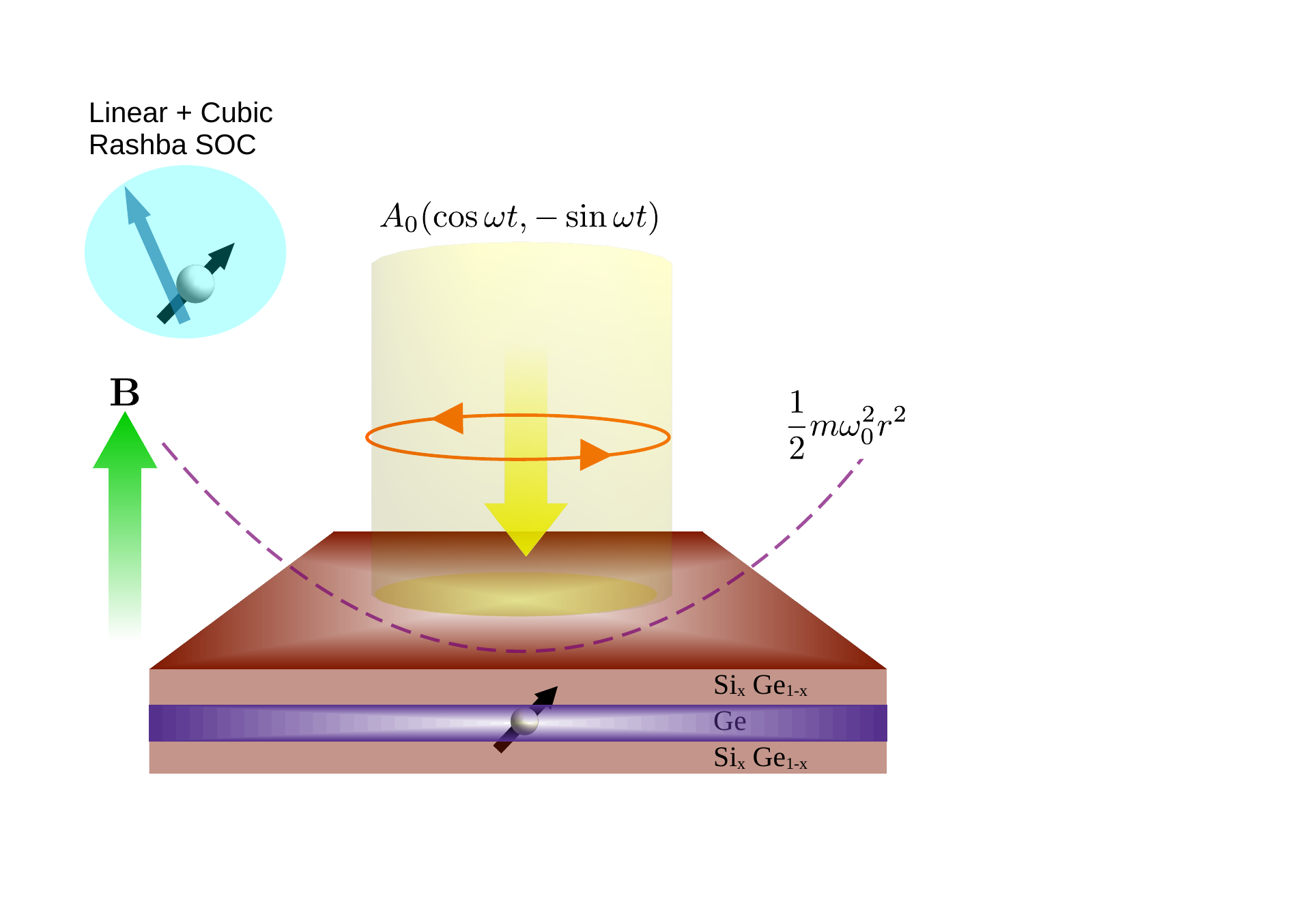}
		\caption{Schematic diagram of the model.}
		\label{model}
	\end{figure}

	
	In group IV and III-V semiconductors, the hole states lie close to the $\Gamma$ point of valence band of these materials and have effective  spin $J=3/2$.  These states can be described by the $4\times4$ Luttinger Hamiltonian \cite{Luttinger}. In 2D quantum wells confined along the growth direction, the heavy-hole (HH) states (spin $\pm3/2$) split apart from the light-hole (LH) ones (spin $\pm1/2$) with a higher energy. The splitting depends on the well thickness (say $d$) and varies as $d^{-2}$. We consider a 2D HH-gas of Ge confined electrostatically in the lateral direction and subjected to a magnetic field perpendicular to the 2D plane (${\bf B}=B \hat{z}$) as shown in Fig.[\ref{model}]. The confinement can be modelled by a parabolic potential $U(x,y)=m\omega_0^2(x^2+y^2)/2$ where  $\omega_0=\hbar/(m l_0^2)$ with $l_0$ being the confinement lengthscale (approximated as radius of the dot) and $m$ being the effective heavy hole mass. Including the SOC effects, the Hamiltonian of the HH states in the out-of-plane magnetic field can be written as
  
	\begin{equation}\label{orig}
		\begin{aligned}
			H_0=&\frac{P^2}{2m}+U(x,y)-\frac{1}{2}g_\perp\mu_B B \sigma_z+i\alpha_c(P_-^3\sigma_+-P_+^3\sigma_-)\\
			&-i\alpha_l(P_-\sigma_+-P_+\sigma_-)
		\end{aligned}
	\end{equation}
	where ${\bf P}={\bf p}-|e|{\bf A}_B{\bf (r)}$, ${\bf A}_B{\bf (r)}=B \l(-y,x\r)/2$, $\sigma_\pm=(\sigma_x\pm i\sigma_y)/2$, $P_\pm=P_x\pm iP_y$. The parameter $\alpha_c=3\gamma_0\alpha_R \langle E_z\rangle/(2\Delta m_e)$ is the cubic Rashba SOC strength (corresponding to the dominant spherically symmetric contribution) which is directly proportional to the average electric field at the interface $\langle E_z\rangle$ and inversely proportional to HH-LH splitting $\Delta$. Here, $m_e$ is the bare electron mass, $\alpha_R$ the coupling constant, $\gamma_0$ a Luttinger parameter and $g_\perp$ the out-of-plane component of the $g$-factor tensor. The parameter $\gamma_0$ is equal to the Luttinger parameters $\gamma_2$ and $\gamma_3$ within the spherical approximation of the Luttinger Hamiltonian i.e. $\gamma_0=\gamma_2\approx\gamma_3$ \cite{spherical}. The parameter $\alpha_l$ denotes the newly claimed  linear Rashba SOC strength \cite{emergent}.  The Rashba couplings can be controlled either by changing the interfacial dc electric field, the well thickness or both. When the Rashba couplings are tuned to zero i.e $\alpha_l=\alpha_c=0$, the Hamiltonian is exactly solvable using the following coordinate transformations~\cite{phonon3}:
	
	\begin{equation}\label{eq1}
		x=\frac{1}{\sqrt{2\Omega}}\left(\sqrt{\omega_1}q_1+\sqrt{\omega_2}q_2\right),
	\end{equation}
	
	\begin{equation}\label{eq2}
		y=\frac{1}{m\sqrt{2\Omega}}\left(\frac{p_1}{\sqrt{\omega_1}}-\frac{p_2}{\sqrt{\omega_2}}\right),
	\end{equation}
	
	\begin{equation}
		p_x=\sqrt{\frac{\Omega}{2}}\left(\frac{p_1}{\sqrt{\omega_1}}+\frac{p_2}{\sqrt{\omega_2}}\right),
	\end{equation}
	
	\begin{equation}
		p_y=m\sqrt{\frac{\Omega}{2}}\left(-\sqrt{\omega_1}q_1+\sqrt{\omega_2}q_2\right),
	\end{equation}
	where $[q_i,q_j]=[p_i,p_j]=0$, $[q_i,p_j]=i\hbar \delta_{i,j}$ and the constants are defined as:  $\omega_{1,2}=\Omega\pm\omega_c/2$, $\Omega=\sqrt{\omega_0^2+\omega_c^2/4}$, $\omega_c=|e|B/m$, $\omega_z=g_\perp\mu_B B/\hbar$.
	It is to be noted that for holes, $\omega_{1,2}$ have their signs exchanged when compared to those defined in  Ref.~\cite{phonon3}. 
 Upon transforming, the SOC-free Hamiltonian in the new coordinates can be written as a sum of two uncoupled harmonic oscillators with Zeeman coupling
	\begin{equation}
		H_\text{FD}=\frac{p_1^2}{2m}+\frac{1}{2}m\omega_1^2q_1^2+\frac{p_2^2}{2m}+\frac{1}{2}m\omega_2^2q_2^2-\frac{\hbar \omega_z}{2}\sigma_z.
	\end{equation}
	Using the ladder operators
	\begin{equation}\label{ladder1}
		\hat{a}_{1,2}=\sqrt{\frac{m\omega_{1,2}}{2\hbar}}\left(q_{1,2}+i\frac{p_{1,2}}{m\omega_{1,2}}\right)
	\end{equation}
	and
	\begin{equation}\label{ladder2}
		\hat{a}_{1,2}^\dagger=\sqrt{\frac{m\omega_{1,2}}{2\hbar}}\left(q_{1,2}-i\frac{p_{1,2}}{m\omega_{1,2}}\right),
	\end{equation}
	the Hamiltonian can be cast as
	\begin{equation}\label{H-orig}
		H_\text{FD}=\hbar\omega_1\left( \hat{n}_1+\frac{1}{2}\right)+\hbar\omega_2\left(\hat{n}_2+\frac{1}{2}\right)-\frac{\hbar \omega_z}{2}\sigma_z,
	\end{equation}
 where $\hat{n}_i=\hat{a}_i^\dagger \hat{a}_i$ is the number operator. Its eigenstates are the  well-known Fock-Darwin (FD) levels with energies $E_{n_1,n_2,s}=\hbar\omega_1\l(n_1+1/2\r)+\hbar\omega_2\l(n_2+1/2\r)-\hbar\omega_z s/3$ and eigenvectors $|n_1,n_2,s\rangle=\Phi_{n_1}(q_1\sqrt{m\omega_1/\hbar})\Phi_{n_2}(q_2\sqrt{m\omega_2/\hbar})|s\rangle$ where $s=\pm 3/2$. Here, $\Phi_n(...)$ is the $n_{}$th excited state of the harmonic oscillator in the specified coordinates. Therefore, on tuning $\alpha_{c,l}\to0$, we get pure localized spin states which can be used as qubits. 
 
 For finite $\alpha_{c,l}$, the last two terms of Eq. (\ref{orig}), representing the Rashba interaction $H_R$,  can be rewritten as
 \begin{equation}\label{Rashba-H}
    \begin{aligned}
        H_R=H_R^{\text{(C)}}+H_R^{\text{(L)}},
    \end{aligned} 
 \end{equation}
 where 
 \begin{equation}\label{Rashba-C}
    \begin{aligned}
        H_R^{\text{(C)}}=\alpha_c(m\hbar )^{3/2}\l[(f_+ \hat{a}_1^\dagger+f_- \hat{a}_2)^3\sigma_+ + \text{h.c.}\r]
    \end{aligned} 
 \end{equation}
and 
\begin{equation}\label{Rashba-L}
    H_R^{\text{(L)}}=\alpha_l\sqrt{m\hbar }\l[(f_+ \hat{a}_1^\dagger+f_- \hat{a}_2)\sigma_+ + \text{h.c.}\r]
\end{equation}
with
	\begin{equation}
		f_\pm=\pm\sqrt{\Omega}+\frac{\omega_c}{2\sqrt{\Omega}}.
	\end{equation}

Hence, the total Hamiltonian is $H_0=H_\text{FD}+ H_R^{\text{(C)}}+H_R^{\text{(L)}}$. The superscripts `C' and `L' denote cubic and linear Rashba couplings respectively. The Rashba interactions couple the FD levels with different spin quantum numbers and hence spin is no longer a conserved quantity. The exact eigenstates of the system in presence of the Rashba coupling(s) are unknown and hence perturbation theory is often used to study the physics of these systems. For very small Rashba strengths, the FD levels can still be considered as approximate energy levels of the system as the first order correction in energy due to the Rashba couplings are zero. The first order corrections to the energy eigenstates are however non-zero and states with opposite spins get mixed. For example, the cubic Rashba coupling mixes $|0,0,-3/2\rangle$ with $|3,0,3/2\rangle$ upto first order in $\alpha_c$ while the linear Rashba coupling mixes $|0,0,-3/2\rangle$ with $|1,0,3/2\rangle$ upto first order in $\alpha_l$. However, for time spans much shorter than the coupling time scale, spin is approximately a good quantum number.

\textit{Symmetries: }The system also has some continuous symmetries which can be seen from the commutator relations of angular momentum.  In terms of the number operators, the orbital angular momentum operator can be written as $L_z=\hbar(\hat{n}_2-\hat{n}_1)$. Since $[L_z,H_\text{FD}]$=0, the orbital angular momentum is conserved in absence of the Rashba couplings. On defining operators
\begin{equation}\label{JzL}
    \hat{J}^{(\text{L})}_z=L_z+\hbar \sigma_z/2~~ \text{for} ~~\alpha_l\neq0, \alpha_c=0, 
\end{equation}
and 
\begin{equation}\label{JzC}
    \hat{J}^{(\text{C})}_z=L_z+3\hbar \sigma_z/2~~ \text{for} ~~  \alpha_c\neq0,\alpha_l=0
\end{equation}
we get $[\hat{J}^{(\text{L})}_z, H_\text{FD}+H_R^{(\text{L})}]=0$ and $[\hat{J}^{(\text{C})}_z, H_\text{FD}+H_R^{(\text{C})}]=0$. This implies the presence of a rotational symmetry about $z$-axis in the total Hilbert space (spin plus orbital) when either of the two Rashba couplings are present. As we show later, this symmetry of the Hamiltonian can be exploited to obtain the time-evolution of the system under circular driving.

	\section{Circular drive by laser}\label{drive}

    A coherent laser beam of circularly polarized radiation is shone upon the hole gas normally [Fig. \ref{model}]. We treat the laser beam classically by modelling it as a plane wave with electric and magnetic fields given by ${\bf E}({\bf r}, t)=E_0[\sin(\omega t+kz),\cos(\omega t+kz),0]$ and ${\bf B}({\bf r}, t)=-(E_0/c )[-\cos (\omega t+ k z), \sin (\omega t+ k z), 0]$ respectively. The effect of the electromagnetic field is incorporated into the Hamiltonian(\ref{orig}) through the vector and scalar potentials in two different gauges viz. velocity gauge and length gauge. It is to be noted that the ac electric field is in-plane and also averages out to zero over the time scales of qubit operations. Hence, it has no influence on the Rashba strength ($\propto \langle E_z \rangle$) in this model.
    
     \subsection{Velocity gauge}
     The most natural choice of gauge to describe plane wave radiation is the velocity gauge. In this gauge, the beam can be represented by a vector potential ${\bf A}_r({\bf r}, t)=A_0 [\cos (\omega t+ k z), -\sin (\omega t +k z), 0]$ and $\phi ({\bf r}, t)=0$, where $A_0=E_0/\omega$ with $E_0$ being the electric field amplitude. The Hamiltonian (\ref{orig}) becomes periodic in time through Peierls substitution i.e. $H(t)= H_0 ({\bf P}-|e|{\bf A}_r({\bf r},t))$. So, in this gauge, the coupling with radiation is only through the vector potential. Since $z=0$ for the hole gas, ${\bf A}_r({\bf r}, t)\equiv{\bf A}_r(t)=A_0 (\cos \omega t, -\sin \omega t, 0)$. The driven Hamiltonian can be decomposed as $H(t)=H_\text{FD}+H_R+V(t)$  where
	\begin{equation}\label{Vt}
		V(t)=\sum_{n=-3}^{3}V_n \e^{i n \omega t}
	\end{equation}
	where $V_n$ are the Fourier components given by whose matrix elements are given by:
	\begin{equation}
			V_0= \frac{e^2A_0^2}{2m},	
	\end{equation}
	\begin{equation}\label{v1}
		\begin{aligned} 
			V_1=& 3 i\alpha_c |e|A_0 m\hbar \l(f_+ \hat{a}_1^\dagger+f_- \hat{a}_2\r)^2\sigma_+ \\
			&i \alpha_l |e| A_0\sigma_+ +i\frac{|e|A_0}{2m}\sqrt{m\hbar }(f_+ \hat{a}_1 +f_- \hat{a}_2^\dagger ),    
		\end{aligned}  
	\end{equation}
	\begin{equation}\label{v2}
		V_2= -3\alpha_c\sqrt{m\hbar }|e|^2A_0^2\l(f_+ \hat{a}_1^\dagger+f_- \hat{a}_2 \r)\sigma_+
	\end{equation}
	and 
	\begin{equation}
		V_3= -i\alpha_c |e|^3 A_0^3\sigma_+~~.
	\end{equation}
	Due to Hermiticity, the Fourier components are related as $V_{-n}=V_n^\dagger$. The forms of the matrix elements of $V_n$ and $H_R$ can be found in Appendix [\ref{app-matrix-elem}]. Since the magnetic vector lies in-plane, it does not have any Zeeman interaction with the hole spins because the $2\times 2$ HH submatrices fulfill the property: $\mathcal{J}_x=\mathcal{J}_y=0$ and $\mathcal{J}_z=3\sigma_z/2$ \cite{jxjy0,hole-loss}.
 
 The second term of $V_1$ and the term $V_3$ couple spins with the same orbital quantum numbers. It shows that Rabi transitions $|3/2\rangle\Longleftrightarrow|-3/2\rangle$ may be induced within the same orbital sector $(\Delta n_1=\Delta n_2=0)$ using circularly polarized light if the higher levels are decoupled. To begin with, let us consider the block of the two lowest lying FD states viz. $|0,0,3/2\rangle$ and  $|0,0,-3/2\rangle$ that have opposite spins:
	\begin{equation} \label{h22}
		H_{2\times2}=\l(\begin{array}{cc}
			\frac{e^2A_0^2}{2m}+\hbar \Omega -\frac{\hbar \omega_z}{2}   & \kappa(t) \\
			\kappa^* (t) & \frac{e^2A_0^2}{2m}+\hbar \Omega +\frac{\hbar \omega_z}{2} 
		\end{array}\r)
	\end{equation}
	where $\kappa(t)=i \alpha_l |e| A_0 \e^{i \omega t} -i \alpha_c |e|^3 A_0^3 \e^{3i \omega t}$.
	The lowest block resembles a 2-level system driven by harmonic modes of frequencies $3\omega$ and $\omega$ corresponding to cubic and linear Rashba SOCs respectively. 
 For $\omega=\omega_z/3$ and $\omega=\omega_z$, the Rabi frequencies are $2\alpha_c |e|^3 E_0^3/(\hbar\omega^3)$ and $2\alpha_l |e| E_0/(\hbar\omega)$ respectively. This shows that the vector potential of the coherent radiation can cause hole-spin resonance ($\Delta n_1=0,\Delta n_2=0,\Delta s=3$) in presence of Rashba SOC. However, the Rabi oscillations are  killed by the coupling of this block with the higher energy levels and hence the 2-level picture does not capture the physics of the complete system.  



 \subsection{Length gauge}
 
 We may also choose another gauge where ${\bf A}^\prime_r({\bf r},t)=(0,0,E_0/c[x\sin(\omega t+ k z)+y \cos(\omega t+ k z)])$ and $\phi^\prime_r({\bf r},t)=-E_0[x\sin(\omega t+ k z)+y \cos(\omega t+ k z)]$. The two gauges are related as:  ${\bf A}^\prime_r({\bf r},t)={\bf A}_r({\bf r},t)+\grad \Lambda ({\bf r},t)$ and $\phi^\prime_r({\bf r},t)=\phi_r({\bf r},t)-\partial_t\Lambda ({\bf r},t)$ where 
 \begin{equation}\label{Lambda}
     \Lambda({\bf r},t)\equiv\Lambda=(E_0/\omega)[-x\cos(\omega t+ k z)+y \sin(\omega t+ k z)].
 \end{equation}
 Since $P_z$ is absent, $\tilde{A}_z$ does not couple with the static Hamiltonian. So, the coupling with radiation is only through the scalar potential $\phi^\prime_r({\bf r},t)=-E_0(x\sin\omega t+y \cos \omega t)$ at $z=0$ i.e. $H^\prime(t)= H_\text{FD} + H_R+ |e|\phi^\prime_r({\bf r},t)$. This is called the length gauge.

The Hamiltonian in the length gauge is identical to that of an EDSR setup. For EDSR, a circularly rotating electric field ${\bf E}=E_0(\sin \omega t,\cos \omega t)$ is applied across the dot using two perpendicular pairs of gates. Then, the interaction of the heavy holes with the field can be written as
\begin{equation}\label{Vg}
    V_g(t)=-|e|\int_{}^{\bf r}{\bf E}\cdot d{\bf r^\prime}=-F_0 (x\sin \omega t+y \cos \omega t)
\end{equation}
where $F_0=|e|E_0$. Hence, $V_g(t)=|e|{red}\phi^\prime_r({\bf r},t)$. It is to be noted that although the oscillating electric field also produces a magnetic field, its magnitude is $1/c$ times smaller the electric field and would have negligible effect on the spin dynamics. Hence, we can safely ignore the magnetic vector potential in this case \cite{edsr-loss-orig}. 

Using the transformations (\ref{eq1}) and (\ref{eq2}) in (\ref{Vg}), we get
\begin{equation}\label{Vgladder}
\begin{aligned}
   V_g(t)=i\frac{F_0}{2}\sqrt{\frac{\hbar}{m \Omega}}\big(\hat{a}_1 \e^{i\omega t} - \hat{a}_1^\dagger \e^{-i\omega t}-\hat{a}_2 \e^{-i\omega t} + \hat{a}_2^\dagger \e^{i\omega t}\big)  
\end{aligned}
\end{equation}

The total Hamiltonian of the driven dot is $H^\prime(t)=H_\text{FD}+H_R+V_g(t)$ where $H_\text{FD}$ is the exactly solvable part and $H_R+V_g(t)$ is to be treated as perturbation. The perturbation does not couple the spins in the lowest energy block which are the Zeeman-split ground states (i.e. $n_1=n_2=0$ orbital sector). So, spin rotations can be achieved only through the higher order transitions. For a drive of the form (\ref{Vgladder}), only the linear Rashba coupling supports EDSR. This can be explained as follows. At resonance $\omega=\omega_z$,  $V_g (t)$ can cause a virtual transition  with no spin flip ($\Delta n_i=\pm 1, \Delta s=0$) followed by another virtual transition accompanied by a spin flip ($\Delta n_i=\pm 1, \Delta s=\pm 3$)  mediated by the linear Rashba coupling in (\ref{Rashba-H}). This brings about the desired Rabi oscillations in the system even in absence of a rotating magnetic field. The cubic Rashba coupling cannot cause EDSR because the cubic terms not couple levels with $\Delta n_i\neq\pm 1$ and hence the $x$-linear drive cannot cause virtual transitions back to the original level. Thus, the length gauge provides a better picture of the Rabi oscillations as compared to the velocity gauge.

 \section{Time evolution}

\subsection{Time evolution in different gauges}\label{evol-gauge}

The Hamiltonians and the solutions of the time-dependent Schr{\"o}dinger equation (TDSE) in the two gauges are related as

\begin{equation}
    H^\prime(t)=\e^{i|e|\Lambda/\hbar} H(t)\e^{-i|e|\Lambda/\hbar}-|e|\partial_t \Lambda
\end{equation}
and 
\begin{equation}\label{psigauge}
    |\psi^\prime(t)\rangle=\e^{i|e|\Lambda/\hbar}|\psi(t)\rangle,
\end{equation}
respectively, where $\Lambda$ is defined in Eq. (\ref{Lambda}). From (\ref{psigauge}), it follows that
\begin{equation}
    U^\prime(t,0)=\e^{i|e|\Lambda({\bf r},t)/\hbar} U(t,0) \e^{-i|e|\Lambda({\bf r},0)/\hbar}.
\end{equation}
Clearly, the time evolution is gauge-dependent (as the Hamiltonian is gauge-dependent). To render the transition amplitudes gauge-invariant, the initial $(|i\rangle )$ and final $(|f\rangle)$ states must be gauge transformed in the following way \cite{gauge}:
\begin{equation}
    |i^\prime\rangle=\e^{i|e|\Lambda({\bf r},0)/\hbar}|i\rangle,
\end{equation}
\begin{equation}
    |f^\prime\rangle=\e^{i|e|\Lambda({\bf r},t)/\hbar}|f\rangle.
\end{equation} Then, we have $\langle f^\prime| U^\prime(t,0) |i^\prime\rangle=\langle f | U(t,0) |i\rangle$. It is easier to study the dynamics in the length gauge because the interaction in this gauge has lesser number of terms (\ref{Vgladder}) than that in the velocity gauge (\ref{Vt}). So, it is to be noted that the numerical results presented in this paper are obtained using the length gauge only.

The exact analytical solutions of the TDSE cannot be obtained for this system in either of the two gauges. Since we are interested in Rabi oscillations, firstly we obtain an approximate analytical expression of the Rabi frequency by treating the Rashba coupling(s) and the drive perturbatively. Secondly, we compute the numerical solutions and the Rabi frequencies using the methods of unitary transformation and Floquet theory.

\subsection{Analytical formalism}\label{evol-anal}

An approximate analytical form of the Rabi frequency can also be obtained in the length gauge using perturbation theory. The linear Rashba coupling is off-diagonal in the FD basis as it couples blocks with  orbital quantum number differing by 1 i.e. $\Delta n_i=\pm1$. For small Rashba strengths ($ \frac{\alpha_l \sqrt{m \hbar \omega_0 }}{ \hbar \omega_0}=\tilde{\alpha}_l\ll1$) as compared to the confinement energy scale $\hbar \omega_0$,  we can perform a  Schrieffer-Wolff (SW) transformation\cite{edsr-loss-orig,phonon2,spin-decay,Fernandez-Fernandez_2023} to diagonalize the Hamiltonian of the dot such that the off-diagonal couplings are removed upto linear order in $\tilde{\alpha}_l$. Using the transformation, the effective 2-level Hamiltonian for this system can be written as [see Appendix \ref{app-SW} for details]
\begin{equation}\label{heff}
\begin{aligned}
H_\text{eff}(t)
    &=\l(\begin{array}{cc}
      E_1 &  -i \gamma \e^{i\omega t}\\
        i \gamma \e^{-i\omega t} &  E_2
    \end{array}\r)
\end{aligned}
\end{equation}
where
\begin{equation}
    E_1= -\frac{\hbar\omega_z}{2} -\frac{\alpha_l^2 m f_-^2}{\omega_2+\omega_z},
\end{equation}
\begin{equation}
    E_2= \frac{\hbar\omega_z}{2} -\frac{\alpha_l^2 m f_+^2}{\omega_1-\omega_z}
\end{equation}
and
\begin{equation}
    \gamma=\frac{F_0\alpha_l}{2\sqrt{\Omega}}\l(\frac{f_+}{\omega_1-\omega_z}+\frac{f_-}{\omega_2+\omega_z}\r).
\end{equation}
The Hamiltonian (\ref{heff}) is equivalent to that of a two-level Rabi problem with oscillations in occupation probabilities given by
\begin{equation}
    P_1(t)=P_\text{max} \sin^2(\omega_R t), ~~~P_0(t)=1-P_1(t).
\end{equation}
where the amplitude, Rabi frequency and the level separation are given by
\begin{equation}\label{P}
    P_\text{max}=\frac{\gamma^2}{\hbar^2 \omega_R^2},
\end{equation}
\begin{equation}
    \omega_R=\l[\frac{\gamma^2}{\hbar^2}+\frac{(\omega-\omega_{21})^2}{4}\r]^{1/2}
\end{equation} and
\begin{equation}
    \omega_{21}=\frac{E_2-E_1}{\hbar}=\omega_z+\frac{\alpha_l^2 m}{\hbar}\l(\frac{f_-^2}{\omega_2+\omega_z}-\frac{f_+^2}{\omega_1-\omega_z}\r),
\end{equation}
respectively. Hence, the new resonance condition is $\omega=\omega_{21}$ due to the energy correction second order in $\alpha_l$. The resonant Rabi frequency is $ \omega_\text{res}=2\gamma/\hbar$ and is hence, linearly proportional to both $\alpha_l$ and $F_0$.  

\begin{figure}
		\centering
		\hspace{-0.3cm}\includegraphics[trim={0.8cm 0cm 0cm 0cm},clip,width=9cm]{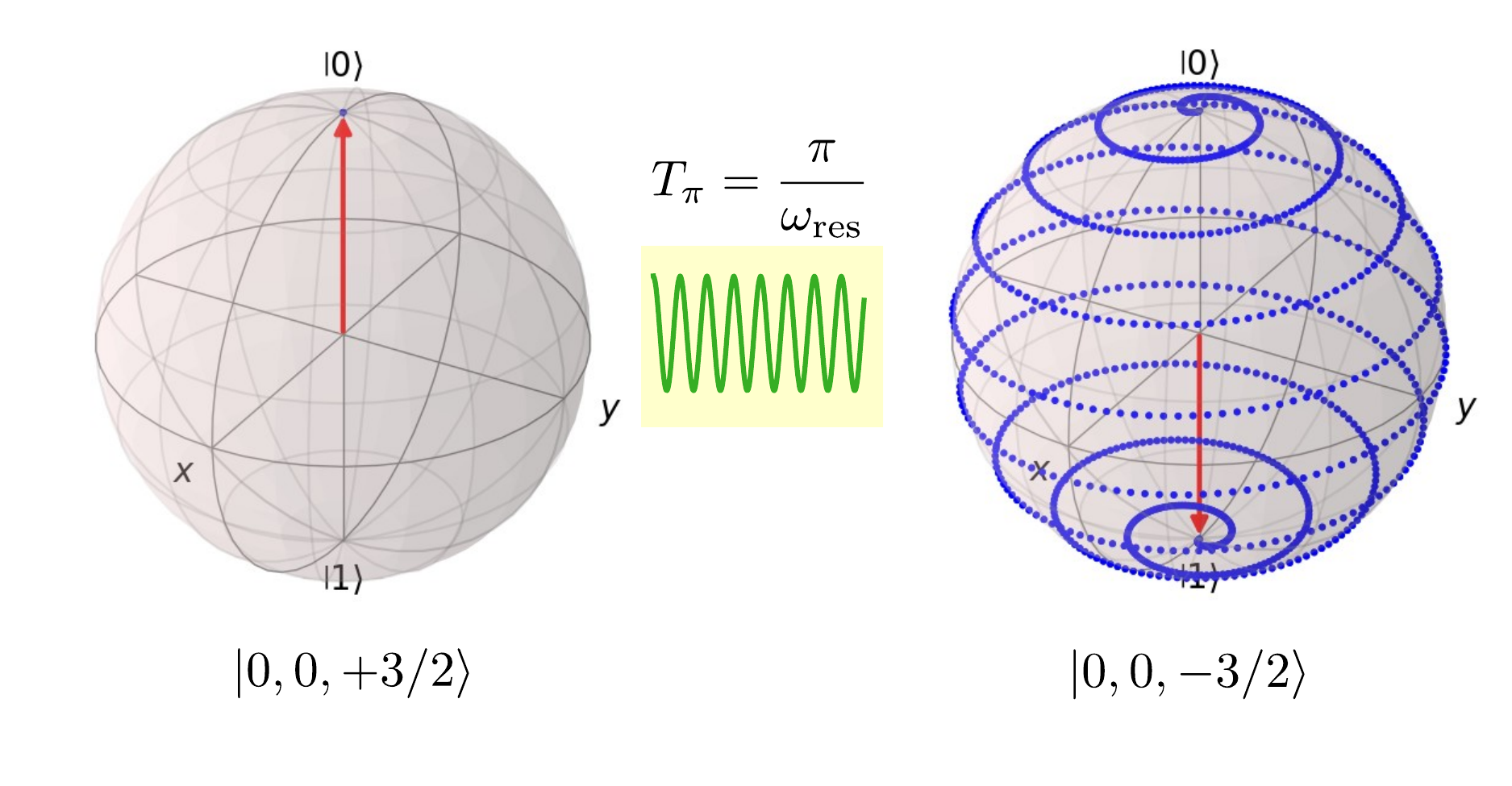}
		\caption{ Bloch sphere dynamics of the qubit on application of the laser beam for half Rabi cycle i.e. $T_\pi=\pi/\omega_\text{res}$. We use the following system parameters: $\tilde{F}_0=0.02\times300$, $B=0.5$ T, $\tilde{\alpha}_l=0.0047,~ \tilde{\omega}_c=0.303,~ \tilde{\omega}_z=0.214,~ \omega=\omega_\text{21}$ which are defined in Sec. \ref{results}.  The spin-up state spirals down to the spin-down state on the Bloch sphere because the driving frequency is much larger than the Rabi frequency. Here, $\tilde{F}_0$ has been magnified 300 times the actual value for better visualization of the rotations (although a stronger drive leads to other effects which is discussed later). As a result, the spin vector rotates $\sim 9$ times about the $z$-axis during the process of spin-flip.}
		\label{bloch-qubit}
	\end{figure}
 
At resonance, the time-evolved state is given by
\begin{equation}
    |\psi(t)\rangle=\l(\begin{array}{c}
        \e^{-i E_1 t/\hbar} \cos (\gamma ~t/\hbar)  \\
         \e^{-i E_2 t/\hbar} \sin (\gamma ~t/\hbar)
    \end{array}\r). 
\end{equation}
The expectation value of the spin vector follows the following trajectory:
\begin{equation}
    \langle\boldsymbol{\sigma}(t)\rangle=[\cos \omega t \sin (2\gamma t/\hbar), -\sin \omega t \sin (2\gamma t/\hbar), \cos (2\gamma t/\hbar) ].
\end{equation}
The Bloch sphere dynamics through a half Rabi cycle $(T_\pi=\pi/\omega_\text{res})$ is shown in Fig. [\ref{bloch-qubit}] when the system is initialized in spin-up state.
The spirals occur due to the finite $\langle\sigma_x(t)\rangle$ and $\langle\sigma_y(t)\rangle$ which oscillate with a frequency of $\omega$. It can be shown that the number of rotations about the $z$-axis is $\omega/(2\omega_\text{res})$. Since $\omega\gg\omega_\text{res}$ in our case, the spin vector $\langle\boldsymbol{\sigma}(t)\rangle $ makes a large number of  rotations/precessions $\sim2710$ about the $z$-axis before flipping down completely at $T=\pi/\omega_\text{res}$. Hence, the Rabi oscillation is not simply a `rotation' of the spin about $x$- or $y$-axis, but a spiralling-down motion.
The irradiation by laser for half Rabi cycle acts as a $\pi$-pulse with possible applications in designing a quantum NOT gate.

\subsection{Numerical formalism}\label{evol-numerical}

\subsubsection{Unitary transformation}

We know that the TDSE for a simple two-level system under circular driving can be solved exactly by purging the time-dependence of the Hamiltonian through an appropriate unitary transformation. Using a similar approach and the fact that $[J^{(\text{L})}_z, H_\text{FD}+H_R^{(\text{L})}]=0$ and $[J^{(\text{C})}_z, H_\text{FD}+H_R^{(\text{C})}]=0$, we can deduce a static Hamiltonian $\mathcal{H}^{(\text{L})}$ or $\mathcal{H}^{(\text{C})}$ when either of the two Rashba couplings (linear or cubic) is present in the system [see Appendix \ref{app-unitary} for details]. For the velocity gauge, we deduce
\begin{equation}
    \mathcal{H}^{\text{(L)/(C)}}=\mathcal{H}_0+\mathcal{H}_R^{\text{(L)/(C)}}
\end{equation}
where
\begin{equation}
\begin{aligned}
    &\mathcal{H}_0=\hbar \Omega (\hat{n}_1+\hat{n}_2+1)+\hbar\l(\omega-\frac{\omega_c}{2}\r)(\hat{n}_2-\hat{n}_1)\\
    &+\frac{|e|^2 A_0^2}{2m}-\frac{i |e|A_0\sqrt{m\hbar}}{2m}\l(f_+(\hat{a}_1^\dagger-\hat{a}_1)-f_-(\hat{a}_2^\dagger-\hat{a}_2)\r),
\end{aligned}
\end{equation}
\begin{equation}
\begin{aligned}
    &\mathcal{H}_R^{(\text{L})}=\frac{\hbar}{2}(\omega-\omega_z)\sigma_z \\
    &+\alpha_l\bigg[\l(\sqrt{m\hbar}(f_+ \hat{a}_1^\dagger+f_- \hat{a}_2) +i|e|A_0 \r)\sigma_+ + \text{h.c.}\bigg],
\end{aligned}
\end{equation}
\begin{equation}
\begin{aligned}
    &\mathcal{H}_R^{(\text{C})}=\frac{\hbar}{2}(3\omega-\omega_z)\sigma_z \\
    &+\alpha_c\bigg[\l(\sqrt{m\hbar}(f_+ \hat{a}_1^\dagger+f_- \hat{a}_2) +i|e|A_0 \r)^3\sigma_+ + \text{h.c.}\bigg].
\end{aligned}
\end{equation}

Similarly, for the length gauge, we get
\begin{equation}\label{Halpha-ladder}
    \mathcal{H}^{\prime\text{(L)/(C)}}=\mathcal{H}^\prime_0+\mathcal{H}_R^{\prime\text{(L)/(C)}}
\end{equation}
where
\begin{equation}
\begin{aligned}
    &\mathcal{H}^\prime_0=\hbar \Omega (\hat{n}_1+\hat{n}_2+1)+\hbar\l(\omega-\frac{\omega_c}{2}\r)(\hat{n}_2-\hat{n}_1)\\
    &+\frac{i F_0}{2}\sqrt{\frac{\hbar}{m\Omega}}\l(\hat{a}_1-\hat{a}_1^\dagger-\hat{a}_2+\hat{a}_2^\dagger\r),
\end{aligned}
\end{equation}
\begin{equation}
\begin{aligned}
    \mathcal{H}_R^{\prime (\text{L})}=\frac{\hbar}{2}(\omega-\omega_z)\sigma_z + H_R^{(\text{L})},
\end{aligned}
\end{equation}
\begin{equation}
\begin{aligned}
    \mathcal{H}_R^{\prime(\text{C})}=\frac{\hbar}{2}(3\omega-\omega_z)\sigma_z + H_R^{(\text{C})}
\end{aligned}
\end{equation}

and $H_R^{(\text{C})}$ and $H_R^{(\text{L})}$ are defined in equations (\ref{Rashba-C}) and (\ref{Rashba-L}) respectively.

In this gauge, the time evolution operator of the system can be written as the ordered product of two unitary  operators,
\begin{equation}\label{evolnew}
  U^{\prime\text{(L)/(C)}}(t)=\e^{i \hat{J}_z^{\text{(L)/(C)}}\omega t/\hbar}\e^{-i \mathcal{H}^{\prime\text{(L)/(C)}}} t/\hbar.  
\end{equation}
 The first factor accounts for the unitary transformation  and the second, containing the static Hamiltonian, gives the dynamical phase in the transformed frame. For the velocity gauge, $\mathcal{H}^{\prime\text{(L)/(C)}}$ is simply replaced by $\mathcal{H}^{\text{(L)/(C)}}$ in the second exponential. If the system is initialized in a FD state $|n_1^i,n_2^i,s^i\rangle$, then the transition amplitude to a state $|n_1^f,n_2^f,s^f\rangle$ is
\begin{equation}\label{aft}
\begin{aligned}
    &a_f(t)=\e^{i(n_2^f-n_1^f+s^{f\text{(L)/(C)}}})\omega t\times \\&
    \sum_{m}^{}\e^{-i\varepsilon_m^{\prime } t/\hbar}\langle n_1^f,n_2^f,s^f|\varepsilon_m^{\prime}\rangle \langle \varepsilon_m^{\prime} |n_1^i,n_2^i,s^i\rangle,
\end{aligned}  
\end{equation}
where $\mathcal{H}^{\prime\text{(L)/(C)}} |\varepsilon_m^{\prime}\rangle =\varepsilon_m^{\prime} |\varepsilon_m^{\prime }\rangle$, 
    $s^{f\text{(L)}}=s^f/3$ and $s^{f\text{(C)}}=s^f$ with $s^f=\pm3/2$.
The eigenvectors of $\mathcal{H}^{\prime\text{(L)/(C)}}$ can be obtained numerically by truncating its matrix upto a sufficiently large number of  FD levels. With $|0,0,3/2\rangle$ as the initial state, we then obtain the occupation probabilities of the states $|0,0,\pm3/2\rangle$ as a function of time by computing $|a_f(t)|^2$ using Eq. (\ref{aft}).

    \subsubsection{Floquet theory}

	When both the Rashba couplings are present, we do not have a suitable unitary transformation to make the Hamiltonian time-independent. On periodic driving, the basis of Floquet states is more relevant to work with as they states behave like static states in an extended Hilbert space of the driven system. They evolve similar to the static energy eigenstates but with a sum of quasienergy values and ``$n$'' multiples of photon energies contained in their dynamical phases. Since the Hamiltonian of the driven dot is periodic in time, the dynamics can be studied using Floquet theory. By Floquet's theorem, following solutions to the TDSE exist:
	\begin{equation}
		|u(t)\rangle=\e^{-i\epsilon_\eta t/\hbar} |\phi_\eta(t)\rangle
	\end{equation}
	where $\epsilon_\eta$ are the real-valued quasienergies and $|\phi_\eta(t)\rangle$ are the corresponding Floquet modes periodic in time. Considering the FD states as $|n_1,n_2,s\rangle\equiv|l\rangle$, the transition amplitude ($i\rightarrow f$) can be written as [See Appendices \ref{appA} and \ref{appB} for detailed derivation], 
	\begin{equation}\label{floq-amp}
		a_f(t)=\sum_{\eta}^{}\l(\sum_{n^\prime}(c_{i,\eta}^{n^\prime})^*\r)\l(\sum_{n} c_{f,\eta}^{n}\e^{-i (\epsilon_\eta-n\hbar\omega) t/\hbar} \r)
	\end{equation}
	where $l_\text{max}$ is the maximum number of FD levels considered in the problem and $c_{l,\eta}^{n}$ is defined in equation (\ref{fourierbasis}). The number of independent Floquet modes is equal to
the number of FD levels considered in the calculation.
The Floquet modes can be obtained by numerical diag-
onalization of the Floquet Hamiltonian truncated upto
a large number of FD levels and the occupation prob-
abilities $|a_f(t)|^2$ of the states $|0,0,\pm 3/2 \rangle$ can hence be
calculated using Eq. (\ref{floq-amp}).

 \section{Results and Discussion}\label{results}

 \begin{figure}
		\centering
		\hspace{-0.2cm}\includegraphics[trim={1cm 0cm 0cm 0cm},clip,width=9cm]{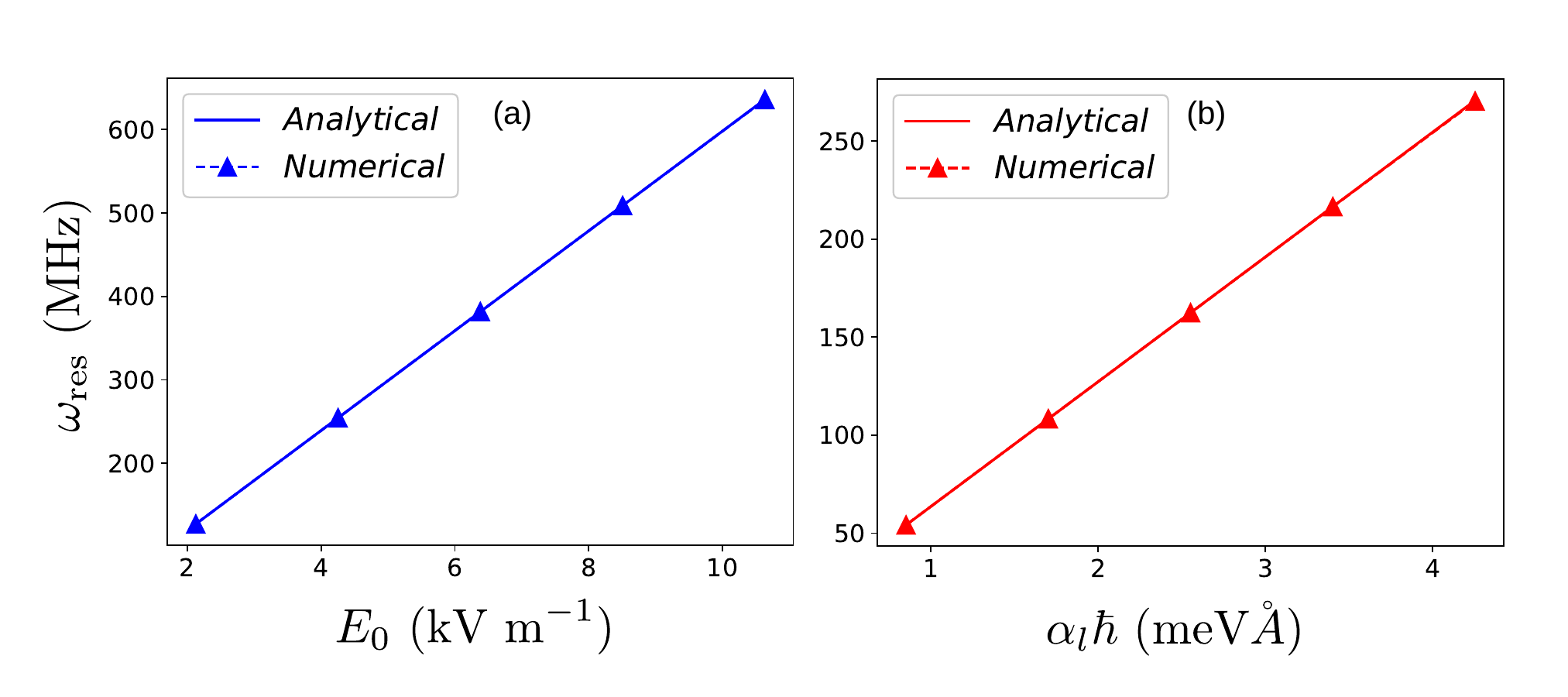}
		\caption{Comparison between analytical and numerical values of resonant Rabi frequencies: (a) Variation of $\omega_\text{res}$ with the radiation amplitude $E_0$ for $\alpha_l =2.01$ meV A$/\hbar$ and $B=0.5$ T. (b) Variation of $\omega_\text{res}$ with $\alpha_l$ for $E_0=2.172$ kV/m and $B=0.5$ T. }
		\label{comparison-neu}
	\end{figure}

\begin{figure}
		\centering
		\includegraphics[trim={0cm 9cm 0cm 0cm},clip,width=9cm]{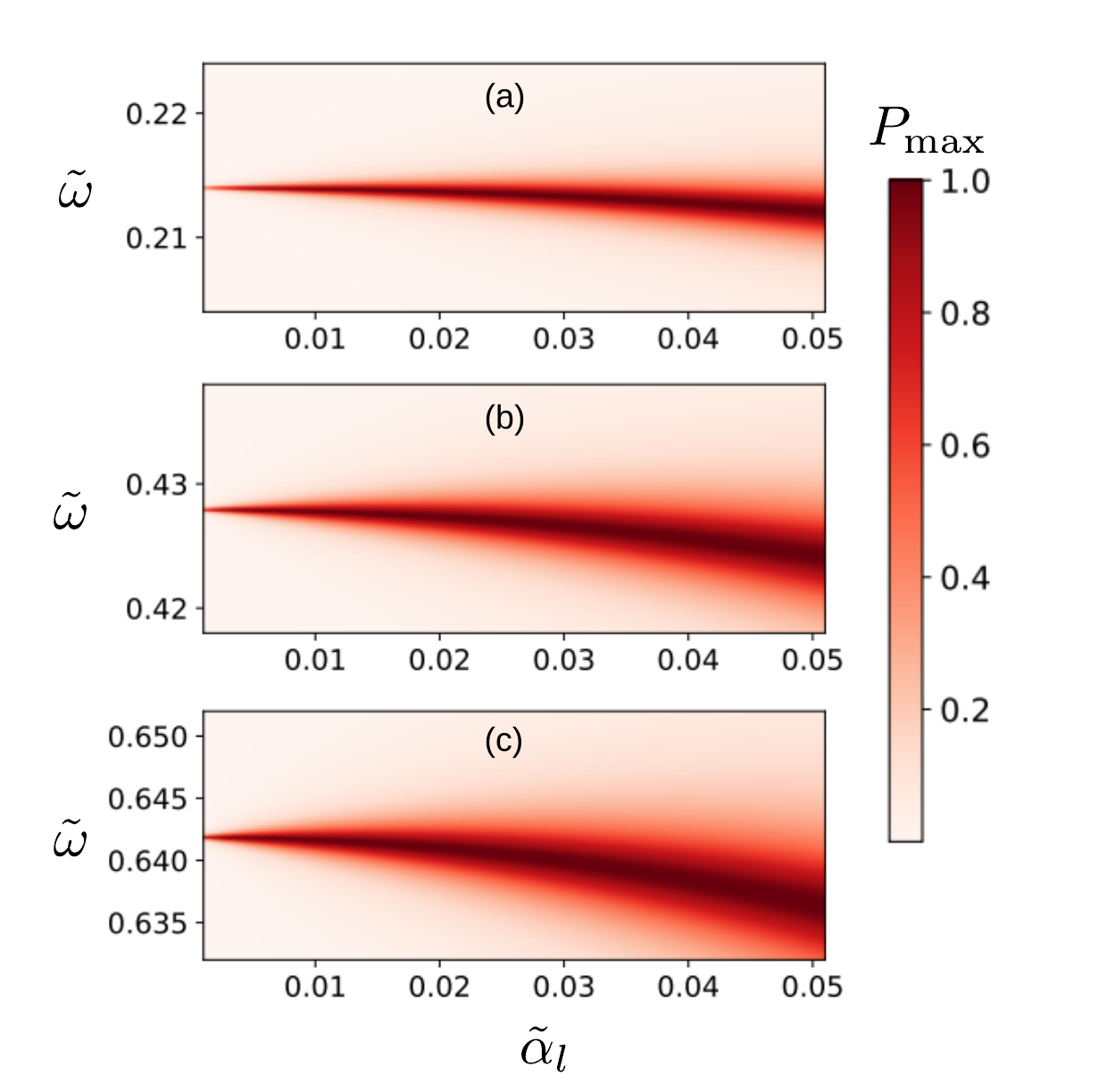}
		\caption{Plot of maximum transition probability as a function of $\tilde{\omega}$ and $\tilde{\alpha}_l$ for a fixed radiation amplitude $\tilde{F}_0=0.08$ and different values of magnetic field -- (a) $B=0.5$ T, (b) $B=1.0$ T and (c) $B=1.5$ T. The dark curves indicate the resonances. The width of the resonances increase with $B$ and $\alpha_l$.}
		\label{reso}
	\end{figure}

Let us define dimensionless quantities as $\tilde{\omega}_z=\omega_z/\omega_0$, $\tilde{\omega}_c=\omega_c/\omega_0$, $\tilde{\omega}=\omega/\omega_0$, $\tilde{\alpha}_c=\alpha_c p_0^3/(\hbar \omega_0)$, $\tilde{\alpha}_l=\alpha_l p_0/(\hbar \omega_0)$ and $\tilde{F}_0=F_0 /(p_0\omega_0)$ where $p_0=\sqrt{\hbar m \omega_0}$.  For a confinement length $l_0=20$ nm and using known values of parameters for Ge/Si quantum wells~\cite{cubic-rashba-paper,emergence,emergent} i.e. $m\sim0.09~m_e$, $g_\perp\approx15.7$, $\alpha_c=2.26\times10^{5}$ meV  A$^3/\hbar^3$,
$\alpha_l=2.01$ meV  A/$\hbar$, we get $\tilde{\alpha}_l=0.0047$, $\tilde{\omega}_z=0.428~B$ and  $\tilde{\omega}_c=0.606~B$ where $B$ is the magnetic field strength in tesla. For all the results that follow, we use these parameters unless stated otherwise. For $B=0.5$ T, the resonant driving frequency is $\omega\approx\omega_z= 6.9\times 10^{11}$ Hz. For $\tilde{F}_0=0.02$, $E_0\approx2127 $ V/m which is well within the attainable limits for modern-day lasers.

\subsection{Analytical results}

For $\tilde{F}_0=0.02$ and $B=0.5$ T and the system parameters mentioned above, we get $\omega_\text{res}=127.24$ MHz. This value can be increased further by applying stronger laser beams. Figure [\ref{comparison-neu}] shows excellent agreement between the analytically and numerically computed values of resonant Rabi frequencies for small values of $\tilde{\alpha}_l$ and $\tilde{F}_0$. Figure [\ref{reso}] shows density plots of the probability amplitude $P_\text{max}$ from Eq. (\ref{P}) as a function of $\tilde{\omega}$ and $\tilde{\alpha}_l$ for a fixed radiation amplitude $\tilde{F}_0=0.08$ and different values of magnetic field -- (a) $B=0.5$ T, (b) $B=1.0$ T and (c) $B=1.5$ T. The dark curves on the plots indicate the resonances in $P_\text{max}$. The width of the resonances $(\propto \gamma)$ gradually increases with $B$ and $\alpha_l$. Since the linear Rashba strength is nearly fixed by the calculations \cite{emergent,emergence}, sharper resonances can be achieved by working at low magnetic fields.

\begin{figure}
		\centering
		\hspace{-0.3cm}\includegraphics[trim={0.8cm 0cm 0cm 0cm},clip,width=9cm]{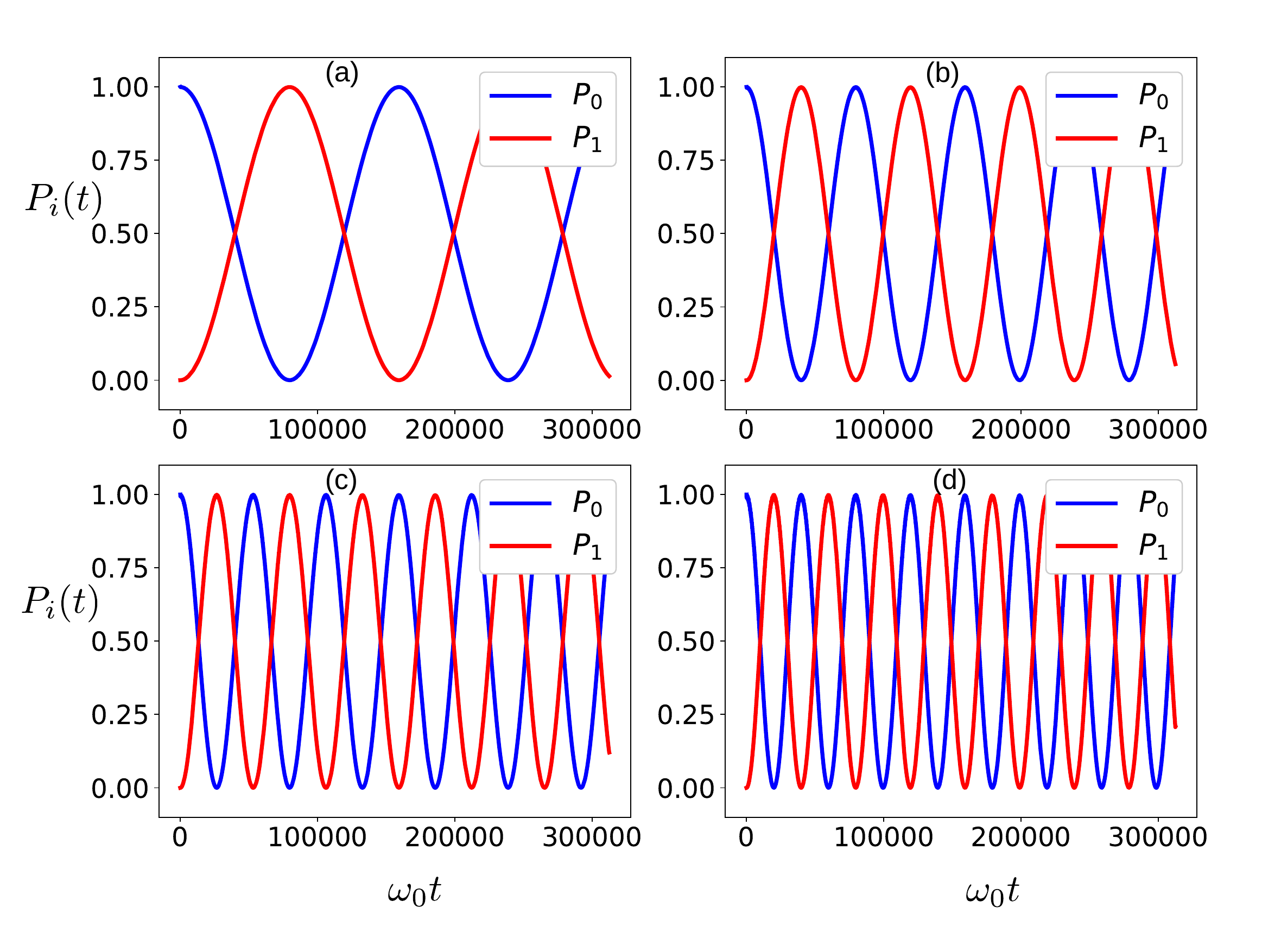}
		\caption{ Plots of occupation probability as a function of time for realistic system parameters (i.e. $B=0.5$ T, $\tilde{\alpha}_l=0.0047,~ \tilde{\omega}_c=0.303,~ \tilde{\omega}_z=0.214,~ \omega=\omega_\text{21}$) and different values of driving strength: (a) $\tilde{F}_0=0.02$,  (b) $\tilde{F}_0=0.04$, (c) $\tilde{F}_0=0.06$ and (d) $\tilde{F}_0=0.08$. The resonant Rabi frequency is an increasing function of the driving strength.}
		\label{Rabi-f}
	\end{figure}

\begin{figure}
		\centering
		\hspace{-0.3cm}\includegraphics[trim={0.5cm 0cm 0cm 0cm},clip,width=9cm]{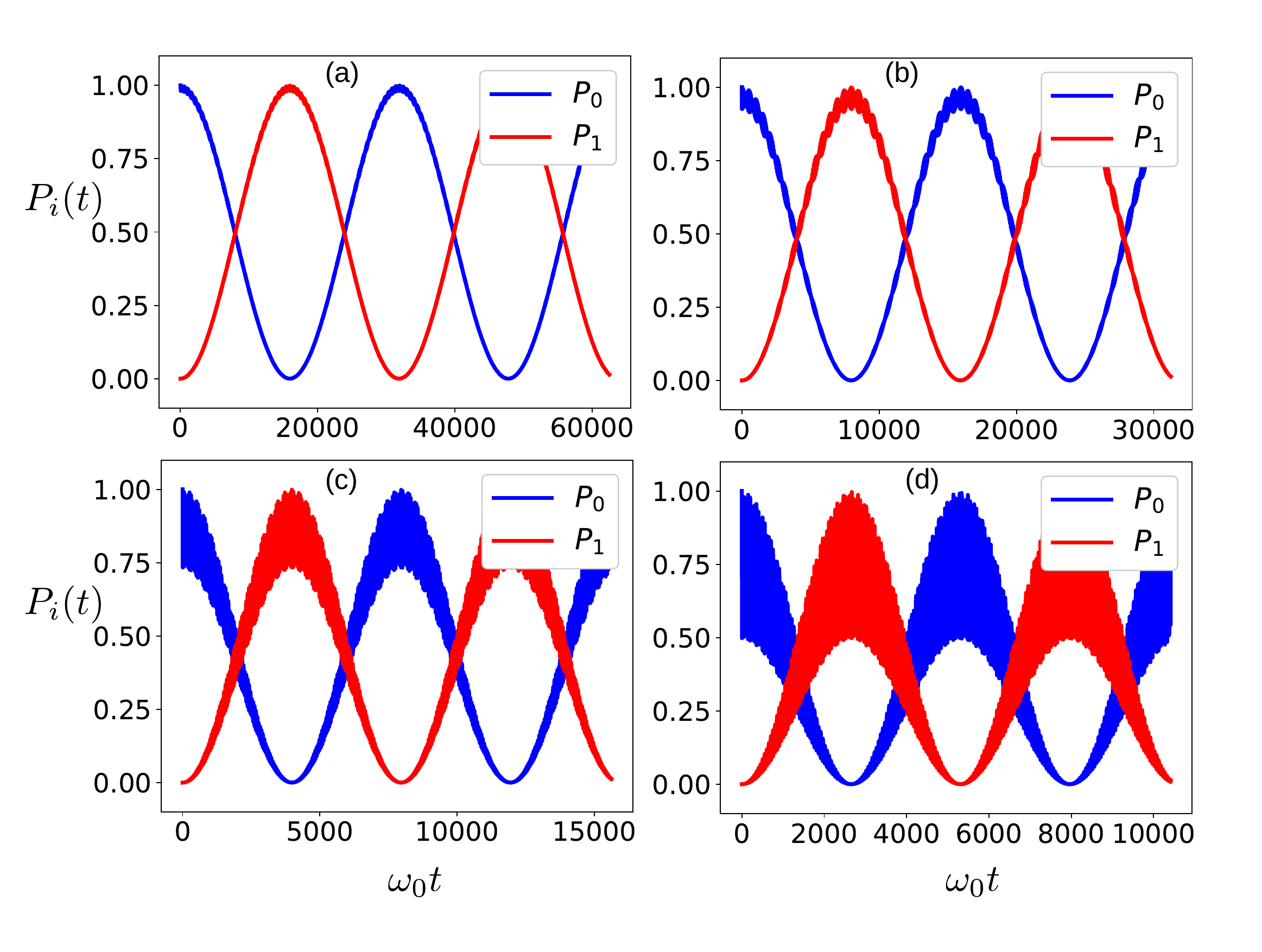}
		\caption{ Plots of occupation probability as a function of time for realistic system parameters (i.e. $B=0.5$ T, $\tilde{\alpha}_l=0.0047,~ \tilde{\omega}_c=0.303,~ \tilde{\omega}_z=0.214,~ \omega=\omega_\text{21}$) and larger values of radiation amplitude: (a) $\tilde{F}_0=0.1$,  (b) $\tilde{F}_0=0.2$, (c) $\tilde{F}_0=0.4$ and (d) $\tilde{F}_0=0.6$.}
		\label{strong}
	\end{figure}

\subsection{Numerical results}

\begin{figure}
		\centering
		\hspace{-0.3cm}\includegraphics[trim={0.5cm 0cm 0cm 0cm},clip,width=8.5cm]{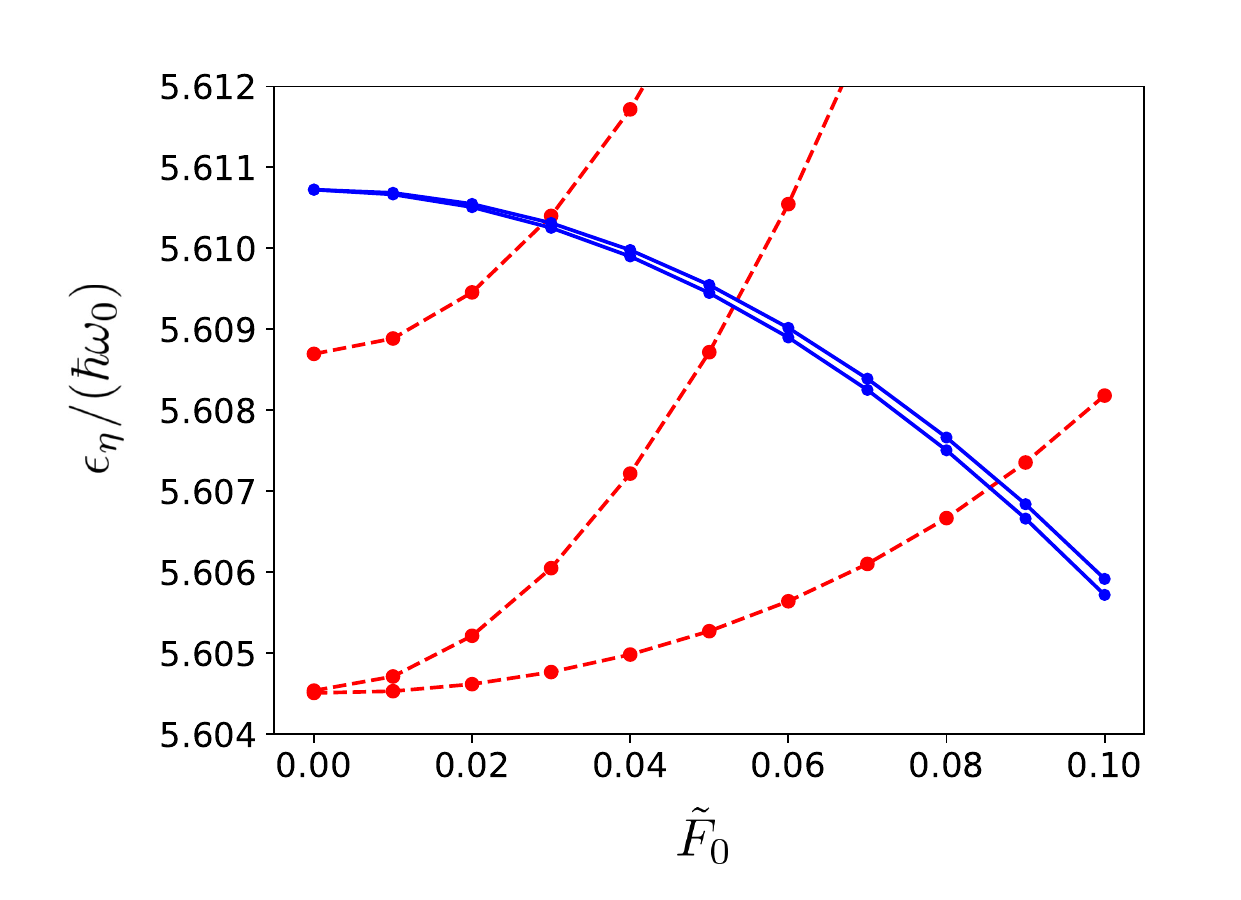}
		\caption{Variation of quasienergies of certain Floquet levels with radiation amplitude for realistic system parameters (i.e. $B=0.5$ T, $\tilde{\alpha}_l=0.0047,~ \tilde{\omega}_c=0.303,~ ,\tilde{\omega}_z=0.214, \omega=\omega_{21}$). The quasienergies of the levels represented by the red dotted curves decrease with $F_0$ and also do not contribute to the Rabi oscillations for the given initial and final states. The levels represented by the blue solid curves participate in the Rabi rotations and the Rabi frequency, equal to the gap between them, increases (linearly) with $F_0$.}
		\label{quasi-plots}
	\end{figure}
 
 First, we use the method of unitary transformation to obtain the Rabi oscillations when $\alpha_l\neq0,~\alpha_c=0$. Figure [\ref{Rabi-f}] shows variation of occupation probabilities of the ground ($|0,0,3/2\rangle$) and first excited ($|0,0,-3/2\rangle$) states of the dot, labelled as $P_0(t)$ and $P_1(t)$ respectively, with time for different values of radiation amplitude when the system is initialized in the ground state and the resonance condition is satisfied. The resonant Rabi frequency clearly increases with the radiation amplitude, as expected from the expression of $\omega_\text{res}$. Similar behaviour is displayed with respect to the variation of $\alpha_l$. The numerical methods can also be used to study the time evolution of the qubit for stronger electrical drives characterized by larger values of $\tilde{F}_0$. This would incorporate the higher order terms of the perturbation theory discussed in Sec. \ref{evol-anal}. Figure [\ref{strong}] shows the probability oscillations for stronger laser beams. We observe that the resonant Rabi frequency increases but a high frequency noise, whose amplitude grows with  $F_0$, is superimposed on the Rabi oscillations. Its origin can be explained using the results of the Schrieffer-Wolff transformation. On the right hand side of Eq. (\ref{swt}), $H^0_\text{SW}$ represents the undriven diagonal (upto the order of $\alpha_l^2$) Hamiltonian, $V_g(t)$ couples identical spin states with different orbital quantum numbers and $[S,V_g(t)]$ couples opposite spin states with same orbital quantum number. For low driving strengths, effect of $V_g(t)$ is negligible as compared to $[S,V_g(t)]$ and nearly no transitions take place from the ground state to the excited states with higher orbital quantum numbers. For larger driving strengths, $V_g(t)$ competes with $[S,V_g(t)]$ and off-resonant transitions ($|\Delta n_i|\neq0$) with smaller amplitudes but a much larger frequency than the resonant ($n_1=n_2=0,\Delta s=\pm3,$)
oscillations take place in addition to the latter. Those transitions appear as a high-frequency noise overlapped with the original Rabi oscillations since the total probability has to be conserved. This noise would hamper the fidelity of the quantum gate at the cost of faster operations. Hence, a low laser amplitude ($\sim$ 2-8 kV/m) is recommendable to perform qubit rotations in this system with a good accuracy. The inefficiency of EDSR for stronger drives has also been observed in double quantum dots \cite{sherman}.

 \begin{figure}
		\centering
		\hspace{-0.3cm}\includegraphics[trim={0.5cm 0cm 0cm 0cm},clip,width=9cm]{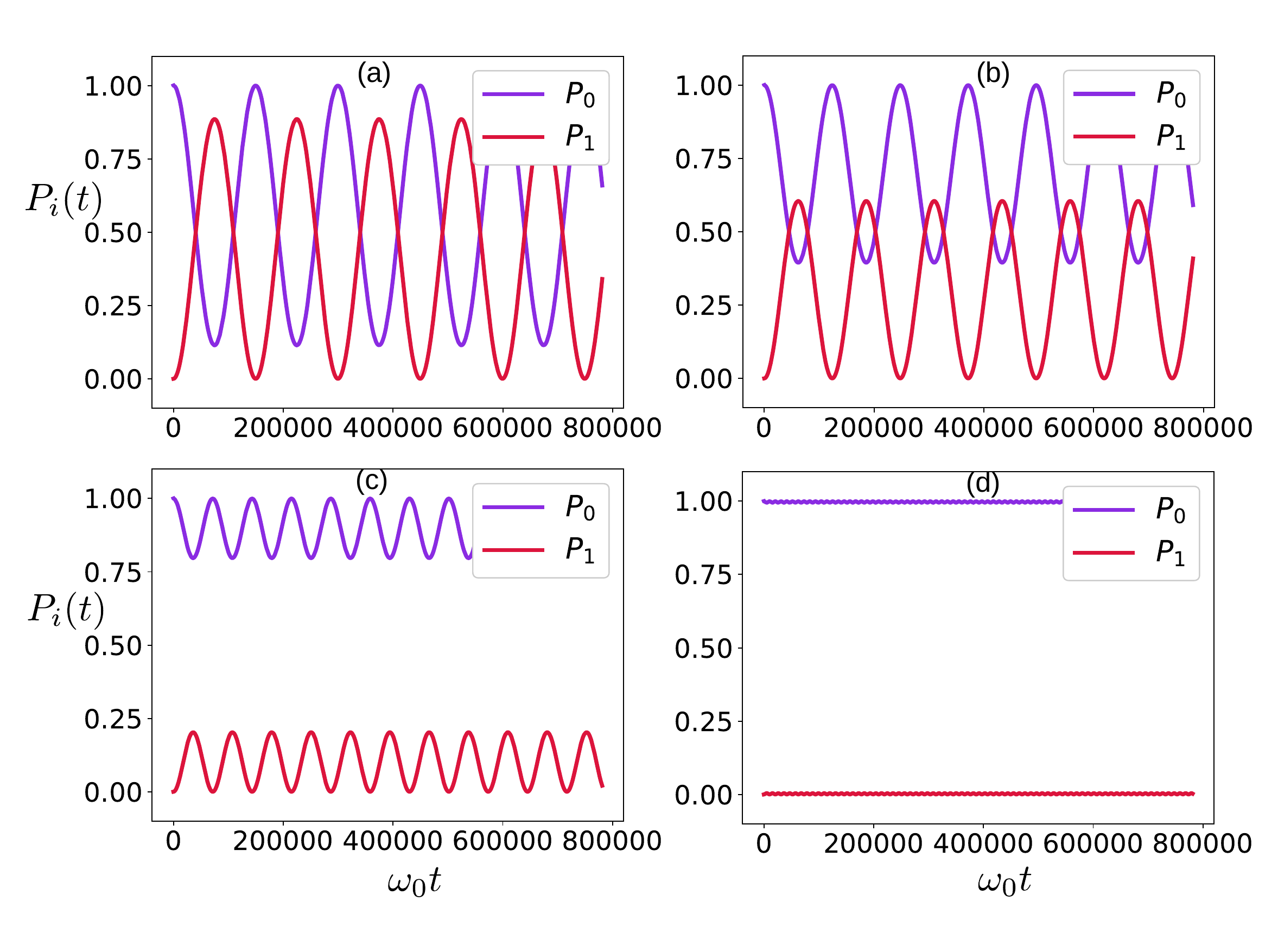}
		\caption{Plots of occupation probability as a function of time for realistic system parameters (i.e. $B=0.5$ T, $\tilde{\alpha}_l=0.0047,~ \tilde{\omega}_c=0.303,~ ,\tilde{\omega}_z=0.214,~\tilde{F}_0=0.02, \omega=\omega_{21}$) and different values of cubic Rashba coupling: (a) $\tilde{\alpha}_c=0.002$,  (b) $\tilde{\alpha}_c=0.003$, (c) $\tilde{\alpha}_c=0.0047$ (equal to $\alpha_l$) and (d) $\tilde{\alpha}_c=0.013$ (typical value). The Rabi oscillations become increasingly non-resonant with increase in $\alpha_c$ and nearly vanish for typical values for $\alpha_c$ for this system.}
		\label{cubic-reso}
	\end{figure}

  As mentioned earlier, no probability oscillations are observed for $\alpha_l=0,~\alpha_c\neq0$ as the cubic Rashba coupling does not support EDSR. When both $\alpha_l\neq0$ and $\alpha_c\neq0$, the method of unitary transformation fails and hence we use Floquet theory to obtain the time dynamics. Firstly, we elaborate how the Floquet theory explains the time evolution.
 In Eq. (\ref{floq-amp}), the term within the first parenthesis represents the projection of the initial state on the $\eta^\text{th}$ Floquet mode. The different $n^\text{th}$-order sidebands of the $\eta^\text{th}$ Floquet mode evolve in time with dynamical phases $\e^{-i (\epsilon_\eta-n\hbar\omega) t/\hbar}$. The term within the second parenthesis denotes the dynamical transition amplitudes from the $\eta^\text{th}$ Floquet mode with ``$-n$" photons (or $n^\text{th}$ order sideband) to the $l^\text{th}$ FD level. For a strictly two-level Rabi problem at resonance, the ground and first excited states have equal magnitudes of projections $(=1/\sqrt{2})$ on each of the Floquet modes and also the first excited state has its projection on the sideband of same photon number $n$ for each $\eta$. As a result, the final transition amplitude from $|0\rangle\to|1\rangle$ is a sum of two  oscillating terms viz. $\e^{-i(\epsilon_0-n\hbar\omega)t/\hbar}$ and $\e^{-i(\epsilon_1-n\hbar\omega)t/\hbar}$ of equal magnitudes, which give rise to Rabi oscillations of frequency equal to the quasienergy difference only i.e. $\omega_\text{res}=|\epsilon_0-\epsilon_1|/\hbar$ and maximum transition probability equal to $2\times(1/\sqrt{2})^2=1$. 
 
 Using (\ref{floq-amp}), the transition probability to the $l^\text{th}$ FD state can be simplified as
	\begin{equation}\label{prob}
		\begin{aligned}
			&P_l(t)=|a_l(t)|^2=\\&\sum_{\{\chi\},\eta>\zeta,n>m}2R^{\eta\zeta}_{lmnm^\prime n^\prime}\cos\l[\frac{(\Delta\epsilon_{\eta\zeta}-\Delta_{nm}\hbar\omega)t}{\hbar}+\theta^{\eta\zeta}_{lmnm^\prime n^\prime}\r]\\&+ \sum_{\{\kappa\},\eta>\zeta}2R^{\eta\zeta}_{lnnm^\prime n^\prime}\cos\l[\Delta \epsilon_{\eta\zeta} t/\hbar+\theta^{\eta\zeta}_{lnnm^\prime n^\prime}\r]\\&+ \sum_{\{\rho\},n>m}2R^{\eta\eta}_{lmnm^\prime n^\prime}\cos\l[-\Delta_{nm}\omega t+\theta^{\eta\eta}_{lmnm^\prime n^\prime}\r]\\&+\sum_{\{\xi\}}|c^n_{l\eta}|^2(c^{n^\prime}_{0,\eta})^* (c_{0,\eta}^{m^\prime})
		\end{aligned}	
	\end{equation}
	where $\{\chi\}=\{\eta, n,n^\prime,\zeta,m,m^\prime\}$, $\{\kappa\}=\{\eta, n,n^\prime,\zeta,m^\prime\}$,
	$\{\rho\}=\{\eta, n,n^\prime,m,m^\prime\}$, $\{\xi\}=\{\eta, n,n^\prime,m^\prime\}$, $\Delta\epsilon_{\eta \zeta}=\epsilon_\eta-\epsilon_\zeta$, $\Delta_{nm}=n-m$, $R^{\eta\zeta}_{lmnm^\prime n^\prime}=|c_{l,\eta}^n (c^{n^\prime}_{0,\eta})^* (c_{l,\zeta}^m)^*c^{m^\prime}_{0,\zeta}|$ and $\theta^{\eta\zeta}_{lmnm^\prime n^\prime}=\text{Arg}[c_{l,\eta}^n (c^{n^\prime}_{0,\eta})^* (c_{l,\zeta}^m)^*c^{m^\prime}_{0,\zeta}]$.
	
	For the multi-level system in consideration, we find that the second summation in Eq. (\ref{prob}) is the dominant contribution to the time-dependent part of the oscillations. This means that two quasienergy levels with an identical photon number contribute to the Rabi oscillations. Figure [\ref{quasi-plots}] shows the variation of quasienergies of some of the Floquet levels with the radiation amplitude at resonance and in absence of cubic Rashba coupling. The levels denoted by the blue curves are the ones which have equal projections on both the initial ($|0,0,+3/2\rangle$) and final ($|0,0,-3/2\rangle$) FD levels. Hence, these are the levels which participate in the Rabi oscillations in Floquet picture and the Rabi frequency is equal to the difference of their quasienergies. The gap between the levels increases (linearly) with the radiation amplitude and is consistent with the values of $\omega_\text{res}$.

In figure [\ref{cubic-reso}], we study the dependence of the resonant Rabi oscillations on the cubic Rashba strength $\alpha_c$ using numerical results of the Floquet theory. We observe a gradual diminishing of the resonant Rabi oscillations on increasing the values of $\alpha_c$. We find that that oscillations nearly disappear for realistic values of $\alpha_c$.

\section{Conclusion}\label{conc}

To conclude, we have studied the time dynamics of a planar Ge hole spin-qubit driven by coherent circularly polarized radiation in presence of an out-of-plane magnetic field and Rashba SOCs. The coherent beam may be supplied by ultrafast laser pulses which have been extensively used in recent years for Floquet engineering. We consider the recently claimed $p$-linear and the dominant spherically symmetric component of the $p-$cubic Rashba SOC found in heavy holes. We show that a laser drive of suitable frequency can be used to perform qubit rotations in presence of linear Rashba coupling (only), an effect similar to EDSR with ac gate-voltages. We have shown how the problem can be solved using two different gauges viz. velocity and length gauges. For small Rashba strength and weak laser beam, we use perturbation theory and make a Schrieffer-Wolff transformation to obtain an analytical form of the Rabi frequency. The Rabi frequency is linearly proportional to the  radiation amplitude and the linear Rashba strength. For a laser beam of electric field amplitude 2.127 kV/m, the Rabi frequency is approximately 127 MHz for realistic system parameters. We observe that the width of resonance increases with increase in strengths of the magnetic field and Rashba coupling. 

For higher radiation amplitude, we employ methods of unitary transformation and Floquet theory independently for numerical computation of the qubit dynamics. The method of unitary transformation deals with transforming to a rotating frame of reference and exploiting the rotational symmetry of the system in the total Hilbert space of spin and orbital degrees of freedom. This method is valid when only one of the two Rashba couplings (linear or cubic) is present. Hence, we use it to study the effect of stronger laser beams in the Rabi oscillations when only linear Rashba coupling is considered. We encounter a high frequency noise in the Rabi oscillations for stronger laser beams thereby rendering the drive unsuitable for qubit manipulation. As already known, no Rabi oscillations occur when only cubic Rashba coupling is present in the system. 

We have used Floquet theory to solve the Schr{\"o}dinger equation when both linear and cubic Rashba couplings are simultaneouly present in the system. The numerical results show that the Rabi oscillations gradually diminish as we increase the cubic Rashba strength for a fixed linear Rashba paramater. This implies that the system is driven out of resonance by the cubic Rashba coupling. For a realistic value of the cubic Rashba parameter, the Rabi oscillations nearly vanish despite the presence of linear Rashba coupling. Hence, the cubic Rashba coupling, which can be controlled by changing the interficial electric field using gate electrodes, has to be highly minimized to observe resonant Rabi oscillations in this system. We have also shown that the Rabi frequency is equal to the quasienergy gap between two Floquet levels which is an increasing function of radiation amplitude.

Our results may be useful to achieve optical control of hole qubits in Ge quantum dots. With photolithography already boosting the semiconductor fabrication industry, the use of coherent laser beams for manipulation of spin-qubits is another attractive option to develop quantum NOT gates with the aid of strong SOC of the heavy-hole states in 2D Ge heterostructures.
\\

\begin{center}
    {\bf ACKNOWLEDGEMENTS}
\end{center}

This work was funded by the Free state of Bavaria through the ``Munich Quantum valley" as part of a lighthouse project named ``Quantum circuits with spin qubits and hybrid Josephson junctions". We thank Jordi Picó-Cortés and Luca Magazzù for useful discussions.\\
 
\appendix

\section{Matrix elements of $H_R$ and $V_n$}\label{app-matrix-elem}

The matrix elements of $H_R$ in the FD basis are

\begin{widetext}
\begin{equation}
			\begin{aligned}
				\langle n_1^\prime,n_2^\prime,s^\prime|H_R|n_1,n_2,s\rangle=f_0(n_1^\prime,n_2^\prime,n_1,n_2)\delta_{s^\prime,3/2}~\delta_{s,-3/2}+f_0(n_1,n_2,n_1^\prime,n_2^\prime)\delta_{s^\prime,-3/2}~\delta_{s,3/2},
			\end{aligned}
		\end{equation}
where 
		\begin{equation}
			\begin{aligned}
				f_0(n_1^\prime,n_2^\prime,n_1,n_2)=&\alpha_c(m\hbar)^{3/2}\Big{[}f_-^3\sqrt{n_2(n_2-1)(n_2-2)}~\delta_{n_1^\prime,n_1}~\delta_{n_2^\prime,n_2-3}+3 f_-^2 f_+ \sqrt{(n_1+1)n_2 (n_2-1)}~\delta_{n_1^\prime,n_1+1}~\delta_{n_2^\prime,n_2-2}\\
				&+3 f_- f_+^2\sqrt{(n_1+1)(n_1+2)n_2}~\delta_{n_1^\prime,n_1+2}~\delta_{n_2^\prime,n_2-1}+f_+^3\sqrt{(n_1+1)(n_1+2)(n_1+3)}~\delta_{n_1^\prime,n_1+3}~\delta_{n_2^\prime,n_2}\Big{]}\\
				&+\alpha_l\sqrt{m\hbar}\Big{[}f_+\sqrt{n_1+1}~\delta_{n_1^\prime,n_1+1}~\delta_{n_2^\prime,n_2}+f_-\sqrt{n_2}~\delta_{n_1^\prime,n_1}~\delta_{n_2^\prime,n_2-1}\Big{]}.
			\end{aligned}
		\end{equation}
The matrix elements of $V_n$ are
		\begin{equation}
			\begin{aligned}
				\langle n_1^\prime,n_2^\prime,s^\prime|V_{0}|n_1,n_2,s\rangle=&\frac{|e|^2A_0^2}{2m}~\delta_{n_1^\prime,n_1}~\delta_{n^\prime_2,n_2}~\delta_{s^\prime,s},
			\end{aligned}
		\end{equation}
		\begin{equation}
			\begin{aligned}
				\langle n_1^\prime,n_2^\prime,s^\prime|V_{1}|n_1,n_2,s\rangle=&3 i\alpha_c |e|A_0 m\hbar \Big{[}f_+^2 \sqrt{(n_1+1)(n_1+2)}~\delta_{n_1^\prime,n_1+2}~\delta_{n_2^\prime,n_2}+2f_+f_-\sqrt{(n_1+1)n_2}~\delta_{n_1^\prime,n_1+1}~\delta_{n_2^\prime,n_2-1}\\
				&+f_-^2 \sqrt{n_2(n_2-1)}~\delta_{n_1^\prime,n_1}~\delta_{n_2^\prime,n_2-2}\Big{]}~\delta_{s^\prime,3/2}~\delta_{s,-3/2}~+~i \alpha_l |e| A_0~\delta_{n_1^\prime,n_1}~\delta_{n_2^\prime,n_2}~\delta_{s^\prime,3/2}~\delta_{s,-3/2}\\
				&i\frac{|e|A_0}{2m}\sqrt{m\hbar }\Big{[}f_+\sqrt{n_1}~\delta_{n_1^\prime,n_1-1}~\delta_{n_2^\prime,n_2}+f_-\sqrt{n_2+1}~\delta_{n_1^\prime,n_1}~\delta_{n_2^\prime,n_2+1}\Big{]}~\delta_{s^\prime,s},
			\end{aligned}
		\end{equation}
		
		\begin{equation}
			\begin{aligned}
				\langle n_1^\prime,n_2^\prime,s^\prime|V_{2}|n_1, n_2,s\rangle =& -3\alpha_c \sqrt{m\hbar}|e|^2A_0^2\Big{[}f_+\sqrt{n_1+1}~\delta_{n_1^\prime,n_1+1}~\delta_{n_2^\prime,n_2}+f_-\sqrt{n_2}~\delta_{n_1^\prime,n_1}~\delta_{n_2^\prime,n_2-1}\Big{]}~\delta_{s^\prime,3/2}~\delta_{s,-3/2}
			\end{aligned}
		\end{equation}
		and
		\begin{equation}
			\begin{aligned}
            \langle n_1^\prime,n_2^\prime,s^\prime|V_{3}|n_1, n_2,s\rangle = -i\alpha_c |e|^3A_0^3~\delta_{n_1^\prime,n_1}~\delta_{n_2^\prime,n_2}~\delta_{s^\prime,3/2}~\delta_{s,-3/2}.
			\end{aligned}
		\end{equation}
\end{widetext}

\section{Schrieffer-Wolff transformation}\label{app-SW}

The SW transformation can be written as 
 \begin{equation}\label{SW}
    H^0_{\text{SW}}=\e^S (H_\text{FD} + H_R) \e^{-S}\approx H_\text{FD}+[S,H_R]/2 
 \end{equation}
where $S^\dagger=-S$ and
\begin{equation}\label{commrelation}
    [H_\text{FD},S]=H_R.
\end{equation}
We consider the following ansatz for $S$:
 \begin{equation}\label{ansatz}
     S=\hat{S}^{(1)}+\hat{S}^{(2)}\sigma_+ -\hat{S}^{{(2)}^\dagger}\sigma_- 
 \end{equation}
 where $\hat{S}^{(i)}=S^{(i)}_{1a} \hat{a}_1+S^{(i)}_{1b} \hat{a}_1^\dagger+S^{(i)}_{2a} \hat{a}_2+S^{(i)}_{2b} \hat{a}_2^\dagger$ with $i=1,2$. Using (\ref{ansatz})  in (\ref{commrelation}) and comparing both the sides, we get $\hat{S}^{(1)}=0$ and
 \begin{equation}\label{Sform}
     \hat{S}^{(2)}=\alpha\sqrt{\frac{m}{\hbar}}\l(\frac{f_+}{\omega_1-\omega_z}\hat{a}_1^\dagger-\frac{f_-}{\omega_2+\omega_z}\hat{a}_2\r).
 \end{equation}
Using (\ref{Sform}) in (\ref{SW}), we get
\begin{equation}
    H^0_{SW}=H_\text{FD}+\hat{\xi}_+ \mathbb{P}_+ +\hat{\xi}_- \mathbb{P}_-,
\end{equation}
where $\mathbb{P}_\pm=(1\pm\sigma_z)/2$ are the spin-projection operators and 
\begin{equation}
\begin{aligned}
     &\hat{\xi}_+=\frac{\alpha_l^2 m}{2}\bigg[\frac{2f_+^2}{\omega_1-\omega_z}\hat{n}_1-\frac{2f_-^2}{\omega_2+\omega_z}(1+\hat{n}_2)\\
    &+f_+ f_-(\hat{a}_1\hat{a}_2+\hat{a}_2^\dagger\hat{a}_1^\dagger)\l(\frac{1}{\omega_1-\omega_z}-\frac{1}{\omega_2+\omega_z}\r)\bigg],
\end{aligned}
\end{equation}

\begin{equation}
\begin{aligned}
    &\hat{\xi}_-=-\frac{\alpha_l^2 m}{2}\bigg[\frac{2f_+^2}{\omega_1-\omega_z}(1+\hat{n}_1)-\frac{2f_-^2}{\omega_2+\omega_z}\hat{n}_2\\
    &+f_+ f_-(\hat{a}_1\hat{a}_2+\hat{a}_2^\dagger\hat{a}_1^\dagger)\l(\frac{1}{\omega_1-\omega_z}-\frac{1}{\omega_2+\omega_z}\r)\bigg],
\end{aligned}
\end{equation}
 
For a weak driving ($\frac{F_0}{ \hbar \omega_0}\sqrt{\frac{\hbar}{m \omega_0}}=\tilde{F}_0\ll1$), the SW Hamiltonian can be written as
\begin{equation}\label{swt}
\begin{aligned}
    H_\text{SW}(t)&=H^0_\text{SW}+\e^S V_g(t) \e^{-S}\\ & \approx H^0_\text{SW}+V_g(t)+[S,V_g(t)]
\end{aligned}
\end{equation}

Then, the lowest energy block of $H_\text{SW}(t)$ (spanned by $|0,0,\pm 3/2\rangle$ states) can be written as
\begin{equation}
\begin{aligned}
H_\text{eff}(t)&=-\frac{\hbar\omega_z}{2}\sigma_z-\frac{\alpha_l^2 m f_-^2}{\omega_2+\omega_z} \mathbb{P}_+ -\frac{\alpha_l^2 m f_+^2}{\omega_1-\omega_z} \mathbb{P}_-\\
&-i \gamma\e^{i\omega t}\sigma_+ +i \gamma\e^{-i\omega t}\sigma_-\\
    &=\l(\begin{array}{cc}
       -\frac{\hbar\omega_z}{2} -\frac{\alpha_l^2 m f_-^2}{\omega_2+\omega_z} &  -i \gamma \e^{i\omega t}\\
        i \gamma \e^{-i\omega t} &  \frac{\hbar\omega_z}{2} -\frac{\alpha_l^2 m f_+^2}{\omega_1-\omega_z}
    \end{array}\r)
\end{aligned}
\end{equation}
where
\begin{equation}\label{Rabi-gamma}
    \gamma=\frac{F_0\alpha_l}{2\sqrt{\Omega}}\l(\frac{f_+}{\omega_1-\omega_z}+\frac{f_-}{\omega_2+\omega_z}\r).
\end{equation}

 \section{Method of Unitary transformation}\label{app-unitary}

We can get a static Hamiltonian through a unitary transformation when atleast one of the Rashba couplings is absent. Imposing this condition, the driven Hamiltonian in the velocity gauge containing exclusively the linear or cubic Rashba coupling is given by
\begin{equation}\label{driven-one}
		\begin{aligned}
			&H^\text{(L)/(C)}(t)=H_{\text{FD}}+H_r(t)+H^{\text{(L)/(C)}}_R(t)
   \end{aligned}
\end{equation} where 
\begin{equation}
    H_r(t)=\frac{|e|^2 A_0^2}{2m}-\frac{|e|A_0}{m}\big(P_x\cos\omega t-P_y \sin\omega t\big),
\end{equation}
\begin{equation}
    H^{\text{(L)}}_R(t)=\alpha_l\big[-i(P_- -|e|A_0 \e^{i \omega t})\sigma_+ + \text{h.c.}\big],
\end{equation}
and
\begin{equation}
    H^{\text{(C)}}_R(t)=\alpha_c\big[i(P_- -|e|A_0 \e^{i \omega t})^3\sigma_+ + \text{h.c.}\big].
\end{equation}

In the length gauge, the driven Hamiltonian is simply

\begin{equation}\label{driven-two}
		H^{\prime\text{(L)/(C)}}(t)=H_{\text{FD}}+H_R^\text{(L)/(C)}+ V_g(t).
\end{equation}

The driven Hamiltonians in either gauge can be reduced to a static one by solving the TDSE in a frame rotating with ${\bf A}_r$. If $|\psi(t)\rangle$ is the solution in the rest frame, then the solution in the rotating frame, having angular speed `$-\omega$' (clockwise) about $z$-axis, is given by

\begin{equation}\label{rotationeq}
     |\Phi(t)\rangle=\mathcal{R}_z(\omega t)|\psi(t)\rangle
 \end{equation}
 where $\mathcal{R}_z$ is the standard rotation operator about $z$-axis. For the driven Rashba Hamiltonian, it can be written as
 \begin{equation}\label{rotationop}
     \mathcal{R}^{\text{(L)/(C)}}_z(\omega t)=\e^{-i \hat{J}_z^{\text{(L)/(C)}} \omega t/\hbar}
 \end{equation}
 where $\hat{J}_z^{\text{(L)/(C)}}$ are defined in Eqs. (\ref{JzL}) and (\ref{JzC}).
 Using (\ref{rotationeq}) in the TDSE and working with the Hamiltonian in length gauge, we get
\begin{equation}
    i \hbar \frac{\partial}{\partial t}|\Phi(t)\rangle=\l(\mathcal{R}_z H^{\prime \text{(L)/(C)}} (t) \mathcal{R}_z^\dagger+i \hbar \mathcal{R}_z^\dagger \frac{\partial }{\partial t} \mathcal{R}_z\r)|\Phi(t)\rangle.
\end{equation}
where $\mathcal{R}_z\equiv\mathcal{R}^\text{(L)/(C)}_z$. 
Using Eq. (\ref{rotationop}), it reduces to
\begin{equation}
     i \hbar \frac{\partial}{\partial t}|\Phi(t)\rangle=\mathcal{H}^{\prime \text{(L)/(C)}}|\Phi(t)\rangle
\end{equation}
where
\begin{equation}\label{Hprime}
    \mathcal{H}^{\prime \text{(L)/(C)}}=H^{\prime \text{(L)/(C)}}(t=0)+\hat{J}_z^\text{(L)/(C)}\omega
\end{equation}
is the time-independent Hamiltonian in the rotating frame. In terms of ladder operators, the transformed Hamiltonian in the velocity gauge can be written as
\begin{equation}
    \mathcal{H}^{\text{(L)/(C)}}=\mathcal{H}_0+\mathcal{H}_R^{\text{(L)/(C)}}
\end{equation}
where
\begin{equation}
\begin{aligned}
    &\mathcal{H}_0=\hbar \Omega (\hat{n}_1+\hat{n}_2+1)+\hbar\l(\omega-\frac{\omega_c}{2}\r)(\hat{n}_2-\hat{n}_1)\\
    &+\frac{|e|^2 A_0^2}{2m}-\frac{i |e|A_0\sqrt{m\hbar}}{2m}\l(f_+(\hat{a}_1^\dagger-\hat{a}_1)-f_-(\hat{a}_2^\dagger-\hat{a}_2)\r),
\end{aligned}
\end{equation}
\begin{equation}
\begin{aligned}
    &\mathcal{H}_R^{(\text{L})}=\frac{\hbar}{2}(\omega-\omega_z)\sigma_z \\
    &+\alpha_l\bigg[\l(\sqrt{m\hbar}(f_+ \hat{a}_1^\dagger+f_- \hat{a}_2) +i|e|A_0 \r)\sigma_+ + \text{h.c.}\bigg],
\end{aligned}
\end{equation}
\begin{equation}
\begin{aligned}
    &\mathcal{H}_R^{(\text{C})}=\frac{\hbar}{2}(3\omega-\omega_z)\sigma_z \\
    &+\alpha_c\bigg[\l(\sqrt{m\hbar}(f_+ \hat{a}_1^\dagger+f_- \hat{a}_2) +i|e|A_0 \r)^3\sigma_+ + \text{h.c.}\bigg].
\end{aligned}
\end{equation}

Similarly, for the length gauge, we get

\begin{equation}\label{Halpha-ladder2}
    \mathcal{H}^{\prime\text{(L)/(C)}}=\mathcal{H}^\prime_0+\mathcal{H}_R^{\prime\text{(L)/(C)}}
\end{equation}
where
\begin{equation}
\begin{aligned}
    &\mathcal{H}^\prime_0=\hbar \Omega (\hat{n}_1+\hat{n}_2+1)+\hbar\l(\omega-\frac{\omega_c}{2}\r)(\hat{n}_2-\hat{n}_1)\\
    &+\frac{i F_0}{2}\sqrt{\frac{\hbar}{m\Omega}}\l(\hat{a}_1-\hat{a}_1^\dagger-\hat{a}_2+\hat{a}_2^\dagger\r),
\end{aligned}
\end{equation}
\begin{equation}
\begin{aligned}
    \mathcal{H}_R^{\prime (\text{L})}=\frac{\hbar}{2}(\omega-\omega_z)\sigma_z + H_R^{(\text{L})},
\end{aligned}
\end{equation}
\begin{equation}
\begin{aligned}
    \mathcal{H}_R^{\prime(\text{C})}=\frac{\hbar}{2}(3\omega-\omega_z)\sigma_z + H_R^{(\text{C})}
\end{aligned}
\end{equation}

and $H_R^{(\text{C})}$ and $H_R^{(\text{L})}$ are defined in equations (\ref{Rashba-C}) and (\ref{Rashba-L}) respectively.

Hence, the time-evolved state in the rest frame
\begin{equation}
    |\psi(t)\rangle=\mathcal{R}_z^\dagger|\Phi(t)\rangle=\e^{i \hat{J}_z^{\text{(L)/(C)}}\omega t/\hbar}\e^{-i \mathcal{H}^{\prime\text{(L)/(C)}} t/\hbar}|\psi(0)\rangle.
\end{equation}
If the system is initialized in an FD state $|n_1^i,n_2^i,s^i\rangle$, then the transition amplitude to a state  $|n_1^f,n_2^f,s^f\rangle$ is
\begin{equation}
\begin{aligned}
    a_f(t)&=\langle n_1^f,n_2^f,s^f|\psi(t)\rangle\\
    &=\e^{i(n_2^f-n_1^f+s^{f\text{(L)/(C)}}})\omega t\sum_{m}^{}\e^{-i\varepsilon_m^\prime t/\hbar }d_{f,m}^\prime (d_{i,m}^\prime)^*~~~,
\end{aligned}  
\end{equation}
where $\mathcal{H}^{\prime\text{(L)/(C)}} |\varepsilon_m^\prime\rangle =\varepsilon_m^\prime |\varepsilon_m^\prime\rangle$, $d_{f,m}^\prime=\langle n_1^f,n_2^f,s^f|\varepsilon_m^\prime\rangle$, $d_{i,m}^\prime=\langle n_1^i,n_2^i,s^i|\varepsilon_m^\prime\rangle$,
$s^{f\text{(L)}}=s^f/3$ and $s^{f\text{(C)}}=s^f$ with $s^f=\pm3/2$. The transition probabilities are 
\begin{equation}
    \begin{aligned}
        P_f(t)&=
    \sum_{m,n}^{}\e^{-i(\varepsilon_m^\prime-\varepsilon_n^\prime) t/\hbar} d_{f,m}^\prime (d_{i,m}^\prime)^* (d_{f,n}^\prime)^* d_{i,n}^\prime,\\
    &= \sum_{m}^{}|d_{f,m}^\prime|^2 |d_{i,m}^\prime|^2
    \\&+\sum_{m\neq n}^{}\e^{-i(\varepsilon_m^\prime-\varepsilon_n^\prime) t/\hbar} d_{f,m}^\prime (d_{i,m}^\prime)^* (d_{f,n}^\prime)^* d_{i,n}^\prime.
    \end{aligned}
\end{equation}

	\section{Computation of Floquet modes}\label{appA}
	
	The dynamics of a quantum system is described by the time-dependent TDSE,
	\begin{equation}\label{tdse}
		i\hbar\frac{\partial}{\partial t}|\psi(t)\rangle=H(t)|\psi(t)\rangle.
	\end{equation}
	If $H(t+T)$=$H(t)$, the Floquet theorem states that there exists solutions to Eq. (\ref{tdse}) called Floquet states given by
	\begin{equation}\label{floq}
		|u_\eta(t)\rangle=\e^{-i \epsilon_\eta t/\hbar}|\phi_\eta(t)\rangle,
	\end{equation}
	where $\epsilon_\eta$ is a real-valued number called the quasienergy and $|\phi_\eta(t)\rangle$ is called the Floquet mode which has the same periodicity as the Hamiltonian, i.e. $|\phi_\eta(t+T)\rangle=|\phi_\eta(t)\rangle$.
	
	On substituting Eq. (\ref{floq}) in Eq. (\ref{tdse}), we get:
	\begin{equation}\label{eig}
		\l(H(t)-i\hbar\frac{\partial}{\partial t}\r)|\phi_\eta(t)\rangle=\epsilon_\eta|\phi_\eta(t)\rangle
	\end{equation}
	So, the quasienergies are eigenvalues of the Floquet quasienergy operator $\l(H(t)-i\hbar\frac{\partial}{\partial t}\r)$ with Floquet modes as the eigenstates.
	
	The Floquet modes can be expanded in the FD basis $\l\{|l\rangle\r\}$ (or any orthonormal basis) as: 
	\begin{equation}\label{psit2}
		|\phi_\eta (t)\rangle=\sum_{l}c_{l,\eta}(t)|l\rangle,
	\end{equation} 
	where $c_{l,\eta}(t)=\langle l|\phi_\eta (t)\rangle$ with $c_{l,\eta}(t)=c_{l,\eta}(t+T)$. Since $c_{l,\eta}(t)$ are time-periodic, they can be expanded in Fourier basis as:
	\begin{equation}\label{psit3}
		c_{l,\eta}(t)=\sum_{n=-\infty}^{\infty}c^n_{l,\eta}\e^{i n \omega t}, 
	\end{equation}
	with $\omega=2\pi/T$ being the angular frequency of the periodic drive. Using (\ref{psit2}) and (\ref{psit3}), the Floquet modes can be rewritten as
	\begin{equation}\label{fourierbasis}
		|\phi_\eta(t)\rangle=\sum_{n=-\infty}^{\infty}\sum_{l} c_{l,\eta}^n|l\rangle \e^{i n\omega t}.
	\end{equation}
	Substituting Eq. (\ref{fourierbasis}) in Eq. (\ref{eig}), multiplying $\e^{-i n^\prime \omega t}\langle l^\prime|$ from left and time-averaging over a period $T$ gives:
	\begin{equation}
		\begin{aligned}
			\sum_{n=-\infty,l}^{\infty}\Bigg(\frac{1}{T}\int_{0}^{T}&\langle l^\prime |H(t)| l \rangle \e^{i(n-n^\prime)\omega t}dt+ n\hbar\omega \delta_{n,n^\prime}\delta_{l,l^\prime}\Bigg)c_{l,\eta}^n\\&=\epsilon_\eta c_{l^\prime,\eta}^{n^\prime}
		\end{aligned}
	\end{equation}
	The above system of equations represents an infinite-dimensional matrix eigenvalue equation:
	\begin{equation}\label{matrix}
		H_F \Phi_\eta= \epsilon_\eta \Phi_\eta
	\end{equation} 
	where $H_F$ is the Floquet Hamiltonian matrix and $\Phi_\eta$ is its eigenvector corresponding to the quasienergy eigenvalue $\epsilon_\eta$. The matrix elements of $H_F$ and $\Phi_\eta$ are given by
	\begin{equation}
		\langle l^\prime,n^\prime |H_F| l,n \rangle=\frac{1}{T}\int_{0}^{T}\langle l^\prime |H(t)| l \rangle \e^{i(n-n^\prime)\omega t}dt+ n\hbar\omega \delta_{n,n^\prime}\delta_{l,l^\prime} 
	\end{equation}
	and $\Phi_\eta=(.~.~c_{l,\eta}^{n}.~.~)^\mathcal{T}$ where $\mathcal{T}$ stands for transpose. The eigenvalues are obtained numerically by truncating the matrix upto a certain order in $n$ depending on the strength of the periodic drive. Once $\Phi_\eta$ is computed, the Floquet mode can be obtained by plugging the coefficients $\{c_{l,\eta}^{n}\}$ back into equation (\ref{fourierbasis}).
	
	\section{Time-evolved state in Floquet picture}\label{appB}
	
	At $t=0$, equation (\ref{fourierbasis}) gives
	\begin{equation}\label{eta0}
		|\phi_\eta(0)\rangle=\sum_{l}\l(\sum_{n^\prime=-\infty}^{\infty} c_{l,\eta}^{n^\prime}\r)|l\rangle.
	\end{equation}
	
	We drop the infinite summation limit of $n^\prime$ from this point as we consider only a finite $n_{\text{max}}$. It can be shown that $\{|\phi_\eta(0)\rangle\}$ form a complete basis i.e. $\sum_{\eta}^{}|\phi_\eta (0)\rangle \langle \phi_\eta(0)|=1$. Hence, if the system is initialized in a FD state, say $|i\rangle$, then 
	\begin{equation}
		\begin{aligned}
			|\psi(0)\rangle=|i\rangle&=\sum_{\eta}^{}|\phi_\eta(0) \rangle \langle \phi_\eta (0) |i\rangle \\
			& =\sum_{\eta}\l(\sum_{n^\prime}(c_{i,\eta}^{n^\prime})^*\r) |\phi_\eta (0)\rangle,
		\end{aligned}
	\end{equation}
	
	To obtain the time-evolved state $|\psi(t)\rangle$, we use the Floquet time-evolution operator 
 
 \begin{equation}\label{Ut}
     U(t)=\mathcal{P}(t)\e^{-i \mathcal{H}_F t/\hbar},
 \end{equation}
 where $\mathcal{P}(t)=\mathcal{P}(t+T)$ is a time-periodic unitary operator (with $\mathcal{P}(0)=1$) and $ \mathcal{H}_F$ is a time-independent Hermitian operator called the Floquet Hamiltonian. The stroboscopic time-evolution operator is $U(T)=\e^{-i \mathcal{H}_F T/\hbar}$. The application of $U(t)$ on $|\psi(0)\rangle$ gives
	\begin{equation}\label{psit1}
		\begin{aligned}
			|\psi(t)\rangle&=\sum_{\eta}^{}\l(\sum_{n^\prime}(c_{i,\eta}^{n^\prime})^*\r) \mathcal{P}(t)\e^{-i \mathcal{H}_F t/\hbar} |\phi_\eta (0)\rangle
		\end{aligned}
	\end{equation}
	Since $|\phi_\eta (0)\rangle$ is an eigenstate of $\mathcal{H}_F$ [see (\ref{matrix})] with eigenvalue $\epsilon_\eta$, we get
	\begin{equation}
		\begin{aligned}
			|\psi(t)\rangle&=\sum_{\eta}^{}\l(\sum_{n^\prime}(c_{i,\eta}^{n^\prime})^*\r) \e^{-i \epsilon_\eta t/\hbar} \mathcal{P}(t) |\phi_\eta (0)\rangle\\
			&=\sum_{\eta}^{}\l(\sum_{n^\prime}(c_{i,\eta}^{n^\prime})^*\r) \e^{-i \epsilon_\eta t/\hbar} |\phi_\eta (t)\rangle\\
			&=\sum_{\eta}^{}\l(\sum_{n^\prime}(c_{i,\eta}^{n^\prime})^*\r)\l(\sum_{l,n} c_{l,\eta}^{n}\e^{-i (\epsilon_\eta-n\hbar\omega) t/\hbar} \r)|l\rangle\\
		\end{aligned}
	\end{equation}
	where we have used Eq. (\ref{fourierbasis}) in the last step. Thus, the probability amplitude of finding the system in state $|f\rangle$ when initialized in state  $|i\rangle$ is
	\begin{equation}\label{if-final}
		\langle f |\psi(t)\rangle=\sum_{\eta}^{}\l(\sum_{n^\prime}(c_{i,\eta}^{n^\prime})^*\r)\l(\sum_{n} c_{f,\eta}^{n}\e^{-i (\epsilon_\eta-n\hbar\omega) t/\hbar} \r)
	\end{equation}

\bibliography{bibliography}

\begin{thebibliography}{76}%
\makeatletter
\providecommand \@ifxundefined [1]{%
 \@ifx{#1\undefined}
}%
\providecommand \@ifnum [1]{%
 \ifnum #1\expandafter \@firstoftwo
 \else \expandafter \@secondoftwo
 \fi
}%
\providecommand \@ifx [1]{%
 \ifx #1\expandafter \@firstoftwo
 \else \expandafter \@secondoftwo
 \fi
}%
\providecommand \natexlab [1]{#1}%
\providecommand \enquote  [1]{``#1''}%
\providecommand \bibnamefont  [1]{#1}%
\providecommand \bibfnamefont [1]{#1}%
\providecommand \citenamefont [1]{#1}%
\providecommand \href@noop [0]{\@secondoftwo}%
\providecommand \href [0]{\begingroup \@sanitize@url \@href}%
\providecommand \@href[1]{\@@startlink{#1}\@@href}%
\providecommand \@@href[1]{\endgroup#1\@@endlink}%
\providecommand \@sanitize@url [0]{\catcode `\\12\catcode `\$12\catcode
  `\&12\catcode `\#12\catcode `\^12\catcode `\_12\catcode `\%12\relax}%
\providecommand \@@startlink[1]{}%
\providecommand \@@endlink[0]{}%
\providecommand \url  [0]{\begingroup\@sanitize@url \@url }%
\providecommand \@url [1]{\endgroup\@href {#1}{\urlprefix }}%
\providecommand \urlprefix  [0]{URL }%
\providecommand \Eprint [0]{\href }%
\providecommand \doibase [0]{https://doi.org/}%
\providecommand \selectlanguage [0]{\@gobble}%
\providecommand \bibinfo  [0]{\@secondoftwo}%
\providecommand \bibfield  [0]{\@secondoftwo}%
\providecommand \translation [1]{[#1]}%
\providecommand \BibitemOpen [0]{}%
\providecommand \bibitemStop [0]{}%
\providecommand \bibitemNoStop [0]{.\EOS\space}%
\providecommand \EOS [0]{\spacefactor3000\relax}%
\providecommand \BibitemShut  [1]{\csname bibitem#1\endcsname}%
\let\auto@bib@innerbib\@empty
\bibitem [{\citenamefont {Burkard}\ \emph {et~al.}(2023)\citenamefont
  {Burkard}, \citenamefont {Ladd}, \citenamefont {Pan}, \citenamefont
  {Nichol},\ and\ \citenamefont {Petta}}]{buckard-rmp}%
  \BibitemOpen
  \bibfield  {author} {\bibinfo {author} {\bibfnamefont {G.}~\bibnamefont
  {Burkard}}, \bibinfo {author} {\bibfnamefont {T.~D.}\ \bibnamefont {Ladd}},
  \bibinfo {author} {\bibfnamefont {A.}~\bibnamefont {Pan}}, \bibinfo {author}
  {\bibfnamefont {J.~M.}\ \bibnamefont {Nichol}},\ and\ \bibinfo {author}
  {\bibfnamefont {J.~R.}\ \bibnamefont {Petta}},\ }\bibfield  {title} {\bibinfo
  {title} {Semiconductor spin qubits},\ }\href
  {https://doi.org/10.1103/RevModPhys.95.025003} {\bibfield  {journal}
  {\bibinfo  {journal} {Rev. Mod. Phys.}\ }\textbf {\bibinfo {volume} {95}},\
  \bibinfo {pages} {025003} (\bibinfo {year} {2023})}\BibitemShut {NoStop}%
\bibitem [{\citenamefont {Chatterjee}\ \emph {et~al.}(2021)\citenamefont
  {Chatterjee}, \citenamefont {Stevenson}, \citenamefont {Franceschi},
  \citenamefont {Morello}, \citenamefont {P.de},\ and\ \citenamefont
  {Kuemmeth}}]{chatterjee-review}%
  \BibitemOpen
  \bibfield  {author} {\bibinfo {author} {\bibfnamefont {A.}~\bibnamefont
  {Chatterjee}}, \bibinfo {author} {\bibfnamefont {P.}~\bibnamefont
  {Stevenson}}, \bibinfo {author} {\bibfnamefont {S.~D.}\ \bibnamefont
  {Franceschi}}, \bibinfo {author} {\bibfnamefont {A.}~\bibnamefont {Morello}},
  \bibinfo {author} {\bibfnamefont {L.~N.}\ \bibnamefont {P.de}},\ and\
  \bibinfo {author} {\bibfnamefont {F.}~\bibnamefont {Kuemmeth}},\ }\bibfield
  {title} {\bibinfo {title} {Semiconductor qubits in practice},\ }\href
  {https://doi.org/10.1038/s42254-021-00283-9} {\bibfield  {journal} {\bibinfo
  {journal} {Nat. Rev. Phys.}\ }\textbf {\bibinfo {volume} {3}},\ \bibinfo
  {pages} {157} (\bibinfo {year} {2021})}\BibitemShut {NoStop}%
\bibitem [{\citenamefont {Awschalom}\ \emph {et~al.}(2013)\citenamefont
  {Awschalom}, \citenamefont {Bassett}, \citenamefont {Dzurak}, \citenamefont
  {Hu},\ and\ \citenamefont {Petta}}]{david-spintronics}%
  \BibitemOpen
  \bibfield  {author} {\bibinfo {author} {\bibfnamefont {D.~D.}\ \bibnamefont
  {Awschalom}}, \bibinfo {author} {\bibfnamefont {L.~C.}\ \bibnamefont
  {Bassett}}, \bibinfo {author} {\bibfnamefont {A.~S.}\ \bibnamefont {Dzurak}},
  \bibinfo {author} {\bibfnamefont {E.~L.}\ \bibnamefont {Hu}},\ and\ \bibinfo
  {author} {\bibfnamefont {J.~R.}\ \bibnamefont {Petta}},\ }\bibfield  {title}
  {\bibinfo {title} {Quantum spintronics: Engineering and manipulating
  atom-like spins in semiconductors},\ }\href
  {https://doi.org/10.1126/science.1231364} {\bibfield  {journal} {\bibinfo
  {journal} {Science}\ }\textbf {\bibinfo {volume} {339}},\ \bibinfo {pages}
  {1174} (\bibinfo {year} {2013})}\BibitemShut {NoStop}%
\bibitem [{\citenamefont {Zhang}\ \emph {et~al.}(2019)\citenamefont {Zhang},
  \citenamefont {Li}, \citenamefont {Cao}, \citenamefont {Xiao}, \citenamefont
  {Guo},\ and\ \citenamefont {Guo}}]{zhang-rev}%
  \BibitemOpen
  \bibfield  {author} {\bibinfo {author} {\bibfnamefont {X.}~\bibnamefont
  {Zhang}}, \bibinfo {author} {\bibfnamefont {H.-O.}\ \bibnamefont {Li}},
  \bibinfo {author} {\bibfnamefont {G.}~\bibnamefont {Cao}}, \bibinfo {author}
  {\bibfnamefont {M.}~\bibnamefont {Xiao}}, \bibinfo {author} {\bibfnamefont
  {G.-C.}\ \bibnamefont {Guo}},\ and\ \bibinfo {author} {\bibfnamefont {G.~P.}\
  \bibnamefont {Guo}},\ }\bibfield  {title} {\bibinfo {title} {Semiconductor
  quantum computation},\ }\href {https://doi.org/10.1093/nsr/nwy153} {\bibfield
   {journal} {\bibinfo  {journal} {Nat. Science Rev.}\ }\textbf {\bibinfo
  {volume} {6}},\ \bibinfo {pages} {32} (\bibinfo {year} {2019})}\BibitemShut
  {NoStop}%
\bibitem [{\citenamefont {Kane}(1998)}]{kane}%
  \BibitemOpen
  \bibfield  {author} {\bibinfo {author} {\bibfnamefont {B.~E.}\ \bibnamefont
  {Kane}},\ }\bibfield  {title} {\bibinfo {title} {A silicon-based nuclear spin
  quantum computer},\ }\href {https://doi.org/10.1038/30156} {\bibfield
  {journal} {\bibinfo  {journal} {Nat.}\ }\textbf {\bibinfo {volume} {393}},\
  \bibinfo {pages} {133} (\bibinfo {year} {1998})}\BibitemShut {NoStop}%
\bibitem [{\citenamefont {Fowler}\ \emph {et~al.}(2012)\citenamefont {Fowler},
  \citenamefont {Mariantoni}, \citenamefont {Martinis},\ and\ \citenamefont
  {Cleland}}]{fowler}%
  \BibitemOpen
  \bibfield  {author} {\bibinfo {author} {\bibfnamefont {A.~G.}\ \bibnamefont
  {Fowler}}, \bibinfo {author} {\bibfnamefont {M.}~\bibnamefont {Mariantoni}},
  \bibinfo {author} {\bibfnamefont {J.~M.}\ \bibnamefont {Martinis}},\ and\
  \bibinfo {author} {\bibfnamefont {A.~N.}\ \bibnamefont {Cleland}},\
  }\bibfield  {title} {\bibinfo {title} {Surface codes: Towards practical
  large-scale quantum computation},\ }\href
  {https://doi.org/10.1103/PhysRevA.86.032324} {\bibfield  {journal} {\bibinfo
  {journal} {Phys. Rev. A}\ }\textbf {\bibinfo {volume} {86}},\ \bibinfo
  {pages} {032324} (\bibinfo {year} {2012})}\BibitemShut {NoStop}%
\bibitem [{\citenamefont {Loss}\ and\ \citenamefont {DiVincenzo}(1998)}]{LD}%
  \BibitemOpen
  \bibfield  {author} {\bibinfo {author} {\bibfnamefont {D.}~\bibnamefont
  {Loss}}\ and\ \bibinfo {author} {\bibfnamefont {D.~P.}\ \bibnamefont
  {DiVincenzo}},\ }\bibfield  {title} {\bibinfo {title} {Quantum computation
  with quantum dots},\ }\href {https://doi.org/10.1103/PhysRevA.57.120}
  {\bibfield  {journal} {\bibinfo  {journal} {Phys. Rev. A}\ }\textbf {\bibinfo
  {volume} {57}},\ \bibinfo {pages} {120} (\bibinfo {year} {1998})}\BibitemShut
  {NoStop}%
\bibitem [{\citenamefont {Koppens}\ \emph {et~al.}(2006)\citenamefont
  {Koppens}, \citenamefont {Buizert}, \citenamefont {Tielrooij}, \citenamefont
  {Vink}, \citenamefont {Nowack}, \citenamefont {Meunier}, \citenamefont
  {Kouwenhoven},\ and\ \citenamefont {Vandersypen}}]{esr-koppens}%
  \BibitemOpen
  \bibfield  {author} {\bibinfo {author} {\bibfnamefont {F.~H.~L.}\
  \bibnamefont {Koppens}}, \bibinfo {author} {\bibfnamefont {C.}~\bibnamefont
  {Buizert}}, \bibinfo {author} {\bibfnamefont {K.~J.}\ \bibnamefont
  {Tielrooij}}, \bibinfo {author} {\bibfnamefont {I.~T.}\ \bibnamefont {Vink}},
  \bibinfo {author} {\bibfnamefont {K.~C.}\ \bibnamefont {Nowack}}, \bibinfo
  {author} {\bibfnamefont {T.}~\bibnamefont {Meunier}}, \bibinfo {author}
  {\bibfnamefont {L.~P.}\ \bibnamefont {Kouwenhoven}},\ and\ \bibinfo {author}
  {\bibfnamefont {L.~M.~K.}\ \bibnamefont {Vandersypen}},\ }\bibfield  {title}
  {\bibinfo {title} {Driven coherent oscillations of a single electron spin in
  a quantum dot},\ }\href {https://doi.org/10.1038/nature05065} {\bibfield
  {journal} {\bibinfo  {journal} {Nat.}\ }\textbf {\bibinfo {volume} {442}},\
  \bibinfo {pages} {766} (\bibinfo {year} {2006})}\BibitemShut {NoStop}%
\bibitem [{\citenamefont {Koppens}\ \emph {et~al.}(2008)\citenamefont
  {Koppens}, \citenamefont {Nowack},\ and\ \citenamefont
  {Vandersypen}}]{koppens-echo}%
  \BibitemOpen
  \bibfield  {author} {\bibinfo {author} {\bibfnamefont {F.~H.~L.}\
  \bibnamefont {Koppens}}, \bibinfo {author} {\bibfnamefont {K.~C.}\
  \bibnamefont {Nowack}},\ and\ \bibinfo {author} {\bibfnamefont {L.~M.~K.}\
  \bibnamefont {Vandersypen}},\ }\bibfield  {title} {\bibinfo {title} {Spin
  echo of a single electron spin in a quantum dot},\ }\href
  {https://doi.org/10.1103/PhysRevLett.100.236802} {\bibfield  {journal}
  {\bibinfo  {journal} {Phys. Rev. Lett.}\ }\textbf {\bibinfo {volume} {100}},\
  \bibinfo {pages} {236802} (\bibinfo {year} {2008})}\BibitemShut {NoStop}%
\bibitem [{\citenamefont {Pla}\ \emph {et~al.}(2012)\citenamefont {Pla},
  \citenamefont {Tan}, \citenamefont {Dehollain}, \citenamefont {Lim},
  \citenamefont {Morton}, \citenamefont {Jamieson}, \citenamefont {Dzurak},\
  and\ \citenamefont {Morello}}]{esr-pla}%
  \BibitemOpen
  \bibfield  {author} {\bibinfo {author} {\bibfnamefont {J.~J.}\ \bibnamefont
  {Pla}}, \bibinfo {author} {\bibfnamefont {K.~Y.}\ \bibnamefont {Tan}},
  \bibinfo {author} {\bibfnamefont {J.~P.}\ \bibnamefont {Dehollain}}, \bibinfo
  {author} {\bibfnamefont {W.~H.}\ \bibnamefont {Lim}}, \bibinfo {author}
  {\bibfnamefont {J.~J.}\ \bibnamefont {Morton}}, \bibinfo {author}
  {\bibfnamefont {D.~N.}\ \bibnamefont {Jamieson}}, \bibinfo {author}
  {\bibfnamefont {A.~S.}\ \bibnamefont {Dzurak}},\ and\ \bibinfo {author}
  {\bibfnamefont {A.}~\bibnamefont {Morello}},\ }\bibfield  {title} {\bibinfo
  {title} {A single-atom electron spin qubit in silicon},\ }\href
  {https://doi.org/10.1038/nature11449} {\bibfield  {journal} {\bibinfo
  {journal} {Nat.}\ }\textbf {\bibinfo {volume} {489}},\ \bibinfo {pages} {541}
  (\bibinfo {year} {2012})}\BibitemShut {NoStop}%
\bibitem [{\citenamefont {Veldhorst}\ \emph {et~al.}(2014)\citenamefont
  {Veldhorst}, \citenamefont {Hwang}, \citenamefont {Yang}, \citenamefont
  {Leenstra}, \citenamefont {Ronde}, \citenamefont {Dehollain}, \citenamefont
  {Muhonen}, \citenamefont {Hudson}, \citenamefont {Itoh}, \citenamefont
  {Morello},\ and\ \citenamefont {Dzurak}}]{esr-veld1}%
  \BibitemOpen
  \bibfield  {author} {\bibinfo {author} {\bibfnamefont {M.}~\bibnamefont
  {Veldhorst}}, \bibinfo {author} {\bibfnamefont {J.~C.~C.}\ \bibnamefont
  {Hwang}}, \bibinfo {author} {\bibfnamefont {C.~H.}\ \bibnamefont {Yang}},
  \bibinfo {author} {\bibfnamefont {A.~W.}\ \bibnamefont {Leenstra}}, \bibinfo
  {author} {\bibfnamefont {B.~d.}\ \bibnamefont {Ronde}}, \bibinfo {author}
  {\bibfnamefont {J.~P.}\ \bibnamefont {Dehollain}}, \bibinfo {author}
  {\bibfnamefont {J.~T.}\ \bibnamefont {Muhonen}}, \bibinfo {author}
  {\bibfnamefont {F.~E.}\ \bibnamefont {Hudson}}, \bibinfo {author}
  {\bibfnamefont {K.~M.}\ \bibnamefont {Itoh}}, \bibinfo {author}
  {\bibfnamefont {A.}~\bibnamefont {Morello}},\ and\ \bibinfo {author}
  {\bibfnamefont {A.~S.}\ \bibnamefont {Dzurak}},\ }\bibfield  {title}
  {\bibinfo {title} {An addressable quantum dot qubit with fault-tolerant
  control-fidelity},\ }\href {https://doi.org/10.1038/nnano.2014.216}
  {\bibfield  {journal} {\bibinfo  {journal} {Nat. Nanotech.}\ }\textbf
  {\bibinfo {volume} {9}},\ \bibinfo {pages} {981} (\bibinfo {year}
  {2014})}\BibitemShut {NoStop}%
\bibitem [{\citenamefont {Veldhorst}\ \emph {et~al.}(2015)\citenamefont
  {Veldhorst}, \citenamefont {Yang}, \citenamefont {Hwang}, \citenamefont
  {Huang}, \citenamefont {Dehollain}, \citenamefont {Muhonen}, \citenamefont
  {Simmons}, \citenamefont {Laucht}, \citenamefont {Hudson}, \citenamefont
  {Itoh}, \citenamefont {Morello},\ and\ \citenamefont {Dzurak}}]{esr-veld2}%
  \BibitemOpen
  \bibfield  {author} {\bibinfo {author} {\bibfnamefont {M.}~\bibnamefont
  {Veldhorst}}, \bibinfo {author} {\bibfnamefont {C.~H.}\ \bibnamefont {Yang}},
  \bibinfo {author} {\bibfnamefont {J.~C.~C.}\ \bibnamefont {Hwang}}, \bibinfo
  {author} {\bibfnamefont {W.}~\bibnamefont {Huang}}, \bibinfo {author}
  {\bibfnamefont {J.~P.}\ \bibnamefont {Dehollain}}, \bibinfo {author}
  {\bibfnamefont {J.~T.}\ \bibnamefont {Muhonen}}, \bibinfo {author}
  {\bibfnamefont {V.}~\bibnamefont {Simmons}}, \bibinfo {author} {\bibfnamefont
  {A.}~\bibnamefont {Laucht}}, \bibinfo {author} {\bibfnamefont {F.~E.}\
  \bibnamefont {Hudson}}, \bibinfo {author} {\bibfnamefont {K.~M.}\
  \bibnamefont {Itoh}}, \bibinfo {author} {\bibfnamefont {A.}~\bibnamefont
  {Morello}},\ and\ \bibinfo {author} {\bibfnamefont {A.~S.}\ \bibnamefont
  {Dzurak}},\ }\bibfield  {title} {\bibinfo {title} {A two-qubit logic gate in
  silicon},\ }\href {https://doi.org/10.1038/nature15263} {\bibfield  {journal}
  {\bibinfo  {journal} {Nat.}\ }\textbf {\bibinfo {volume} {526}},\ \bibinfo
  {pages} {410} (\bibinfo {year} {2015})}\BibitemShut {NoStop}%
\bibitem [{\citenamefont {Kato}\ \emph {et~al.}(2003)\citenamefont {Kato},
  \citenamefont {Myers}, \citenamefont {Driscoll}, \citenamefont {Gossard},
  \citenamefont {Levy},\ and\ \citenamefont {Awschalom}}]{kato-gfactor}%
  \BibitemOpen
  \bibfield  {author} {\bibinfo {author} {\bibfnamefont {Y.}~\bibnamefont
  {Kato}}, \bibinfo {author} {\bibfnamefont {R.~C.}\ \bibnamefont {Myers}},
  \bibinfo {author} {\bibfnamefont {D.~C.}\ \bibnamefont {Driscoll}}, \bibinfo
  {author} {\bibfnamefont {A.~C.}\ \bibnamefont {Gossard}}, \bibinfo {author}
  {\bibfnamefont {J.}~\bibnamefont {Levy}},\ and\ \bibinfo {author}
  {\bibfnamefont {D.~D.}\ \bibnamefont {Awschalom}},\ }\bibfield  {title}
  {\bibinfo {title} {Gigahertz electron spin manipulation using
  voltage-controlled g-tensor modulation},\ }\href
  {https://doi.org/10.1126/science.1080880} {\bibfield  {journal} {\bibinfo
  {journal} {Science}\ }\textbf {\bibinfo {volume} {299}},\ \bibinfo {pages}
  {1201} (\bibinfo {year} {2003})}\BibitemShut {NoStop}%
\bibitem [{\citenamefont {Salis}\ \emph {et~al.}(2001)\citenamefont {Salis},
  \citenamefont {Kato}, \citenamefont {Ensslin}, \citenamefont {Driscoll},
  \citenamefont {Gossard},\ and\ \citenamefont
  {Awschalom}}]{salis-gfactor-coherence}%
  \BibitemOpen
  \bibfield  {author} {\bibinfo {author} {\bibfnamefont {G.}~\bibnamefont
  {Salis}}, \bibinfo {author} {\bibfnamefont {Y.}~\bibnamefont {Kato}},
  \bibinfo {author} {\bibfnamefont {K.}~\bibnamefont {Ensslin}}, \bibinfo
  {author} {\bibfnamefont {D.~C.}\ \bibnamefont {Driscoll}}, \bibinfo {author}
  {\bibfnamefont {A.~C.}\ \bibnamefont {Gossard}},\ and\ \bibinfo {author}
  {\bibfnamefont {D.~D.}\ \bibnamefont {Awschalom}},\ }\bibfield  {title}
  {\bibinfo {title} {Electrical control of spin coherence in semiconductor
  nanostructures},\ }\href {https://doi.org/10.1038/414619a} {\bibfield
  {journal} {\bibinfo  {journal} {Nature}\ }\textbf {\bibinfo {volume} {414}},\
  \bibinfo {pages} {619} (\bibinfo {year} {2001})}\BibitemShut {NoStop}%
\bibitem [{\citenamefont {Deacon}\ \emph {et~al.}(2011)\citenamefont {Deacon},
  \citenamefont {Kanai}, \citenamefont {Takahashi}, \citenamefont {Oiwa},
  \citenamefont {Yoshida}, \citenamefont {Shibata}, \citenamefont {Hirakawa},
  \citenamefont {Tokura},\ and\ \citenamefont {Tarucha}}]{deacon-gfactor-InAs}%
  \BibitemOpen
  \bibfield  {author} {\bibinfo {author} {\bibfnamefont {R.~S.}\ \bibnamefont
  {Deacon}}, \bibinfo {author} {\bibfnamefont {Y.}~\bibnamefont {Kanai}},
  \bibinfo {author} {\bibfnamefont {S.}~\bibnamefont {Takahashi}}, \bibinfo
  {author} {\bibfnamefont {A.}~\bibnamefont {Oiwa}}, \bibinfo {author}
  {\bibfnamefont {K.}~\bibnamefont {Yoshida}}, \bibinfo {author} {\bibfnamefont
  {K.}~\bibnamefont {Shibata}}, \bibinfo {author} {\bibfnamefont
  {K.}~\bibnamefont {Hirakawa}}, \bibinfo {author} {\bibfnamefont
  {Y.}~\bibnamefont {Tokura}},\ and\ \bibinfo {author} {\bibfnamefont
  {S.}~\bibnamefont {Tarucha}},\ }\bibfield  {title} {\bibinfo {title}
  {Electrically tuned $g$ tensor in an $\text{InAs}$ self-assembled quantum
  dot},\ }\href {https://doi.org/10.1103/PhysRevB.84.041302} {\bibfield
  {journal} {\bibinfo  {journal} {Phys. Rev. B}\ }\textbf {\bibinfo {volume}
  {84}},\ \bibinfo {pages} {041302} (\bibinfo {year} {2011})}\BibitemShut
  {NoStop}%
\bibitem [{\citenamefont {Pingenot}\ \emph {et~al.}(2011)\citenamefont
  {Pingenot}, \citenamefont {Pryor},\ and\ \citenamefont
  {Flatt\'e}}]{pingenot-gfactor-InGaAs}%
  \BibitemOpen
  \bibfield  {author} {\bibinfo {author} {\bibfnamefont {J.}~\bibnamefont
  {Pingenot}}, \bibinfo {author} {\bibfnamefont {C.~E.}\ \bibnamefont
  {Pryor}},\ and\ \bibinfo {author} {\bibfnamefont {M.~E.}\ \bibnamefont
  {Flatt\'e}},\ }\bibfield  {title} {\bibinfo {title} {Electric-field
  manipulation of the land\'e $g$ tensor of a hole in an
  $\text{In}_{0.5}\text{Ga}_{0.5}\text{As/GaAs}$ self-assembled quantum dot},\
  }\href {https://doi.org/10.1103/PhysRevB.84.195403} {\bibfield  {journal}
  {\bibinfo  {journal} {Phys. Rev. B}\ }\textbf {\bibinfo {volume} {84}},\
  \bibinfo {pages} {195403} (\bibinfo {year} {2011})}\BibitemShut {NoStop}%
\bibitem [{\citenamefont {Ferrón}\ \emph {et~al.}(2019)\citenamefont
  {Ferrón}, \citenamefont {Rodriguez}, \citenamefont {Gómez}, \citenamefont
  {Lado},\ and\ \citenamefont {Fernández-Rossier}}]{ferr-gfactor-aniso}%
  \BibitemOpen
  \bibfield  {author} {\bibinfo {author} {\bibfnamefont {A.}~\bibnamefont
  {Ferrón}}, \bibinfo {author} {\bibfnamefont {S.~A.}\ \bibnamefont
  {Rodriguez}}, \bibinfo {author} {\bibfnamefont {S.~S.}\ \bibnamefont
  {Gómez}}, \bibinfo {author} {\bibfnamefont {J.~L.}\ \bibnamefont {Lado}},\
  and\ \bibinfo {author} {\bibfnamefont {J.}~\bibnamefont
  {Fernández-Rossier}},\ }\bibfield  {title} {\bibinfo {title} {Single spin
  resonance driven by electric modulation of the $g$-factor anisotropy},\
  }\href {https://doi.org/10.1103/PhysRevResearch.1.033185} {\bibfield
  {journal} {\bibinfo  {journal} {Phys. Rev. Res.}\ }\textbf {\bibinfo {volume}
  {1}},\ \bibinfo {pages} {033185} (\bibinfo {year} {2019})}\BibitemShut
  {NoStop}%
\bibitem [{\citenamefont {Ladrière}\ \emph {et~al.}(2008)\citenamefont
  {Ladrière}, \citenamefont {Obata}, \citenamefont {Tokura}, \citenamefont
  {Shin}, \citenamefont {Kubo}, \citenamefont {Yoshida}, \citenamefont
  {Taniyama},\ and\ \citenamefont {Tarucha}}]{Ladri-slanting}%
  \BibitemOpen
  \bibfield  {author} {\bibinfo {author} {\bibfnamefont {M.~P.}\ \bibnamefont
  {Ladrière}}, \bibinfo {author} {\bibfnamefont {T.}~\bibnamefont {Obata}},
  \bibinfo {author} {\bibfnamefont {Y.}~\bibnamefont {Tokura}}, \bibinfo
  {author} {\bibfnamefont {Y.-S.}\ \bibnamefont {Shin}}, \bibinfo {author}
  {\bibfnamefont {T.}~\bibnamefont {Kubo}}, \bibinfo {author} {\bibfnamefont
  {K.}~\bibnamefont {Yoshida}}, \bibinfo {author} {\bibfnamefont
  {T.}~\bibnamefont {Taniyama}},\ and\ \bibinfo {author} {\bibfnamefont
  {S.}~\bibnamefont {Tarucha}},\ }\bibfield  {title} {\bibinfo {title}
  {Electrically driven single-electron spin resonance in a slanting
  $\text{Zeeman}$ field},\ }\href {https://doi.org/10.1038/nphys1053}
  {\bibfield  {journal} {\bibinfo  {journal} {Nat. Phys.}\ }\textbf {\bibinfo
  {volume} {4}},\ \bibinfo {pages} {776} (\bibinfo {year} {2008})}\BibitemShut
  {NoStop}%
\bibitem [{\citenamefont {Brunner}\ \emph {et~al.}(2011)\citenamefont
  {Brunner}, \citenamefont {Shin}, \citenamefont {Obata}, \citenamefont
  {Pioro-Ladri\`ere}, \citenamefont {Kubo}, \citenamefont {Yoshida},
  \citenamefont {Taniyama}, \citenamefont {Tokura},\ and\ \citenamefont
  {Tarucha}}]{brunner-2qubit-edsr}%
  \BibitemOpen
  \bibfield  {author} {\bibinfo {author} {\bibfnamefont {R.}~\bibnamefont
  {Brunner}}, \bibinfo {author} {\bibfnamefont {Y.-S.}\ \bibnamefont {Shin}},
  \bibinfo {author} {\bibfnamefont {T.}~\bibnamefont {Obata}}, \bibinfo
  {author} {\bibfnamefont {M.}~\bibnamefont {Pioro-Ladri\`ere}}, \bibinfo
  {author} {\bibfnamefont {T.}~\bibnamefont {Kubo}}, \bibinfo {author}
  {\bibfnamefont {K.}~\bibnamefont {Yoshida}}, \bibinfo {author} {\bibfnamefont
  {T.}~\bibnamefont {Taniyama}}, \bibinfo {author} {\bibfnamefont
  {Y.}~\bibnamefont {Tokura}},\ and\ \bibinfo {author} {\bibfnamefont
  {S.}~\bibnamefont {Tarucha}},\ }\bibfield  {title} {\bibinfo {title}
  {Two-qubit gate of combined single-spin rotation and interdot spin exchange
  in a double quantum dot},\ }\href
  {https://doi.org/10.1103/PhysRevLett.107.146801} {\bibfield  {journal}
  {\bibinfo  {journal} {Phys. Rev. Lett.}\ }\textbf {\bibinfo {volume} {107}},\
  \bibinfo {pages} {146801} (\bibinfo {year} {2011})}\BibitemShut {NoStop}%
\bibitem [{\citenamefont {Yoneda}\ \emph {et~al.}(2018)\citenamefont {Yoneda},
  \citenamefont {Takeda}, \citenamefont {Otsuka}, \citenamefont {Nakajima},
  \citenamefont {Delbecq}, \citenamefont {Allison}, \citenamefont {Honda},
  \citenamefont {Kodera}, \citenamefont {Oda}, \citenamefont {Hoshi},
  \citenamefont {Usami}, \citenamefont {Itoh},\ and\ \citenamefont
  {Tarucha}}]{yoneda-99pc}%
  \BibitemOpen
  \bibfield  {author} {\bibinfo {author} {\bibfnamefont {J.}~\bibnamefont
  {Yoneda}}, \bibinfo {author} {\bibfnamefont {K.}~\bibnamefont {Takeda}},
  \bibinfo {author} {\bibfnamefont {T.}~\bibnamefont {Otsuka}}, \bibinfo
  {author} {\bibfnamefont {T.}~\bibnamefont {Nakajima}}, \bibinfo {author}
  {\bibfnamefont {M.~R.}\ \bibnamefont {Delbecq}}, \bibinfo {author}
  {\bibfnamefont {G.}~\bibnamefont {Allison}}, \bibinfo {author} {\bibfnamefont
  {T.}~\bibnamefont {Honda}}, \bibinfo {author} {\bibfnamefont
  {T.}~\bibnamefont {Kodera}}, \bibinfo {author} {\bibfnamefont
  {S.}~\bibnamefont {Oda}}, \bibinfo {author} {\bibfnamefont {Y.}~\bibnamefont
  {Hoshi}}, \bibinfo {author} {\bibfnamefont {N.}~\bibnamefont {Usami}},
  \bibinfo {author} {\bibfnamefont {K.~M.}\ \bibnamefont {Itoh}},\ and\
  \bibinfo {author} {\bibfnamefont {S.}~\bibnamefont {Tarucha}},\ }\bibfield
  {title} {\bibinfo {title} {A quantum-dot spin qubit with coherence limited by
  charge noise and fidelity higher than 99.9 percent},\ }\href
  {https://doi.org/10.1038/s41565-017-0014-x} {\bibfield  {journal} {\bibinfo
  {journal} {Nat. Nanotech.}\ }\textbf {\bibinfo {volume} {13}},\ \bibinfo
  {pages} {102} (\bibinfo {year} {2018})}\BibitemShut {NoStop}%
\bibitem [{\citenamefont {Zajac}\ \emph {et~al.}(2018)\citenamefont {Zajac},
  \citenamefont {Sigillito}, \citenamefont {Russ}, \citenamefont {Borjans},
  \citenamefont {Taylor}, \citenamefont {Burkard},\ and\ \citenamefont
  {Petta}}]{zajac-cnot}%
  \BibitemOpen
  \bibfield  {author} {\bibinfo {author} {\bibfnamefont {D.~M.}\ \bibnamefont
  {Zajac}}, \bibinfo {author} {\bibfnamefont {A.~J.}\ \bibnamefont
  {Sigillito}}, \bibinfo {author} {\bibfnamefont {M.}~\bibnamefont {Russ}},
  \bibinfo {author} {\bibfnamefont {F.}~\bibnamefont {Borjans}}, \bibinfo
  {author} {\bibfnamefont {J.~M.}\ \bibnamefont {Taylor}}, \bibinfo {author}
  {\bibfnamefont {G.}~\bibnamefont {Burkard}},\ and\ \bibinfo {author}
  {\bibfnamefont {J.~R.}\ \bibnamefont {Petta}},\ }\bibfield  {title} {\bibinfo
  {title} {Resonantly driven $\text{CNOT}$ gate for electron spins},\ }\href
  {https://doi.org/10.1126/science.aao5965} {\bibfield  {journal} {\bibinfo
  {journal} {Science}\ }\textbf {\bibinfo {volume} {359}},\ \bibinfo {pages}
  {439} (\bibinfo {year} {2018})}\BibitemShut {NoStop}%
\bibitem [{\citenamefont {Nowack}\ \emph {et~al.}(2007)\citenamefont {Nowack},
  \citenamefont {Koppens}, \citenamefont {Nazarov},\ and\ \citenamefont
  {Vandersypen}}]{nowack-electric}%
  \BibitemOpen
  \bibfield  {author} {\bibinfo {author} {\bibfnamefont {K.~C.}\ \bibnamefont
  {Nowack}}, \bibinfo {author} {\bibfnamefont {F.~H.~L.}\ \bibnamefont
  {Koppens}}, \bibinfo {author} {\bibfnamefont {Y.~V.}\ \bibnamefont
  {Nazarov}},\ and\ \bibinfo {author} {\bibfnamefont {L.~M.~K.}\ \bibnamefont
  {Vandersypen}},\ }\bibfield  {title} {\bibinfo {title} {Coherent control of a
  single electron spin with electric fields},\ }\href
  {https://doi.org/10.1126/science.1148092} {\bibfield  {journal} {\bibinfo
  {journal} {Science}\ }\textbf {\bibinfo {volume} {318}},\ \bibinfo {pages}
  {1430} (\bibinfo {year} {2007})}\BibitemShut {NoStop}%
\bibitem [{\citenamefont {Perge}\ \emph {et~al.}(2010)\citenamefont {Perge},
  \citenamefont {Frolov}, \citenamefont {Bakkers},\ and\ \citenamefont
  {Kouwenhoven}}]{Nadj-electric}%
  \BibitemOpen
  \bibfield  {author} {\bibinfo {author} {\bibfnamefont {S.~N.}\ \bibnamefont
  {Perge}}, \bibinfo {author} {\bibfnamefont {S.~M.}\ \bibnamefont {Frolov}},
  \bibinfo {author} {\bibfnamefont {E.~P. A.~M.}\ \bibnamefont {Bakkers}},\
  and\ \bibinfo {author} {\bibfnamefont {L.~P.}\ \bibnamefont {Kouwenhoven}},\
  }\bibfield  {title} {\bibinfo {title} {Spin–orbit qubit in a semiconductor
  nanowire},\ }\href {https://doi.org/10.1038/nature09682} {\bibfield
  {journal} {\bibinfo  {journal} {Nat.}\ }\textbf {\bibinfo {volume} {468}},\
  \bibinfo {pages} {1084} (\bibinfo {year} {2010})}\BibitemShut {NoStop}%
\bibitem [{\citenamefont {Rashba}\ and\ \citenamefont
  {Efros}(2003)}]{rashba-edsr}%
  \BibitemOpen
  \bibfield  {author} {\bibinfo {author} {\bibfnamefont {E.~I.}\ \bibnamefont
  {Rashba}}\ and\ \bibinfo {author} {\bibfnamefont {A.~L.}\ \bibnamefont
  {Efros}},\ }\bibfield  {title} {\bibinfo {title} {Orbital mechanisms of
  electron-spin manipulation by an electric field},\ }\href
  {https://doi.org/10.1103/PhysRevLett.91.126405} {\bibfield  {journal}
  {\bibinfo  {journal} {Phys. Rev. Lett.}\ }\textbf {\bibinfo {volume} {91}},\
  \bibinfo {pages} {126405} (\bibinfo {year} {2003})}\BibitemShut {NoStop}%
\bibitem [{\citenamefont {Golovach}\ \emph {et~al.}(2006)\citenamefont
  {Golovach}, \citenamefont {Borhani},\ and\ \citenamefont
  {Loss}}]{edsr-loss-orig}%
  \BibitemOpen
  \bibfield  {author} {\bibinfo {author} {\bibfnamefont {V.~N.}\ \bibnamefont
  {Golovach}}, \bibinfo {author} {\bibfnamefont {M.}~\bibnamefont {Borhani}},\
  and\ \bibinfo {author} {\bibfnamefont {D.}~\bibnamefont {Loss}},\ }\bibfield
  {title} {\bibinfo {title} {Electric-dipole-induced spin resonance in quantum
  dots},\ }\href {https://doi.org/10.1103/PhysRevB.74.165319} {\bibfield
  {journal} {\bibinfo  {journal} {Phys. Rev. B}\ }\textbf {\bibinfo {volume}
  {74}},\ \bibinfo {pages} {165319} (\bibinfo {year} {2006})}\BibitemShut
  {NoStop}%
\bibitem [{\citenamefont {Bulaev}\ and\ \citenamefont
  {Loss}(2007)}]{edsr-loss-hole}%
  \BibitemOpen
  \bibfield  {author} {\bibinfo {author} {\bibfnamefont {D.~V.}\ \bibnamefont
  {Bulaev}}\ and\ \bibinfo {author} {\bibfnamefont {D.}~\bibnamefont {Loss}},\
  }\bibfield  {title} {\bibinfo {title} {Electric dipole spin resonance for
  heavy holes in quantum dots},\ }\href
  {https://doi.org/10.1103/PhysRevLett.98.097202} {\bibfield  {journal}
  {\bibinfo  {journal} {Phys. Rev. Lett.}\ }\textbf {\bibinfo {volume} {98}},\
  \bibinfo {pages} {097202} (\bibinfo {year} {2007})}\BibitemShut {NoStop}%
\bibitem [{\citenamefont {Brooks}\ and\ \citenamefont
  {Burkard}(2020)}]{edsr-mos2}%
  \BibitemOpen
  \bibfield  {author} {\bibinfo {author} {\bibfnamefont {M.}~\bibnamefont
  {Brooks}}\ and\ \bibinfo {author} {\bibfnamefont {G.}~\bibnamefont
  {Burkard}},\ }\bibfield  {title} {\bibinfo {title} {Electric dipole spin
  resonance of two-dimensional semiconductor spin qubits},\ }\href
  {https://doi.org/10.1103/PhysRevB.101.035204} {\bibfield  {journal} {\bibinfo
   {journal} {Phys. Rev. B}\ }\textbf {\bibinfo {volume} {101}},\ \bibinfo
  {pages} {035204} (\bibinfo {year} {2020})}\BibitemShut {NoStop}%
\bibitem [{\citenamefont {Khomitsky}\ \emph {et~al.}(2012)\citenamefont
  {Khomitsky}, \citenamefont {Gulyaev},\ and\ \citenamefont
  {Sherman}}]{sherman}%
  \BibitemOpen
  \bibfield  {author} {\bibinfo {author} {\bibfnamefont {D.~V.}\ \bibnamefont
  {Khomitsky}}, \bibinfo {author} {\bibfnamefont {L.~V.}\ \bibnamefont
  {Gulyaev}},\ and\ \bibinfo {author} {\bibfnamefont {E.~Y.}\ \bibnamefont
  {Sherman}},\ }\bibfield  {title} {\bibinfo {title} {Spin dynamics in a
  strongly driven system: Very slow rabi oscillations},\ }\href
  {https://doi.org/10.1103/PhysRevB.85.125312} {\bibfield  {journal} {\bibinfo
  {journal} {Phys. Rev. B}\ }\textbf {\bibinfo {volume} {85}},\ \bibinfo
  {pages} {125312} (\bibinfo {year} {2012})}\BibitemShut {NoStop}%
\bibitem [{\citenamefont {Khaetskii}\ and\ \citenamefont
  {Nazarov}(2001)}]{phonon1}%
  \BibitemOpen
  \bibfield  {author} {\bibinfo {author} {\bibfnamefont {A.~V.}\ \bibnamefont
  {Khaetskii}}\ and\ \bibinfo {author} {\bibfnamefont {Y.~V.}\ \bibnamefont
  {Nazarov}},\ }\bibfield  {title} {\bibinfo {title} {Spin-flip transitions
  between $\text{Zeeman}$ sublevels in semiconductor quantum dots},\ }\href
  {https://doi.org/10.1103/PhysRevB.64.125316} {\bibfield  {journal} {\bibinfo
  {journal} {Phys. Rev. B}\ }\textbf {\bibinfo {volume} {64}},\ \bibinfo
  {pages} {125316} (\bibinfo {year} {2001})}\BibitemShut {NoStop}%
\bibitem [{\citenamefont {Golovach}\ \emph {et~al.}(2004)\citenamefont
  {Golovach}, \citenamefont {Khaetskii},\ and\ \citenamefont {Loss}}]{phonon2}%
  \BibitemOpen
  \bibfield  {author} {\bibinfo {author} {\bibfnamefont {V.~N.}\ \bibnamefont
  {Golovach}}, \bibinfo {author} {\bibfnamefont {A.}~\bibnamefont
  {Khaetskii}},\ and\ \bibinfo {author} {\bibfnamefont {D.}~\bibnamefont
  {Loss}},\ }\bibfield  {title} {\bibinfo {title} {Phonon-induced decay of the
  electron spin in quantum dots},\ }\href
  {https://doi.org/10.1103/PhysRevLett.93.016601} {\bibfield  {journal}
  {\bibinfo  {journal} {Phys. Rev. Lett.}\ }\textbf {\bibinfo {volume} {93}},\
  \bibinfo {pages} {016601} (\bibinfo {year} {2004})}\BibitemShut {NoStop}%
\bibitem [{\citenamefont {Bulaev}\ and\ \citenamefont
  {Loss}(2005{\natexlab{a}})}]{phonon3}%
  \BibitemOpen
  \bibfield  {author} {\bibinfo {author} {\bibfnamefont {D.~V.}\ \bibnamefont
  {Bulaev}}\ and\ \bibinfo {author} {\bibfnamefont {D.}~\bibnamefont {Loss}},\
  }\bibfield  {title} {\bibinfo {title} {Spin relaxation and anticrossing in
  quantum dots: Rashba versus $\text{Dresselhaus}$ spin-orbit coupling},\
  }\href {https://doi.org/10.1103/PhysRevB.71.205324} {\bibfield  {journal}
  {\bibinfo  {journal} {Phys. Rev. B}\ }\textbf {\bibinfo {volume} {71}},\
  \bibinfo {pages} {205324} (\bibinfo {year} {2005}{\natexlab{a}})}\BibitemShut
  {NoStop}%
\bibitem [{\citenamefont {Fal'ko}\ \emph {et~al.}(2005)\citenamefont {Fal'ko},
  \citenamefont {Altshuler},\ and\ \citenamefont {Tsyplyatyev}}]{phonon4}%
  \BibitemOpen
  \bibfield  {author} {\bibinfo {author} {\bibfnamefont {V.~I.}\ \bibnamefont
  {Fal'ko}}, \bibinfo {author} {\bibfnamefont {B.~L.}\ \bibnamefont
  {Altshuler}},\ and\ \bibinfo {author} {\bibfnamefont {O.}~\bibnamefont
  {Tsyplyatyev}},\ }\bibfield  {title} {\bibinfo {title} {Anisotropy of spin
  splitting and spin relaxation in lateral quantum dots},\ }\href
  {https://doi.org/10.1103/PhysRevLett.95.076603} {\bibfield  {journal}
  {\bibinfo  {journal} {Phys. Rev. Lett.}\ }\textbf {\bibinfo {volume} {95}},\
  \bibinfo {pages} {076603} (\bibinfo {year} {2005})}\BibitemShut {NoStop}%
\bibitem [{\citenamefont {Erlingsson}\ and\ \citenamefont
  {Nazarov}(2002)}]{hyperfine1}%
  \BibitemOpen
  \bibfield  {author} {\bibinfo {author} {\bibfnamefont {S.~I.}\ \bibnamefont
  {Erlingsson}}\ and\ \bibinfo {author} {\bibfnamefont {Y.~V.}\ \bibnamefont
  {Nazarov}},\ }\bibfield  {title} {\bibinfo {title} {Hyperfine-mediated
  transitions between a $\text{Zeeman}$ split doublet in gaas quantum dots: The
  role of the internal field},\ }\href
  {https://doi.org/10.1103/PhysRevB.66.155327} {\bibfield  {journal} {\bibinfo
  {journal} {Phys. Rev. B}\ }\textbf {\bibinfo {volume} {66}},\ \bibinfo
  {pages} {155327} (\bibinfo {year} {2002})}\BibitemShut {NoStop}%
\bibitem [{\citenamefont {Khaetskii}\ \emph {et~al.}(2002)\citenamefont
  {Khaetskii}, \citenamefont {Loss},\ and\ \citenamefont
  {Glazman}}]{hyperfine2}%
  \BibitemOpen
  \bibfield  {author} {\bibinfo {author} {\bibfnamefont {A.~V.}\ \bibnamefont
  {Khaetskii}}, \bibinfo {author} {\bibfnamefont {D.}~\bibnamefont {Loss}},\
  and\ \bibinfo {author} {\bibfnamefont {L.}~\bibnamefont {Glazman}},\
  }\bibfield  {title} {\bibinfo {title} {Electron spin decoherence in quantum
  dots due to interaction with nuclei},\ }\href
  {https://doi.org/10.1103/PhysRevLett.88.186802} {\bibfield  {journal}
  {\bibinfo  {journal} {Phys. Rev. Lett.}\ }\textbf {\bibinfo {volume} {88}},\
  \bibinfo {pages} {186802} (\bibinfo {year} {2002})}\BibitemShut {NoStop}%
\bibitem [{\citenamefont {Coish}\ and\ \citenamefont
  {Loss}(2004)}]{hyperfine3}%
  \BibitemOpen
  \bibfield  {author} {\bibinfo {author} {\bibfnamefont {W.~A.}\ \bibnamefont
  {Coish}}\ and\ \bibinfo {author} {\bibfnamefont {D.}~\bibnamefont {Loss}},\
  }\bibfield  {title} {\bibinfo {title} {Hyperfine interaction in a quantum
  dot: Non-markovian electron spin dynamics},\ }\href
  {https://doi.org/10.1103/PhysRevB.70.195340} {\bibfield  {journal} {\bibinfo
  {journal} {Phys. Rev. B}\ }\textbf {\bibinfo {volume} {70}},\ \bibinfo
  {pages} {195340} (\bibinfo {year} {2004})}\BibitemShut {NoStop}%
\bibitem [{\citenamefont {Becker}\ \emph {et~al.}(2010)\citenamefont {Becker},
  \citenamefont {Pohl}, \citenamefont {Riemann},\ and\ \citenamefont
  {Abrosimov}}]{Si-isotope}%
  \BibitemOpen
  \bibfield  {author} {\bibinfo {author} {\bibfnamefont {P.}~\bibnamefont
  {Becker}}, \bibinfo {author} {\bibfnamefont {H.-J.}\ \bibnamefont {Pohl}},
  \bibinfo {author} {\bibfnamefont {H.}~\bibnamefont {Riemann}},\ and\ \bibinfo
  {author} {\bibfnamefont {N.}~\bibnamefont {Abrosimov}},\ }\bibfield  {title}
  {\bibinfo {title} {Enrichment of silicon for a better kilogram},\ }\href
  {https://doi.org/10.1002/pssa.200925148} {\bibfield  {journal} {\bibinfo
  {journal} {Phys. Status Solidi A}\ }\textbf {\bibinfo {volume} {207}},\
  \bibinfo {pages} {49} (\bibinfo {year} {2010})}\BibitemShut {NoStop}%
\bibitem [{\citenamefont {Tyryshkin}\ \emph {et~al.}(2012)\citenamefont
  {Tyryshkin}, \citenamefont {Tojo}, \citenamefont {Morton}, \citenamefont
  {Riemann}, \citenamefont {Abrosimov}, \citenamefont {Becker}, \citenamefont
  {Pohl}, \citenamefont {Schenkel}, \citenamefont {Thewalt}, \citenamefont
  {M.},\ and\ \citenamefont {Lyon}}]{Si-isotope2}%
  \BibitemOpen
  \bibfield  {author} {\bibinfo {author} {\bibfnamefont {A.~M.}\ \bibnamefont
  {Tyryshkin}}, \bibinfo {author} {\bibfnamefont {S.}~\bibnamefont {Tojo}},
  \bibinfo {author} {\bibfnamefont {J.~J.~L.}\ \bibnamefont {Morton}}, \bibinfo
  {author} {\bibfnamefont {H.}~\bibnamefont {Riemann}}, \bibinfo {author}
  {\bibfnamefont {N.~V.}\ \bibnamefont {Abrosimov}}, \bibinfo {author}
  {\bibfnamefont {P.}~\bibnamefont {Becker}}, \bibinfo {author} {\bibfnamefont
  {H.-J.}\ \bibnamefont {Pohl}}, \bibinfo {author} {\bibfnamefont
  {T.}~\bibnamefont {Schenkel}}, \bibinfo {author} {\bibfnamefont {M.~L.~W.}\
  \bibnamefont {Thewalt}}, \bibinfo {author} {\bibfnamefont {I.~K.}\
  \bibnamefont {M.}},\ and\ \bibinfo {author} {\bibfnamefont {S.~A.}\
  \bibnamefont {Lyon}},\ }\bibfield  {title} {\bibinfo {title} {Electron spin
  coherence exceeding seconds in high-purity silicon},\ }\href
  {https://doi.org/10.1038/nmat3182} {\bibfield  {journal} {\bibinfo  {journal}
  {Nat. Mater.}\ }\textbf {\bibinfo {volume} {11}},\ \bibinfo {pages} {143}
  (\bibinfo {year} {2012})}\BibitemShut {NoStop}%
\bibitem [{\citenamefont {Itoh}\ \emph {et~al.}(1993)\citenamefont {Itoh},
  \citenamefont {Hansen}, \citenamefont {Haller}, \citenamefont {Farmer},
  \citenamefont {Ozhogin}, \citenamefont {Rudnev},\ and\ \citenamefont
  {Tikhomirov}}]{Ge-isotope}%
  \BibitemOpen
  \bibfield  {author} {\bibinfo {author} {\bibfnamefont {K.}~\bibnamefont
  {Itoh}}, \bibinfo {author} {\bibfnamefont {W.~L.}\ \bibnamefont {Hansen}},
  \bibinfo {author} {\bibfnamefont {E.~E.}\ \bibnamefont {Haller}}, \bibinfo
  {author} {\bibfnamefont {J.~W.}\ \bibnamefont {Farmer}}, \bibinfo {author}
  {\bibfnamefont {V.~I.}\ \bibnamefont {Ozhogin}}, \bibinfo {author}
  {\bibfnamefont {A.}~\bibnamefont {Rudnev}},\ and\ \bibinfo {author}
  {\bibfnamefont {A.}~\bibnamefont {Tikhomirov}},\ }\bibfield  {title}
  {\bibinfo {title} {High purity isotopically enriched ${}^{70}\text{Ge}$ and
  ${}^{74}\text{Ge}$ single crystals: Isotope separation, growth, and
  properties},\ }\href {https://doi.org/10.1557/JMR.1993.1341} {\bibfield
  {journal} {\bibinfo  {journal} {J. Mater. Res.}\ }\textbf {\bibinfo {volume}
  {8}},\ \bibinfo {pages} {1341} (\bibinfo {year} {1993})}\BibitemShut
  {NoStop}%
\bibitem [{\citenamefont {Fang}\ \emph {et~al.}(2023)\citenamefont {Fang},
  \citenamefont {Philippopoulos}, \citenamefont {Culcer}, \citenamefont
  {Coish},\ and\ \citenamefont {Chesi}}]{hole-review}%
  \BibitemOpen
  \bibfield  {author} {\bibinfo {author} {\bibfnamefont {Y.}~\bibnamefont
  {Fang}}, \bibinfo {author} {\bibfnamefont {P.}~\bibnamefont
  {Philippopoulos}}, \bibinfo {author} {\bibfnamefont {D.}~\bibnamefont
  {Culcer}}, \bibinfo {author} {\bibfnamefont {W.~A.}\ \bibnamefont {Coish}},\
  and\ \bibinfo {author} {\bibfnamefont {S.}~\bibnamefont {Chesi}},\ }\bibfield
   {title} {\bibinfo {title} {Recent advances in hole-spin qubits},\ }\href
  {https://doi.org/10.1088/2633-4356/acb87e} {\bibfield  {journal} {\bibinfo
  {journal} {Mater. Quantum. Technol.}\ }\textbf {\bibinfo {volume} {3}},\
  \bibinfo {pages} {012003} (\bibinfo {year} {2023})}\BibitemShut {NoStop}%
\bibitem [{\citenamefont {Chekhovich}\ \emph {et~al.}(2012)\citenamefont
  {Chekhovich}, \citenamefont {Glazov}, \citenamefont {Krysa}, \citenamefont
  {Hopkinson}, \citenamefont {Senellart}, \citenamefont {Lemaitre},
  \citenamefont {Skolnick},\ and\ \citenamefont
  {Tartakovskii}}]{hole-hyperfine1}%
  \BibitemOpen
  \bibfield  {author} {\bibinfo {author} {\bibfnamefont {E.~A.}\ \bibnamefont
  {Chekhovich}}, \bibinfo {author} {\bibfnamefont {M.~M.}\ \bibnamefont
  {Glazov}}, \bibinfo {author} {\bibfnamefont {A.~B.}\ \bibnamefont {Krysa}},
  \bibinfo {author} {\bibfnamefont {M.}~\bibnamefont {Hopkinson}}, \bibinfo
  {author} {\bibfnamefont {P.}~\bibnamefont {Senellart}}, \bibinfo {author}
  {\bibfnamefont {A.}~\bibnamefont {Lemaitre}}, \bibinfo {author}
  {\bibfnamefont {M.~S.}\ \bibnamefont {Skolnick}},\ and\ \bibinfo {author}
  {\bibfnamefont {A.~I.}\ \bibnamefont {Tartakovskii}},\ }\bibfield  {title}
  {\bibinfo {title} {Element-sensitive measurement of the hole-nuclear spin
  interaction in quantum dots},\ }\href {https://doi.org/10.1038/nphys2514}
  {\bibfield  {journal} {\bibinfo  {journal} {Nat. Phys.}\ }\textbf {\bibinfo
  {volume} {9}},\ \bibinfo {pages} {74} (\bibinfo {year} {2012})}\BibitemShut
  {NoStop}%
\bibitem [{\citenamefont {Fischer}\ \emph {et~al.}(2008)\citenamefont
  {Fischer}, \citenamefont {Coish}, \citenamefont {Bulaev},\ and\ \citenamefont
  {Loss}}]{hole-hyperfine2}%
  \BibitemOpen
  \bibfield  {author} {\bibinfo {author} {\bibfnamefont {J.}~\bibnamefont
  {Fischer}}, \bibinfo {author} {\bibfnamefont {W.~A.}\ \bibnamefont {Coish}},
  \bibinfo {author} {\bibfnamefont {D.~V.}\ \bibnamefont {Bulaev}},\ and\
  \bibinfo {author} {\bibfnamefont {D.}~\bibnamefont {Loss}},\ }\bibfield
  {title} {\bibinfo {title} {Spin decoherence of a heavy hole coupled to
  nuclear spins in a quantum dot},\ }\href
  {https://doi.org/10.1103/PhysRevB.78.155329} {\bibfield  {journal} {\bibinfo
  {journal} {Phys. Rev. B}\ }\textbf {\bibinfo {volume} {78}},\ \bibinfo
  {pages} {155329} (\bibinfo {year} {2008})}\BibitemShut {NoStop}%
\bibitem [{\citenamefont {Vidal}\ \emph {et~al.}(2016)\citenamefont {Vidal},
  \citenamefont {Durnev}, \citenamefont {Bouet}, \citenamefont {Amand},
  \citenamefont {Glazov}, \citenamefont {Ivchenko}, \citenamefont {Zhou},
  \citenamefont {Wang}, \citenamefont {Mano}, \citenamefont {Kuroda},
  \citenamefont {Marie}, \citenamefont {Sakoda},\ and\ \citenamefont
  {Urbaszek}}]{hole-hyperfine3}%
  \BibitemOpen
  \bibfield  {author} {\bibinfo {author} {\bibfnamefont {M.}~\bibnamefont
  {Vidal}}, \bibinfo {author} {\bibfnamefont {M.~V.}\ \bibnamefont {Durnev}},
  \bibinfo {author} {\bibfnamefont {L.}~\bibnamefont {Bouet}}, \bibinfo
  {author} {\bibfnamefont {T.}~\bibnamefont {Amand}}, \bibinfo {author}
  {\bibfnamefont {M.~M.}\ \bibnamefont {Glazov}}, \bibinfo {author}
  {\bibfnamefont {E.~L.}\ \bibnamefont {Ivchenko}}, \bibinfo {author}
  {\bibfnamefont {P.}~\bibnamefont {Zhou}}, \bibinfo {author} {\bibfnamefont
  {G.}~\bibnamefont {Wang}}, \bibinfo {author} {\bibfnamefont {T.}~\bibnamefont
  {Mano}}, \bibinfo {author} {\bibfnamefont {T.}~\bibnamefont {Kuroda}},
  \bibinfo {author} {\bibfnamefont {X.}~\bibnamefont {Marie}}, \bibinfo
  {author} {\bibfnamefont {K.}~\bibnamefont {Sakoda}},\ and\ \bibinfo {author}
  {\bibfnamefont {B.}~\bibnamefont {Urbaszek}},\ }\bibfield  {title} {\bibinfo
  {title} {Hyperfine coupling of hole and nuclear spins in symmetric
  (111)-grown $\text{GaAs}$ quantum dots},\ }\href
  {https://doi.org/10.1103/PhysRevB.94.121302} {\bibfield  {journal} {\bibinfo
  {journal} {Phys. Rev. B}\ }\textbf {\bibinfo {volume} {94}},\ \bibinfo
  {pages} {121302} (\bibinfo {year} {2016})}\BibitemShut {NoStop}%
\bibitem [{\citenamefont {Prechtel}\ \emph {et~al.}(2016)\citenamefont
  {Prechtel}, \citenamefont {Kuhlmann}, \citenamefont {Houel}, \citenamefont
  {Ludwig}, \citenamefont {Valentin}, \citenamefont {Wieck},\ and\
  \citenamefont {Warburton}}]{hole-hyperfine4}%
  \BibitemOpen
  \bibfield  {author} {\bibinfo {author} {\bibfnamefont {J.~H.}\ \bibnamefont
  {Prechtel}}, \bibinfo {author} {\bibfnamefont {A.~V.}\ \bibnamefont
  {Kuhlmann}}, \bibinfo {author} {\bibfnamefont {J.}~\bibnamefont {Houel}},
  \bibinfo {author} {\bibfnamefont {A.}~\bibnamefont {Ludwig}}, \bibinfo
  {author} {\bibfnamefont {S.~R.}\ \bibnamefont {Valentin}}, \bibinfo {author}
  {\bibfnamefont {A.~D.}\ \bibnamefont {Wieck}},\ and\ \bibinfo {author}
  {\bibfnamefont {R.~J.}\ \bibnamefont {Warburton}},\ }\bibfield  {title}
  {\bibinfo {title} {Decoupling a hole spin qubit from the nuclear spins},\
  }\href {https://doi.org/10.1038/nmat4704} {\bibfield  {journal} {\bibinfo
  {journal} {Nat. Mater.}\ }\textbf {\bibinfo {volume} {15}},\ \bibinfo {pages}
  {981} (\bibinfo {year} {2016})}\BibitemShut {NoStop}%
\bibitem [{\citenamefont {Zhang}\ \emph {et~al.}(2013)\citenamefont {Zhang},
  \citenamefont {Luo}, \citenamefont {Saraiva}, \citenamefont {Koiller},\ and\
  \citenamefont {Zunger}}]{valley}%
  \BibitemOpen
  \bibfield  {author} {\bibinfo {author} {\bibfnamefont {L.}~\bibnamefont
  {Zhang}}, \bibinfo {author} {\bibfnamefont {J.-W.}\ \bibnamefont {Luo}},
  \bibinfo {author} {\bibfnamefont {A.}~\bibnamefont {Saraiva}}, \bibinfo
  {author} {\bibfnamefont {B.}~\bibnamefont {Koiller}},\ and\ \bibinfo {author}
  {\bibfnamefont {A.}~\bibnamefont {Zunger}},\ }\bibfield  {title} {\bibinfo
  {title} {Genetic design of enhanced valley splitting towards a spin qubit in
  silicon},\ }\href {https://doi.org/10.1038/ncomms3396} {\bibfield  {journal}
  {\bibinfo  {journal} {Nat. Comm.}\ }\textbf {\bibinfo {volume} {4}},\
  \bibinfo {pages} {2396} (\bibinfo {year} {2013})}\BibitemShut {NoStop}%
\bibitem [{\citenamefont {Bulaev}\ and\ \citenamefont
  {Loss}(2005{\natexlab{b}})}]{hole-loss}%
  \BibitemOpen
  \bibfield  {author} {\bibinfo {author} {\bibfnamefont {D.~V.}\ \bibnamefont
  {Bulaev}}\ and\ \bibinfo {author} {\bibfnamefont {D.}~\bibnamefont {Loss}},\
  }\bibfield  {title} {\bibinfo {title} {Spin relaxation and decoherence of
  holes in quantum dots},\ }\href
  {https://doi.org/10.1103/PhysRevLett.95.076805} {\bibfield  {journal}
  {\bibinfo  {journal} {Phys. Rev. Lett.}\ }\textbf {\bibinfo {volume} {95}},\
  \bibinfo {pages} {076805} (\bibinfo {year} {2005}{\natexlab{b}})}\BibitemShut
  {NoStop}%
\bibitem [{\citenamefont {Scappucci}\ \emph {et~al.}(2021)\citenamefont
  {Scappucci}, \citenamefont {Kloeffel}, \citenamefont {Zwanenburg},
  \citenamefont {Loss}, \citenamefont {Myronov}, \citenamefont {Zhang},
  \citenamefont {Franceschi}, \citenamefont {Katsaros},\ and\ \citenamefont
  {Veldhorst}}]{Ge-review}%
  \BibitemOpen
  \bibfield  {author} {\bibinfo {author} {\bibfnamefont {G.}~\bibnamefont
  {Scappucci}}, \bibinfo {author} {\bibfnamefont {C.}~\bibnamefont {Kloeffel}},
  \bibinfo {author} {\bibfnamefont {F.~A.}\ \bibnamefont {Zwanenburg}},
  \bibinfo {author} {\bibfnamefont {D.}~\bibnamefont {Loss}}, \bibinfo {author}
  {\bibfnamefont {M.}~\bibnamefont {Myronov}}, \bibinfo {author} {\bibfnamefont
  {J.-J.}\ \bibnamefont {Zhang}}, \bibinfo {author} {\bibfnamefont {S.~D.}\
  \bibnamefont {Franceschi}}, \bibinfo {author} {\bibfnamefont
  {G.}~\bibnamefont {Katsaros}},\ and\ \bibinfo {author} {\bibfnamefont
  {M.}~\bibnamefont {Veldhorst}},\ }\bibfield  {title} {\bibinfo {title} {The
  germanium quantum information route},\ }\href
  {https://doi.org/10.1038/s41578-020-00262-z} {\bibfield  {journal} {\bibinfo
  {journal} {Nat. Rev. Mater.}\ }\textbf {\bibinfo {volume} {6}},\ \bibinfo
  {pages} {926} (\bibinfo {year} {2021})}\BibitemShut {NoStop}%
\bibitem [{\citenamefont {Lodari}\ \emph {et~al.}(2019)\citenamefont {Lodari},
  \citenamefont {Tosato}, \citenamefont {Sabbagh}, \citenamefont {Schubert},
  \citenamefont {Capellini}, \citenamefont {Sammak}, \citenamefont
  {Veldhorst},\ and\ \citenamefont {Scappucci}}]{hole-mass}%
  \BibitemOpen
  \bibfield  {author} {\bibinfo {author} {\bibfnamefont {M.}~\bibnamefont
  {Lodari}}, \bibinfo {author} {\bibfnamefont {A.}~\bibnamefont {Tosato}},
  \bibinfo {author} {\bibfnamefont {D.}~\bibnamefont {Sabbagh}}, \bibinfo
  {author} {\bibfnamefont {M.~A.}\ \bibnamefont {Schubert}}, \bibinfo {author}
  {\bibfnamefont {G.}~\bibnamefont {Capellini}}, \bibinfo {author}
  {\bibfnamefont {A.}~\bibnamefont {Sammak}}, \bibinfo {author} {\bibfnamefont
  {M.}~\bibnamefont {Veldhorst}},\ and\ \bibinfo {author} {\bibfnamefont
  {G.}~\bibnamefont {Scappucci}},\ }\bibfield  {title} {\bibinfo {title} {Light
  effective hole mass in undoped $\text{Ge/SiGe}$ quantum wells},\ }\href
  {https://doi.org/10.1103/PhysRevB.100.041304} {\bibfield  {journal} {\bibinfo
   {journal} {Phys. Rev. B}\ }\textbf {\bibinfo {volume} {100}},\ \bibinfo
  {pages} {041304} (\bibinfo {year} {2019})}\BibitemShut {NoStop}%
\bibitem [{\citenamefont {Luo}\ \emph {et~al.}(2017)\citenamefont {Luo},
  \citenamefont {Li},\ and\ \citenamefont {Zunger}}]{ge-heavy}%
  \BibitemOpen
  \bibfield  {author} {\bibinfo {author} {\bibfnamefont {J.-W.}\ \bibnamefont
  {Luo}}, \bibinfo {author} {\bibfnamefont {S.-S.}\ \bibnamefont {Li}},\ and\
  \bibinfo {author} {\bibfnamefont {A.}~\bibnamefont {Zunger}},\ }\bibfield
  {title} {\bibinfo {title} {Rapid transition of the hole $\text{Rashba}$
  effect from strong field dependence to saturation in semiconductor
  nanowires},\ }\href {https://doi.org/10.1103/PhysRevLett.119.126401}
  {\bibfield  {journal} {\bibinfo  {journal} {Phys. Rev. Lett.}\ }\textbf
  {\bibinfo {volume} {119}},\ \bibinfo {pages} {126401} (\bibinfo {year}
  {2017})}\BibitemShut {NoStop}%
\bibitem [{\citenamefont {Watzinger}\ \emph {et~al.}(2018)\citenamefont
  {Watzinger}, \citenamefont {Kukučka}, \citenamefont {Vukušić},
  \citenamefont {Gao}, \citenamefont {Wang}, \citenamefont {Schäffler},
  \citenamefont {Zhang},\ and\ \citenamefont {Katsaros}}]{ge-hole-exp1}%
  \BibitemOpen
  \bibfield  {author} {\bibinfo {author} {\bibfnamefont {H.}~\bibnamefont
  {Watzinger}}, \bibinfo {author} {\bibfnamefont {K.}~\bibnamefont {Kukučka}},
  \bibinfo {author} {\bibfnamefont {L.}~\bibnamefont {Vukušić}}, \bibinfo
  {author} {\bibfnamefont {F.}~\bibnamefont {Gao}}, \bibinfo {author}
  {\bibfnamefont {T.}~\bibnamefont {Wang}}, \bibinfo {author} {\bibfnamefont
  {F.}~\bibnamefont {Schäffler}}, \bibinfo {author} {\bibfnamefont {J.-J.}\
  \bibnamefont {Zhang}},\ and\ \bibinfo {author} {\bibfnamefont
  {G.}~\bibnamefont {Katsaros}},\ }\bibfield  {title} {\bibinfo {title} {A
  germanium hole spin qubit},\ }\href
  {https://doi.org/10.1038/s41467-018-06418-4} {\bibfield  {journal} {\bibinfo
  {journal} {Nat. Commun.}\ }\textbf {\bibinfo {volume} {9}},\ \bibinfo {pages}
  {3092} (\bibinfo {year} {2018})}\BibitemShut {NoStop}%
\bibitem [{\citenamefont {Hendrickx}\ \emph
  {et~al.}(2020{\natexlab{a}})\citenamefont {Hendrickx}, \citenamefont
  {Lawrie}, \citenamefont {Petit}, \citenamefont {Sammak}, \citenamefont
  {Scappucci},\ and\ \citenamefont {Veldhorst}}]{ge-hole-exp2}%
  \BibitemOpen
  \bibfield  {author} {\bibinfo {author} {\bibfnamefont {N.~W.}\ \bibnamefont
  {Hendrickx}}, \bibinfo {author} {\bibfnamefont {W.~I.~L.}\ \bibnamefont
  {Lawrie}}, \bibinfo {author} {\bibfnamefont {L.}~\bibnamefont {Petit}},
  \bibinfo {author} {\bibfnamefont {A.}~\bibnamefont {Sammak}}, \bibinfo
  {author} {\bibfnamefont {G.}~\bibnamefont {Scappucci}},\ and\ \bibinfo
  {author} {\bibfnamefont {M.}~\bibnamefont {Veldhorst}},\ }\bibfield  {title}
  {\bibinfo {title} {A single-hole spin qubit},\ }\href
  {https://doi.org/10.1038/s41467-020-17211-7} {\bibfield  {journal} {\bibinfo
  {journal} {Nat. Commun.}\ }\textbf {\bibinfo {volume} {11}},\ \bibinfo
  {pages} {3478} (\bibinfo {year} {2020}{\natexlab{a}})}\BibitemShut {NoStop}%
\bibitem [{\citenamefont {Hendrickx}\ \emph
  {et~al.}(2020{\natexlab{b}})\citenamefont {Hendrickx}, \citenamefont
  {Franke}, \citenamefont {Sammak}, \citenamefont {Scappucci},\ and\
  \citenamefont {Veldhorst}}]{ge-hole-exp3}%
  \BibitemOpen
  \bibfield  {author} {\bibinfo {author} {\bibfnamefont {N.~W.}\ \bibnamefont
  {Hendrickx}}, \bibinfo {author} {\bibfnamefont {D.~P.}\ \bibnamefont
  {Franke}}, \bibinfo {author} {\bibfnamefont {A.}~\bibnamefont {Sammak}},
  \bibinfo {author} {\bibfnamefont {G.}~\bibnamefont {Scappucci}},\ and\
  \bibinfo {author} {\bibfnamefont {M.}~\bibnamefont {Veldhorst}},\ }\bibfield
  {title} {\bibinfo {title} {Fast two-qubit logic with holes in germanium},\
  }\href {https://doi.org/10.1038/s41586-019-1919-3} {\bibfield  {journal}
  {\bibinfo  {journal} {Nat.}\ }\textbf {\bibinfo {volume} {577}},\ \bibinfo
  {pages} {487} (\bibinfo {year} {2020}{\natexlab{b}})}\BibitemShut {NoStop}%
\bibitem [{\citenamefont {Wang}\ \emph {et~al.}(2022)\citenamefont {Wang},
  \citenamefont {Xu}, \citenamefont {Gao}, \citenamefont {Liu}, \citenamefont
  {Ma}, \citenamefont {Zhang}, \citenamefont {Wang}, \citenamefont {Cao},
  \citenamefont {Wang}, \citenamefont {Zhang}, \citenamefont {Culcer},
  \citenamefont {Hu}, \citenamefont {Jiang}, \citenamefont {Li}, \citenamefont
  {G.-C.},\ and\ \citenamefont {Guo}}]{ge-hole-exp4}%
  \BibitemOpen
  \bibfield  {author} {\bibinfo {author} {\bibfnamefont {K.}~\bibnamefont
  {Wang}}, \bibinfo {author} {\bibfnamefont {G.}~\bibnamefont {Xu}}, \bibinfo
  {author} {\bibfnamefont {F.}~\bibnamefont {Gao}}, \bibinfo {author}
  {\bibfnamefont {H.}~\bibnamefont {Liu}}, \bibinfo {author} {\bibfnamefont
  {R.-L.}\ \bibnamefont {Ma}}, \bibinfo {author} {\bibfnamefont
  {X.}~\bibnamefont {Zhang}}, \bibinfo {author} {\bibfnamefont
  {Z.}~\bibnamefont {Wang}}, \bibinfo {author} {\bibfnamefont {G.}~\bibnamefont
  {Cao}}, \bibinfo {author} {\bibfnamefont {T.}~\bibnamefont {Wang}}, \bibinfo
  {author} {\bibfnamefont {J.-J.}\ \bibnamefont {Zhang}}, \bibinfo {author}
  {\bibfnamefont {D.}~\bibnamefont {Culcer}}, \bibinfo {author} {\bibfnamefont
  {X.}~\bibnamefont {Hu}}, \bibinfo {author} {\bibfnamefont {H.~W.}\
  \bibnamefont {Jiang}}, \bibinfo {author} {\bibfnamefont {H.-O.}\ \bibnamefont
  {Li}}, \bibinfo {author} {\bibfnamefont {G.}~\bibnamefont {G.-C.}},\ and\
  \bibinfo {author} {\bibfnamefont {G.-P.}\ \bibnamefont {Guo}},\ }\bibfield
  {title} {\bibinfo {title} {Ultrafast coherent control of a hole spin qubit in
  a germanium quantum dot},\ }\href
  {https://doi.org/10.1038/s41467-021-27880-7} {\bibfield  {journal} {\bibinfo
  {journal} {Nat. Commun.}\ }\textbf {\bibinfo {volume} {13}},\ \bibinfo
  {pages} {206} (\bibinfo {year} {2022})}\BibitemShut {NoStop}%
\bibitem [{\citenamefont {Winkler}(2000)}]{winkler}%
  \BibitemOpen
  \bibfield  {author} {\bibinfo {author} {\bibfnamefont {R.}~\bibnamefont
  {Winkler}},\ }\bibfield  {title} {\bibinfo {title} {Rashba spin splitting in
  two-dimensional electron and hole systems},\ }\href
  {https://doi.org/10.1103/PhysRevB.62.4245} {\bibfield  {journal} {\bibinfo
  {journal} {Phys. Rev. B}\ }\textbf {\bibinfo {volume} {62}},\ \bibinfo
  {pages} {4245} (\bibinfo {year} {2000})}\BibitemShut {NoStop}%
\bibitem [{\citenamefont {Marcellina}\ \emph {et~al.}(2017)\citenamefont
  {Marcellina}, \citenamefont {Hamilton}, \citenamefont {Winkler},\ and\
  \citenamefont {Culcer}}]{winkler2}%
  \BibitemOpen
  \bibfield  {author} {\bibinfo {author} {\bibfnamefont {E.}~\bibnamefont
  {Marcellina}}, \bibinfo {author} {\bibfnamefont {A.~R.}\ \bibnamefont
  {Hamilton}}, \bibinfo {author} {\bibfnamefont {R.}~\bibnamefont {Winkler}},\
  and\ \bibinfo {author} {\bibfnamefont {D.}~\bibnamefont {Culcer}},\
  }\bibfield  {title} {\bibinfo {title} {Spin-orbit interactions in
  inversion-asymmetric two-dimensional hole systems: A variational analysis},\
  }\href {https://doi.org/10.1103/PhysRevB.95.075305} {\bibfield  {journal}
  {\bibinfo  {journal} {Phys. Rev. B}\ }\textbf {\bibinfo {volume} {95}},\
  \bibinfo {pages} {075305} (\bibinfo {year} {2017})}\BibitemShut {NoStop}%
\bibitem [{\citenamefont {Terrazos}\ \emph {et~al.}(2021)\citenamefont
  {Terrazos}, \citenamefont {Marcellina}, \citenamefont {Wang}, \citenamefont
  {Coppersmith}, \citenamefont {Friesen}, \citenamefont {Hamilton},
  \citenamefont {Hu}, \citenamefont {Koiller}, \citenamefont {Saraiva},
  \citenamefont {Culcer},\ and\ \citenamefont {Capaz}}]{strained-Ge}%
  \BibitemOpen
  \bibfield  {author} {\bibinfo {author} {\bibfnamefont {L.~A.}\ \bibnamefont
  {Terrazos}}, \bibinfo {author} {\bibfnamefont {E.}~\bibnamefont
  {Marcellina}}, \bibinfo {author} {\bibfnamefont {Z.}~\bibnamefont {Wang}},
  \bibinfo {author} {\bibfnamefont {S.~N.}\ \bibnamefont {Coppersmith}},
  \bibinfo {author} {\bibfnamefont {M.}~\bibnamefont {Friesen}}, \bibinfo
  {author} {\bibfnamefont {A.~R.}\ \bibnamefont {Hamilton}}, \bibinfo {author}
  {\bibfnamefont {X.}~\bibnamefont {Hu}}, \bibinfo {author} {\bibfnamefont
  {B.}~\bibnamefont {Koiller}}, \bibinfo {author} {\bibfnamefont {A.~L.}\
  \bibnamefont {Saraiva}}, \bibinfo {author} {\bibfnamefont {D.}~\bibnamefont
  {Culcer}},\ and\ \bibinfo {author} {\bibfnamefont {R.~B.}\ \bibnamefont
  {Capaz}},\ }\bibfield  {title} {\bibinfo {title} {Theory of hole-spin qubits
  in strained germanium quantum dots},\ }\href
  {https://doi.org/10.1103/PhysRevB.103.125201} {\bibfield  {journal} {\bibinfo
   {journal} {Phys. Rev. B}\ }\textbf {\bibinfo {volume} {103}},\ \bibinfo
  {pages} {125201} (\bibinfo {year} {2021})}\BibitemShut {NoStop}%
\bibitem [{\citenamefont {Wang}\ \emph {et~al.}(2021)\citenamefont {Wang},
  \citenamefont {Marcellina}, \citenamefont {Hamilton}, \citenamefont {Cullen},
  \citenamefont {Rogge}, \citenamefont {Salfi},\ and\ \citenamefont
  {Culcer}}]{optimal-Ge}%
  \BibitemOpen
  \bibfield  {author} {\bibinfo {author} {\bibfnamefont {Z.}~\bibnamefont
  {Wang}}, \bibinfo {author} {\bibfnamefont {E.}~\bibnamefont {Marcellina}},
  \bibinfo {author} {\bibfnamefont {A.~R.}\ \bibnamefont {Hamilton}}, \bibinfo
  {author} {\bibfnamefont {J.~H.}\ \bibnamefont {Cullen}}, \bibinfo {author}
  {\bibfnamefont {S.}~\bibnamefont {Rogge}}, \bibinfo {author} {\bibfnamefont
  {J.}~\bibnamefont {Salfi}},\ and\ \bibinfo {author} {\bibfnamefont
  {D.}~\bibnamefont {Culcer}},\ }\bibfield  {title} {\bibinfo {title} {Optimal
  operation points for ultrafast, highly coherent $\text{Ge}$ hole spin-orbit
  qubits},\ }\href {https://doi.org/10.1038/s41534-021-00386-2} {\bibfield
  {journal} {\bibinfo  {journal} {npj Quantum Inf 7, 54}\ }\textbf {\bibinfo
  {volume} {7}},\ \bibinfo {pages} {54} (\bibinfo {year} {2021})}\BibitemShut
  {NoStop}%
\bibitem [{\citenamefont {Sarkar}\ \emph {et~al.}(2023)\citenamefont {Sarkar},
  \citenamefont {Wang}, \citenamefont {Rendell}, \citenamefont {Hendrickx},
  \citenamefont {Veldhorst}, \citenamefont {Scappucci}, \citenamefont
  {Khalifa}, \citenamefont {Salfi}, \citenamefont {Saraiva}, \citenamefont
  {Dzurak}, \citenamefont {Hamilton},\ and\ \citenamefont {Culcer}}]{in-plane}%
  \BibitemOpen
  \bibfield  {author} {\bibinfo {author} {\bibfnamefont {A.}~\bibnamefont
  {Sarkar}}, \bibinfo {author} {\bibfnamefont {Z.}~\bibnamefont {Wang}},
  \bibinfo {author} {\bibfnamefont {M.}~\bibnamefont {Rendell}}, \bibinfo
  {author} {\bibfnamefont {N.~W.}\ \bibnamefont {Hendrickx}}, \bibinfo {author}
  {\bibfnamefont {M.}~\bibnamefont {Veldhorst}}, \bibinfo {author}
  {\bibfnamefont {G.}~\bibnamefont {Scappucci}}, \bibinfo {author}
  {\bibfnamefont {M.}~\bibnamefont {Khalifa}}, \bibinfo {author} {\bibfnamefont
  {J.}~\bibnamefont {Salfi}}, \bibinfo {author} {\bibfnamefont
  {A.}~\bibnamefont {Saraiva}}, \bibinfo {author} {\bibfnamefont {A.~S.}\
  \bibnamefont {Dzurak}}, \bibinfo {author} {\bibfnamefont {A.~R.}\
  \bibnamefont {Hamilton}},\ and\ \bibinfo {author} {\bibfnamefont
  {D.}~\bibnamefont {Culcer}},\ }\bibfield  {title} {\bibinfo {title}
  {Electrical operation of planar $\text{Ge}$ hole spin qubits in an in-plane
  magnetic field},\ }\href {https://doi.org/10.1103/PhysRevB.108.245301}
  {\bibfield  {journal} {\bibinfo  {journal} {Phys. Rev. B}\ }\textbf {\bibinfo
  {volume} {108}},\ \bibinfo {pages} {245301} (\bibinfo {year}
  {2023})}\BibitemShut {NoStop}%
\bibitem [{\citenamefont {Liu}\ \emph {et~al.}(2022)\citenamefont {Liu},
  \citenamefont {Xiong}, \citenamefont {Wang}, \citenamefont {Ma},
  \citenamefont {Guan}, \citenamefont {Luo},\ and\ \citenamefont
  {Li}}]{emergent}%
  \BibitemOpen
  \bibfield  {author} {\bibinfo {author} {\bibfnamefont {Y.}~\bibnamefont
  {Liu}}, \bibinfo {author} {\bibfnamefont {J.-X.}\ \bibnamefont {Xiong}},
  \bibinfo {author} {\bibfnamefont {Z.}~\bibnamefont {Wang}}, \bibinfo {author}
  {\bibfnamefont {W.-L.}\ \bibnamefont {Ma}}, \bibinfo {author} {\bibfnamefont
  {S.}~\bibnamefont {Guan}}, \bibinfo {author} {\bibfnamefont {J.-W.}\
  \bibnamefont {Luo}},\ and\ \bibinfo {author} {\bibfnamefont {S.-S.}\
  \bibnamefont {Li}},\ }\bibfield  {title} {\bibinfo {title} {Emergent linear
  $\text{Rashba}$ spin-orbit coupling offers fast manipulation of hole-spin
  qubits in germanium},\ }\href {https://doi.org/10.1103/PhysRevB.105.075313}
  {\bibfield  {journal} {\bibinfo  {journal} {Phys. Rev. B}\ }\textbf {\bibinfo
  {volume} {105}},\ \bibinfo {pages} {075313} (\bibinfo {year}
  {2022})}\BibitemShut {NoStop}%
\bibitem [{\citenamefont {Ivchenko}\ \emph {et~al.}(1996)\citenamefont
  {Ivchenko}, \citenamefont {Kaminski},\ and\ \citenamefont
  {R\"ossler}}]{PhysRevB.54.5852}%
  \BibitemOpen
  \bibfield  {author} {\bibinfo {author} {\bibfnamefont {E.~L.}\ \bibnamefont
  {Ivchenko}}, \bibinfo {author} {\bibfnamefont {A.~Y.}\ \bibnamefont
  {Kaminski}},\ and\ \bibinfo {author} {\bibfnamefont {U.}~\bibnamefont
  {R\"ossler}},\ }\bibfield  {title} {\bibinfo {title} {Heavy-light hole mixing
  at zinc-blende (001) interfaces under normal incidence},\ }\href
  {https://doi.org/10.1103/PhysRevB.54.5852} {\bibfield  {journal} {\bibinfo
  {journal} {Phys. Rev. B}\ }\textbf {\bibinfo {volume} {54}},\ \bibinfo
  {pages} {5852} (\bibinfo {year} {1996})}\BibitemShut {NoStop}%
\bibitem [{\citenamefont {Luo}\ \emph {et~al.}(2015)\citenamefont {Luo},
  \citenamefont {Bester},\ and\ \citenamefont {Zunger}}]{PhysRevB.92.165301}%
  \BibitemOpen
  \bibfield  {author} {\bibinfo {author} {\bibfnamefont {J.-W.}\ \bibnamefont
  {Luo}}, \bibinfo {author} {\bibfnamefont {G.}~\bibnamefont {Bester}},\ and\
  \bibinfo {author} {\bibfnamefont {A.}~\bibnamefont {Zunger}},\ }\bibfield
  {title} {\bibinfo {title} {Supercoupling between heavy-hole and light-hole
  states in nanostructures},\ }\href
  {https://doi.org/10.1103/PhysRevB.92.165301} {\bibfield  {journal} {\bibinfo
  {journal} {Phys. Rev. B}\ }\textbf {\bibinfo {volume} {92}},\ \bibinfo
  {pages} {165301} (\bibinfo {year} {2015})}\BibitemShut {NoStop}%
\bibitem [{\citenamefont {Golub}\ and\ \citenamefont
  {Ivchenko}(2004)}]{PhysRevB.69.115333}%
  \BibitemOpen
  \bibfield  {author} {\bibinfo {author} {\bibfnamefont {L.~E.}\ \bibnamefont
  {Golub}}\ and\ \bibinfo {author} {\bibfnamefont {E.~L.}\ \bibnamefont
  {Ivchenko}},\ }\bibfield  {title} {\bibinfo {title} {Spin splitting in
  symmetrical $\text{SiGe}$ quantum wells},\ }\href
  {https://doi.org/10.1103/PhysRevB.69.115333} {\bibfield  {journal} {\bibinfo
  {journal} {Phys. Rev. B}\ }\textbf {\bibinfo {volume} {69}},\ \bibinfo
  {pages} {115333} (\bibinfo {year} {2004})}\BibitemShut {NoStop}%
\bibitem [{\citenamefont {Durnev}\ \emph {et~al.}(2014)\citenamefont {Durnev},
  \citenamefont {Glazov},\ and\ \citenamefont {Ivchenko}}]{PhysRevB.89.075430}%
  \BibitemOpen
  \bibfield  {author} {\bibinfo {author} {\bibfnamefont {M.~V.}\ \bibnamefont
  {Durnev}}, \bibinfo {author} {\bibfnamefont {M.~M.}\ \bibnamefont {Glazov}},\
  and\ \bibinfo {author} {\bibfnamefont {E.~L.}\ \bibnamefont {Ivchenko}},\
  }\bibfield  {title} {\bibinfo {title} {Spin-orbit splitting of valence
  subbands in semiconductor nanostructures},\ }\href
  {https://doi.org/10.1103/PhysRevB.89.075430} {\bibfield  {journal} {\bibinfo
  {journal} {Phys. Rev. B}\ }\textbf {\bibinfo {volume} {89}},\ \bibinfo
  {pages} {075430} (\bibinfo {year} {2014})}\BibitemShut {NoStop}%
\bibitem [{\citenamefont {Xiong}\ \emph {et~al.}(2021)\citenamefont {Xiong},
  \citenamefont {Guan}, \citenamefont {Luo},\ and\ \citenamefont
  {Li}}]{emergence}%
  \BibitemOpen
  \bibfield  {author} {\bibinfo {author} {\bibfnamefont {J.-X.}\ \bibnamefont
  {Xiong}}, \bibinfo {author} {\bibfnamefont {S.}~\bibnamefont {Guan}},
  \bibinfo {author} {\bibfnamefont {J.-W.}\ \bibnamefont {Luo}},\ and\ \bibinfo
  {author} {\bibfnamefont {S.-S.}\ \bibnamefont {Li}},\ }\bibfield  {title}
  {\bibinfo {title} {Emergence of strong tunable linear $\text{Rashba}$
  spin-orbit coupling in two-dimensional hole gases in semiconductor quantum
  wells},\ }\href {https://doi.org/10.1103/PhysRevB.103.085309} {\bibfield
  {journal} {\bibinfo  {journal} {Phys. Rev. B}\ }\textbf {\bibinfo {volume}
  {103}},\ \bibinfo {pages} {085309} (\bibinfo {year} {2021})}\BibitemShut
  {NoStop}%
\bibitem [{\citenamefont {Rodrgiuez-Mena}\ \emph {et~al.}(2023)\citenamefont
  {Rodrgiuez-Mena}, \citenamefont {Abadillo-Uriel}, \citenamefont {Veste},
  \citenamefont {Martinez}, \citenamefont {Li}, \citenamefont {Sklénard},\
  and\ \citenamefont {Niquet}}]{Dresselhaus}%
  \BibitemOpen
  \bibfield  {author} {\bibinfo {author} {\bibfnamefont {E.~A.}\ \bibnamefont
  {Rodrgiuez-Mena}}, \bibinfo {author} {\bibfnamefont {J.~C.}\ \bibnamefont
  {Abadillo-Uriel}}, \bibinfo {author} {\bibfnamefont {G.}~\bibnamefont
  {Veste}}, \bibinfo {author} {\bibfnamefont {B.}~\bibnamefont {Martinez}},
  \bibinfo {author} {\bibfnamefont {J.}~\bibnamefont {Li}}, \bibinfo {author}
  {\bibfnamefont {B.}~\bibnamefont {Sklénard}},\ and\ \bibinfo {author}
  {\bibfnamefont {Y.-M.}\ \bibnamefont {Niquet}},\ }\bibfield  {title}
  {\bibinfo {title} {Linear-in-momentum spin orbit interactions in planar
  ge/gesi heterostructures and spin qubits},\ }\href
  {https://doi.org/10.1103/PhysRevB.108.205416} {\bibfield  {journal} {\bibinfo
   {journal} {Phys. Rev. B}\ }\textbf {\bibinfo {volume} {108}},\ \bibinfo
  {pages} {205416} (\bibinfo {year} {2023})}\BibitemShut {NoStop}%
\bibitem [{\citenamefont {Abadillo-Uriel}\ \emph {et~al.}(2023)\citenamefont
  {Abadillo-Uriel}, \citenamefont {Rodriguez-Mena}, \citenamefont {Martinez},\
  and\ \citenamefont {Niquet}}]{special-rashba}%
  \BibitemOpen
  \bibfield  {author} {\bibinfo {author} {\bibfnamefont {J.~C.}\ \bibnamefont
  {Abadillo-Uriel}}, \bibinfo {author} {\bibfnamefont {E.~A.}\ \bibnamefont
  {Rodriguez-Mena}}, \bibinfo {author} {\bibfnamefont {B.}~\bibnamefont
  {Martinez}},\ and\ \bibinfo {author} {\bibfnamefont {Y.-M.}\ \bibnamefont
  {Niquet}},\ }\bibfield  {title} {\bibinfo {title} {Hole-spin driving by
  strain-induced spin-orbit interactions},\ }\href
  {https://doi.org/10.1103/PhysRevLett.131.097002} {\bibfield  {journal}
  {\bibinfo  {journal} {Phys. Rev. Lett.}\ }\textbf {\bibinfo {volume} {131}},\
  \bibinfo {pages} {097002} (\bibinfo {year} {2023})}\BibitemShut {NoStop}%
\bibitem [{\citenamefont {Martinez}\ \emph {et~al.}(2022)\citenamefont
  {Martinez}, \citenamefont {Abadillo-Uriel}, \citenamefont {Rodriguez-Mena},\
  and\ \citenamefont {Niquet}}]{inhomogeneous}%
  \BibitemOpen
  \bibfield  {author} {\bibinfo {author} {\bibfnamefont {B.}~\bibnamefont
  {Martinez}}, \bibinfo {author} {\bibfnamefont {J.~C.}\ \bibnamefont
  {Abadillo-Uriel}}, \bibinfo {author} {\bibfnamefont {E.~A.}\ \bibnamefont
  {Rodriguez-Mena}},\ and\ \bibinfo {author} {\bibfnamefont {Y.-M.}\
  \bibnamefont {Niquet}},\ }\bibfield  {title} {\bibinfo {title} {Hole spin
  manipulation in inhomogeneous and nonseparable electric fields},\ }\href
  {https://doi.org/10.1103/PhysRevB.106.235426} {\bibfield  {journal} {\bibinfo
   {journal} {Phys. Rev. B}\ }\textbf {\bibinfo {volume} {106}},\ \bibinfo
  {pages} {235426} (\bibinfo {year} {2022})}\BibitemShut {NoStop}%
\bibitem [{\citenamefont {Orenstein}(2012)}]{ultrafast1}%
  \BibitemOpen
  \bibfield  {author} {\bibinfo {author} {\bibfnamefont {J.}~\bibnamefont
  {Orenstein}},\ }\bibfield  {title} {\bibinfo {title} {Ultrafast spectroscopy
  of quantum materials},\ }\href {http://dx.doi.org/10.1063/PT.3.1717}
  {\bibfield  {journal} {\bibinfo  {journal} {Phys. Today}\ }\textbf {\bibinfo
  {volume} {65}},\ \bibinfo {pages} {9} (\bibinfo {year} {2012})}\BibitemShut
  {NoStop}%
\bibitem [{\citenamefont {Kobayashi}(2018)}]{ultrafast2}%
  \BibitemOpen
  \bibfield  {author} {\bibinfo {author} {\bibfnamefont {T.}~\bibnamefont
  {Kobayashi}},\ }\bibfield  {title} {\bibinfo {title} {Development of
  ultrashort pulse lasers for ultrafast spectroscopy},\ }\href
  {https://doi.org/10.3390/photonics5030019} {\bibfield  {journal} {\bibinfo
  {journal} {Photonics}\ }\textbf {\bibinfo {volume} {5}},\ \bibinfo {pages}
  {19} (\bibinfo {year} {2018})}\BibitemShut {NoStop}%
\bibitem [{\citenamefont {Zong}\ \emph {et~al.}(2023)\citenamefont {Zong},
  \citenamefont {Nebgen}, \citenamefont {Lin}, \citenamefont {Spies},\ and\
  \citenamefont {Zuerch}}]{ultrafast3}%
  \BibitemOpen
  \bibfield  {author} {\bibinfo {author} {\bibfnamefont {A.}~\bibnamefont
  {Zong}}, \bibinfo {author} {\bibfnamefont {B.~R.}\ \bibnamefont {Nebgen}},
  \bibinfo {author} {\bibfnamefont {S.-C.}\ \bibnamefont {Lin}}, \bibinfo
  {author} {\bibfnamefont {J.~A.}\ \bibnamefont {Spies}},\ and\ \bibinfo
  {author} {\bibfnamefont {M.}~\bibnamefont {Zuerch}},\ }\bibfield  {title}
  {\bibinfo {title} {Emerging ultrafast techniques for studying quantum
  materials},\ }\href {https://doi.org/10.1038/s41578-022-00530-0} {\bibfield
  {journal} {\bibinfo  {journal} {Nat. Rev. Mater.}\ }\textbf {\bibinfo
  {volume} {8}},\ \bibinfo {pages} {224} (\bibinfo {year} {2023})}\BibitemShut
  {NoStop}%
\bibitem [{\citenamefont {Luttinger}(1956)}]{Luttinger}%
  \BibitemOpen
  \bibfield  {author} {\bibinfo {author} {\bibfnamefont {J.~M.}\ \bibnamefont
  {Luttinger}},\ }\bibfield  {title} {\bibinfo {title} {Quantum theory of
  cyclotron resonance in semiconductors: General theory},\ }\href
  {https://doi.org/10.1103/PhysRev.102.1030} {\bibfield  {journal} {\bibinfo
  {journal} {Phys. Rev.}\ }\textbf {\bibinfo {volume} {102}},\ \bibinfo {pages}
  {1030} (\bibinfo {year} {1956})}\BibitemShut {NoStop}%
\bibitem [{\citenamefont {Baldereschi}\ and\ \citenamefont
  {Lipari}(1973)}]{spherical}%
  \BibitemOpen
  \bibfield  {author} {\bibinfo {author} {\bibfnamefont {A.}~\bibnamefont
  {Baldereschi}}\ and\ \bibinfo {author} {\bibfnamefont {N.~O.}\ \bibnamefont
  {Lipari}},\ }\bibfield  {title} {\bibinfo {title} {Spherical model of shallow
  acceptor states in semiconductors},\ }\href
  {https://doi.org/10.1103/PhysRevB.8.2697} {\bibfield  {journal} {\bibinfo
  {journal} {Phys. Rev. B}\ }\textbf {\bibinfo {volume} {8}},\ \bibinfo {pages}
  {2697} (\bibinfo {year} {1973})}\BibitemShut {NoStop}%
\bibitem [{\citenamefont {van Kesteren}\ \emph {et~al.}(1990)\citenamefont {van
  Kesteren}, \citenamefont {Cosman}, \citenamefont {van~der Poel},\ and\
  \citenamefont {Foxon}}]{jxjy0}%
  \BibitemOpen
  \bibfield  {author} {\bibinfo {author} {\bibfnamefont {H.~W.}\ \bibnamefont
  {van Kesteren}}, \bibinfo {author} {\bibfnamefont {E.~C.}\ \bibnamefont
  {Cosman}}, \bibinfo {author} {\bibfnamefont {W.~A. J.~A.}\ \bibnamefont
  {van~der Poel}},\ and\ \bibinfo {author} {\bibfnamefont {C.~T.}\ \bibnamefont
  {Foxon}},\ }\bibfield  {title} {\bibinfo {title} {Fine structure of excitons
  in type-ii $\text{GaAs}$/$\text{AlAs}$ quantum wells},\ }\href
  {https://doi.org/10.1103/PhysRevB.41.5283} {\bibfield  {journal} {\bibinfo
  {journal} {Phys. Rev. B}\ }\textbf {\bibinfo {volume} {41}},\ \bibinfo
  {pages} {5283} (\bibinfo {year} {1990})}\BibitemShut {NoStop}%
\bibitem [{\citenamefont {Kobe}\ and\ \citenamefont {Yang}(1985)}]{gauge}%
  \BibitemOpen
  \bibfield  {author} {\bibinfo {author} {\bibfnamefont {D.~H.}\ \bibnamefont
  {Kobe}}\ and\ \bibinfo {author} {\bibfnamefont {K.-H.}\ \bibnamefont
  {Yang}},\ }\bibfield  {title} {\bibinfo {title} {Gauge transformation of the
  time-evolution operator},\ }\href {https://doi.org/10.1103/PhysRevA.32.952}
  {\bibfield  {journal} {\bibinfo  {journal} {Phys. Rev. A}\ }\textbf {\bibinfo
  {volume} {32}},\ \bibinfo {pages} {952} (\bibinfo {year} {1985})}\BibitemShut
  {NoStop}%
\bibitem [{\citenamefont {Borhani}\ \emph {et~al.}(2006)\citenamefont
  {Borhani}, \citenamefont {Golovach},\ and\ \citenamefont
  {Loss}}]{spin-decay}%
  \BibitemOpen
  \bibfield  {author} {\bibinfo {author} {\bibfnamefont {M.}~\bibnamefont
  {Borhani}}, \bibinfo {author} {\bibfnamefont {V.~N.}\ \bibnamefont
  {Golovach}},\ and\ \bibinfo {author} {\bibfnamefont {D.}~\bibnamefont
  {Loss}},\ }\bibfield  {title} {\bibinfo {title} {Spin decay in a quantum dot
  coupled to a quantum point contact},\ }\href
  {https://doi.org/10.1103/PhysRevB.73.155311} {\bibfield  {journal} {\bibinfo
  {journal} {Phys. Rev. B}\ }\textbf {\bibinfo {volume} {73}},\ \bibinfo
  {pages} {155311} (\bibinfo {year} {2006})}\BibitemShut {NoStop}%
\bibitem [{\citenamefont {F-Fernández}\ \emph {et~al.}(2023)\citenamefont
  {F-Fernández}, \citenamefont {P-Cortés}, \citenamefont {Liñán},\ and\
  \citenamefont {Platero}}]{Fernandez-Fernandez_2023}%
  \BibitemOpen
  \bibfield  {author} {\bibinfo {author} {\bibfnamefont {D.}~\bibnamefont
  {F-Fernández}}, \bibinfo {author} {\bibfnamefont {J.}~\bibnamefont
  {P-Cortés}}, \bibinfo {author} {\bibfnamefont {S.~V.}\ \bibnamefont
  {Liñán}},\ and\ \bibinfo {author} {\bibfnamefont {G.}~\bibnamefont
  {Platero}},\ }\bibfield  {title} {\bibinfo {title} {Photo-assisted spin
  transport in double quantum dots with spin–orbit interaction},\ }\href
  {https://doi.org/10.1088/2515-7639/acd1b7} {\bibfield  {journal} {\bibinfo
  {journal} {Journal of Physics: Materials}\ }\textbf {\bibinfo {volume} {6}},\
  \bibinfo {pages} {034004} (\bibinfo {year} {2023})}\BibitemShut {NoStop}%
\bibitem [{\citenamefont {Moriya}\ \emph {et~al.}(2014)\citenamefont {Moriya},
  \citenamefont {Sawano}, \citenamefont {Hoshi}, \citenamefont {Masubuchi},
  \citenamefont {Shiraki}, \citenamefont {Wild}, \citenamefont {Neumann},
  \citenamefont {Abstreiter}, \citenamefont {Bougeard}, \citenamefont {Koga},\
  and\ \citenamefont {Machida}}]{cubic-rashba-paper}%
  \BibitemOpen
  \bibfield  {author} {\bibinfo {author} {\bibfnamefont {R.}~\bibnamefont
  {Moriya}}, \bibinfo {author} {\bibfnamefont {K.}~\bibnamefont {Sawano}},
  \bibinfo {author} {\bibfnamefont {Y.}~\bibnamefont {Hoshi}}, \bibinfo
  {author} {\bibfnamefont {S.}~\bibnamefont {Masubuchi}}, \bibinfo {author}
  {\bibfnamefont {Y.}~\bibnamefont {Shiraki}}, \bibinfo {author} {\bibfnamefont
  {A.}~\bibnamefont {Wild}}, \bibinfo {author} {\bibfnamefont {C.}~\bibnamefont
  {Neumann}}, \bibinfo {author} {\bibfnamefont {G.}~\bibnamefont {Abstreiter}},
  \bibinfo {author} {\bibfnamefont {D.}~\bibnamefont {Bougeard}}, \bibinfo
  {author} {\bibfnamefont {T.}~\bibnamefont {Koga}},\ and\ \bibinfo {author}
  {\bibfnamefont {T.}~\bibnamefont {Machida}},\ }\bibfield  {title} {\bibinfo
  {title} {Cubic $\text{Rashba}$ spin-orbit interaction of a two-dimensional
  hole gas in a strained-$\mathrm{Ge}/\mathrm{SiGe}$ quantum well},\ }\href
  {https://doi.org/10.1103/PhysRevLett.113.086601} {\bibfield  {journal}
  {\bibinfo  {journal} {Phys. Rev. Lett.}\ }\textbf {\bibinfo {volume} {113}},\
  \bibinfo {pages} {086601} (\bibinfo {year} {2014})}\BibitemShut {NoStop}%
\end{thebibliography}%

\end{document}